\DeclareUnicodeCharacter{2032}{\ensuremath{^{\prime}}}
\documentclass[prd,preprintnumbers,superscriptaddress,nofootinbib,twocolumn]{revtex4-2}
\pdfoutput=1
\usepackage[utf8]{inputenc}
\DeclareUnicodeCharacter{2212}{\textminus}
\usepackage{newunicodechar}
\newunicodechar{η}{\eta}
\usepackage{hyperref}
\usepackage{cancel}
\usepackage{textcomp}
\usepackage[utf8]{inputenc}
\usepackage{graphicx}
\usepackage{enumitem}
\usepackage{booktabs}
\usepackage{makecell}
\usepackage{diagbox}  
\usepackage{graphicx} 
\usepackage{array}    
\usepackage{multirow}
\usepackage{slashed}
\usepackage{tikz-feynman}
\tikzfeynmanset{compat=1.1.0}
\usepackage{orcidlink}
\usepackage{xcolor}
\usepackage{booktabs}
\usepackage{multirow}
\usepackage{float}
\usepackage{comment} 
\usepackage{amsmath, amsfonts, amssymb}
\usepackage{bm, bbm}
\usepackage{lipsum}
\usepackage[caption=false]{subfig}
\usepackage{slashed}
\usepackage{tikz-feynman}
\usepackage[normalem]{ulem}
\usepackage{subfloat}
\usepackage{pifont}

\newcommand{\xmark}{\ding{55}}

\newcommand{\beq}{\begin{equation}\begin{aligned}{}}
\newcommand{\eeq}{\end{aligned}\end{equation}}
\newcommand{\beqa}[1]{\begin{equation}\begin{aligned}{#1}}
\newcommand{\eeqa}{\end{aligned}\end{equation}}

\newcommand{\bea}{\begin{eqnarray}{}}
\newcommand{\eea}{\end{eqnarray}}

\definecolor{rgegray}{RGB}{90,100,110}
\definecolor{lightblue}{RGB}{0,80,160}
\definecolor{MintBg}{HTML}{E5F5E0}
\definecolor{DarkEmerald}{HTML}{1B7837}
\definecolor{darkolive}{RGB}{85,107,47}
\definecolor{huntergreen}{RGB}{53,94,59}
\usepackage{tikz}
\usetikzlibrary{decorations.pathmorphing,arrows.meta,positioning}
\tikzfeynmanset{ with arrow/.style = {
   decoration={
     markings,
     mark=at position 0.5
          with {\arrow[xshift=1mm]{Triangle[black,width=0.8mm,length=1.2mm]}}
     },
   postaction=decorate}
}
\tikzset{
  gluon/.style={decorate, decoration={coil,aspect=0.5,amplitude=2.5pt,segment length=4pt}, thick},
  quark/.style={-{Latex[length=2mm]}, thick},
  axion/.style={dashed, thick},
  vertex/.style={fill=gray!70, draw=black, rectangle, minimum size=4pt, inner sep=0pt},
}
\usetikzlibrary{decorations.pathmorphing}

\tikzset{
	insertion/.pic={kg
		\draw[fill=white, line width=0.8pt] (0,0) circle[radius=0.18];
		\draw[line width=0.8pt] (-0.13,-0.13) -- (0.13,0.13);
		\draw[line width=0.8pt] (-0.13,0.13) -- (0.13,-0.13);
	}
}
\usepackage{hyperref}
\hypersetup{colorlinks=true,citecolor=blue,linkcolor=blue,urlcolor=blue}
\newcommand{\nn}{\nonumber}

\newcommand{\al}{\alpha}
\newcommand{\be}{\beta}
\newcommand{\ga}{\gamma}

\begin{document}
%%%%%%%%%%%%%%%%%%%%%%%%%%%%%%%%
\title{Electric Dipole Moments as Precision Probes of CP-Violating Top-Quark Interactions in SMEFT}

\author{Subhajit Kala\,\orcidlink{/0000-0001-5915-0212}}
\email{s.kala@iitg.ac.in}
\affiliation{%
 Department of Physics, Indian Institute of Technology Guwahati,\\
 North Guwahati, Assam-781039, India.
}%

\author{Soumitra Nandi\,\orcidlink{/0000-0001-6567-0302}}
\email{soumitra.nandi@iitg.ac.in}
\affiliation{%
 Department of Physics, Indian Institute of Technology Guwahati,\\
 North Guwahati, Assam-781039, India.
}%

\author{Sonakshi Saha\,\orcidlink{0009-0000-6640-7792}}
\email{sonakshi.saha@iitg.ac.in}
\affiliation{%
 Department of Physics, Indian Institute of Technology Guwahati,\\
 North Guwahati, Assam-781039, India.
}%

\begin{abstract}
Electric dipole moments (EDMs) provide some of the most sensitive probes of CP-violating interactions beyond the Standard Model (SM). Within the Standard Model Effective Field Theory (SMEFT), we investigate constraints on CP-violating top-quark interactions using current and projected leptonic and hadronic EDM measurements. Our analysis incorporates the complete chain of relevant effects, including one- and two-loop matching, Barr-Zee contributions, Weinberg-operator and heavy-quark threshold corrections, and renormalization-group induced operator mixing. We demonstrate the strong complementarity between leptonic and hadronic EDM observables and show that future proton EDM searches could significantly strengthen constraints on several CP-violating top-quark interactions.
\end{abstract}


\maketitle
%-----------------------------
\section{Introduction}
%----------------------------

The observed matter-antimatter asymmetry of the Universe remains one of the most profound unsolved puzzles in particle physics and cosmology. To dynamically generate a net baryon asymmetry through Baryogenesis, CP-violating processes are fundamentally required, as dictated by the Sakharov conditions \cite{Sakharov:1967dj}. While the Standard Model (SM) accommodates $CP$ violation through the irreducible complex phase of the Cabibbo-Kobayashi-Maskawa (CKM) mixing matrix \cite{Kobayashi:1973fv}, however this source quantitatively is insufficient by many orders of magnitude to describe the observed cosmological abundance. This mismatch provides a compelling indication that additional sources of $CP$ violation may exist beyond the SM. Consequently, identifying and constraining all experimentally accessible sources of $CP$ violation is an important component of the search for physics beyond the Standard Model (BSM).

A particularly compelling possibility is that new sources of $CP$ violation reside in interactions involving the heaviest SM particles. In this regard, the top quark, owing to its large mass and intimate connection to electroweak symmetry breaking, constitutes a prime probe of physics beyond the SM. New physics (NP) can generate $CP$-violating interactions of the top quark with gauge bosons, the Higgs boson, and other fermions, encoded in the $CP$-odd components of effective couplings. Such effects are conveniently described within the SMEFT \cite{Grzadkowski:2010es}, where heavy new-physics effects are parameterized by higher-dimensional operators. 
% Consequently, precision probes of $CP$-violating top-quark interactions, particularly through Electric Dipole Moment (EDM) observables, offer a powerful and model-independent means of uncovering sources of $CP$ violation beyond the CKM mechanism.
Consequently, searching for such interactions can therefore provide an important avenue for uncovering new sources of $CP$ violation beyond the CKM mechanism. 

To probe these elusive phases, low-energy precision observables offer a powerful and complementary alternative. In particular, Electric Dipole Moments (EDMs) of leptons, nucleons, and atoms provide an exceptionally clean avenue to probe $CP$-violating interactions. The current experimental sensitivity to leptons and neutron EDMs are,
%-------------------------------
\begin{align}
    |d_e|&< 4.1 \times 10^{-30}~e\,\text{cm}~~(90\%~\text{CL})~\text{\cite{Roussy:2022cmp}}\,,\nn\\
    |d_{\mu}|&<1.8 \times 10^{-19}~e\,\text{cm}~~(95\%~\text{CL})~\text{\cite{Muong-2:2008ebm}}\,,\nn\\
    |d_{\tau}|&= (-0.62 \pm 0.63) \times 10^{-17} \mathrm{e.cm}\,\text{\cite{Belle:2021ybo}}\,,\\
    |d_{n}|&<1.8\times 10^{-26} \,\text{cm}~~(90\%~\text{CL})~\text{\cite{Abel:2020pzs}}\,.\nn    
\end{align}
%-------------------------------
These bounds will be tightened further in the near future. However, in the SM, EDMs vanish at the one- and two- loop level and only receive negligible contributions at multi-loop order ($4$- loop level in the lepton sector). In particular, the SM predictions of electron and neutron EDMs are \cite{Pospelov:2013sca, Booth:1993af, Czarnecki:1997bu},
%------------------------------
\begin{align}
    d_e&\sim 10^{-41} e\,\text{cm}\,,\nn\\
    d_n&\sim 10^{-32} e\,\text{cm}\,.
\end{align}
%------------------------------
The resulting tiny SM background makes EDMs particularly sensitive probes of new sources of $CP$ violation, allowing potential $CP$-violating phases induced by NP to be identified and constrained with high precision.

The purpose of the present work is to systematically constrain the $CP$-violating phase of various non-Hermitian top-quark interactions by calculating their virtual contributions to experimentally available, low-energy EDM observables.
In the literature, phenomenological studies have explored the sensitivity of low-energy EDM observables to selected classes of $CP$-violating top-quark interactions. In particular, the impact of flavor-conserving $CP$-violating top-dipole and top-Higgs Yukawa interactions on EDM observables has been investigated in Refs.~\cite{Panico:2018hal,Aebischer:2021uvt}. The sensitivity of neutron-EDM measurements to a set of dimension-six SMEFT operators, restricted to one-loop effects, was studied in Refs.~\cite{Kley:2021yhn,Ardu:2025rqy}. Constraints on flavor-changing top-Higgs and top-dipole couplings from neutron-EDM measurements have also been analyzed in Refs.~\cite{Gorbahn:2014sha,Kala:2026xzo}. While these studies have provided valuable insights into specific classes of top-quark interactions, a systematic framework incorporating the full range of EDM observables and the complete chain of low-energy $CP$-violating operators induced by top-quark interactions remains lacking.

In this work, we perform a comprehensive analysis of $CP$-violating top-quark interactions by exploiting a broad set of complementary leptonic, hadronic, atomic, molecular, and nuclear EDM observables. At the low-energy level, anomalous top-quark interactions generate a variety of $CP$-odd operators, including the electron, muon, and tau EDMs ($d_e$, $d_\mu$, $d_\tau$), the light- and heavy-quark EDMs ($d_u$, $d_d$, $d_s$, $d_c$, $d_b$), the corresponding quark chromoelectric dipole moments (CEDMs) ($\tilde d_u$, $\tilde d_d$, $\tilde d_s$, $\tilde d_c$, $\tilde d_b$), and the purely gluonic Weinberg three-gluon operator ($w$). These effective operators induce a rich set of experimentally accessible observables, including the neutron and proton EDMs ($d_n$, $d_p$), the CP-odd pion-nucleon couplings ($g_0$, $g_1$), diamagnetic atomic EDMs of $^{199}$Hg, $^{129}$Xe, $^{171}$Yb, and $^{225}$Ra, as well as light-nuclear EDMs such as the deuteron and helion EDMs. On the leptonic side, measurements in paramagnetic molecules such as ThO, HfF$^+$, and YbF provide exceptional sensitivity to the electron EDM. By combining all these complementary observables, our analysis simultaneously probes electromagnetic dipole moments, chromoelectric dipole moments, and purely gluonic sources of $CP$ violation generated by anomalous top-quark interactions. We consider flavor-conserving neutral-current as well as flavor-violating charged- and neutral-current interactions.

To consistently connect these precision observables to the underlying SMEFT operators, we incorporate the complete sequence of relevant quantum effects, including one- and two-loop matching contributions, finite Barr-Zee corrections, Weinberg-operator contributions, heavy-quark threshold effects, renormalization-group evolution (RGE), and operator mixing. This global approach allows us to fully exploit the strong complementarity among leptonic, hadronic, atomic, and nuclear EDM measurements and provides one of the most comprehensive probes of the $CP$-violating top-quark sector currently achievable through indirect precision experiments.

The paper is organized as follows. In Sec.~\ref{sec:electric_dipole_moments},
we review the current status of EDM experiments and discuss the relevant
observables associated with EDM measurements. Sec.~\ref{sec:Effective_framework}
presents our theoretical framework, including the operator basis and the renormalisation group evolution
formalism connecting the EDM observables to the effective operators.
In Sec.~\ref{sec:results}, we discuss the impact of the different EDM
inputs on the resulting constraints on the operator basis. 
% In Sec .~\ref {sec:dim8_matching}, we derived the sensitivity of EDM observables on dimension $8$ anomalous top SMEFT Wilson Coefficients (WCs). 
We summarize
our main findings in Sec.~\ref{sec:EDM_summary} and Sec.~\ref{sec:dim8_matching}, and conclude in
Sec.~\ref{sec:conclusion}.

%---------------------------
\section{Electric dipole moments}\label{sec:electric_dipole_moments}
%---------------------------
\subsection*{Non-Relativistic Description and Discrete Symmetries}
In the non-relativistic limit, the interaction of a spin-$1/2$ particle with external, static electric ($\mathbf{E}$) and magnetic ($\mathbf{B}$) fields can be described by the effective Hamiltonian
%---------------
\begin{align} \label{eq:non_rel_hamiltonian}
\mathcal{H}_{\rm NR} = - \mu \, \mathbf{S} \cdot \mathbf{B} - d \, \mathbf{S} \cdot \mathbf{E},
\end{align}
%-----------------
where $\mathbf{S}$ denotes the spin vector of the particle, while $\mu$ and $d$ are its magnetic dipole moment (MDM) and EDM, respectively. The transformation properties of the MDM and EDM under discrete symmetries can be inferred directly from Eq.~\eqref{eq:non_rel_hamiltonian}. The scalar product $\mathbf{S}\cdot\mathbf{B}$ is invariant under both parity ($P$) and time reversal ($T$), since $\mathbf{S}$ and $\mathbf{B}$ transform in the same way under these symmetries. Consequently, the MDM interaction is both $P$- and $T$-even. In contrast, $\mathbf{S}\cdot\mathbf{E}$ changes sign under both $P$ and $T$, since the electric field $\mathbf{E}$ is a polar vector, whereas the spin $\mathbf{S}$ is an axial vector. Therefore, a non-zero EDM interaction violates both $P$ and $T$. Assuming $CPT$ invariance, as required for a local, Lorentz-invariant quantum field theory, $T$ violation implies $CP$ violation. Thus, the observation of a non-zero EDM would provide a direct signal of $CP$ violation beyond the SM.  

\subsection*{Relativistic Formulation}

To seamlessly connect these low-energy observables to high-energy top quark interactions, we must elevate this description to a relativistic framework. For a spin-$1/2$ Dirac fermion $\psi$ with mass $m$, the interactions with the electromagnetic field tensor $F_{\mu\nu}$ can be parametrised via dimension-five effective operators. The effective Lagrangian containing the dipole interactions is given by:
\begin{equation}
    \mathcal{L}_{\text{dipole}} = - \frac{\mu}{2} \bar{\psi} \sigma^{\mu\nu} \psi F_{\mu\nu} - i\frac{ d}{2} \bar{\psi} \sigma^{\mu\nu} \gamma_5 \psi F_{\mu\nu},
    \label{eq:rel_lagrangian}
\end{equation}
where $\sigma^{\mu\nu} = \frac{i}{2}[\gamma^\mu, \gamma^\nu]$. 

The first term represents the $CP$-conserving magnetic dipole interaction, which yields the standard MDM. In the SM, the minimal coupling of a fermion to the electromagnetic field via the Dirac equation inherently predicts a tree-level gyromagnetic ratio of $g=2$. However, quantum loop corrections introduce deviations from this value, known as the anomalous magnetic moment, $a_f = (g-2)/2$. The $CP$-conserving dimension-five operator in Eq.~(\ref{eq:rel_lagrangian}) explicitly captures this anomalous contribution. Therefore, the phenomenological coupling $\mu$ can be expressed in terms of the fermion's fractional charge $Q_f$, fundamental charge $e$, and mass $m_f$ as:
\begin{align}
    \mu = a_f \left( \frac{e Q_f}{2 m_f} \right).
    \label{eq:mdm_structural}
\end{align}
The second term in Eq.~(\ref{eq:rel_lagrangian}), containing the $\gamma_5$ matrix, introduces an additional factor of $i$ to ensure the Hermiticity of the Lagrangian. This term represents the $CP$-violating electric dipole interaction. By analogy to the anomalous magnetic moment, the EDM $d$ is generated purely by quantum loop corrections involving non-Hermitian couplings. 

%----------------------------
\subsection{Leptonic EDMs and Experimental Status}\label{subsec:leptonic_EDMs}
%----------------------------
To interface the phenomenological EDM definition with modern effective field theory frameworks, it is standard practice to express the dipole interactions in a chiral basis. The general dipole Lagrangian can be written in terms of a complex dimension five weak effective coupling (WET) $L_{\ell\gamma}$:
\begin{equation}
    \mathcal{L}_{\text{dipole}} = - \frac{1}{2} \frac{L^{\ell\gamma}_{pp}}{\Lambda} \bar{\ell}_{L_p} \sigma^{\mu\nu} \ell_{R_p} F_{\mu\nu} - \frac{1}{2} \frac{L^{\ell\gamma~*}_{pp}}{\Lambda} \bar{\ell}_{R_p} \sigma^{\mu\nu} \ell_{L_p} F_{\mu\nu}.
    \label{eq:chiral_dipole}
\end{equation}
Here the indices $p$ represents three generations of leptons.
Comparing Eq.~(\ref{eq:chiral_dipole}) with Eq.~(\ref{eq:rel_lagrangian}), we can readily identify the physical dipole moments as the real and imaginary components of this effective coupling. 
%----------------------------
\begin{align}\label{eq:lepton_EDM_matching}
    a_{\ell}&=\frac{2 m_{\ell}}{e Q_{\ell}}\frac{1}{\Lambda}\mathrm{Re}[L^{\ell \gamma}_{pp}]\,,& d_{\ell}&=\frac{1}{\Lambda}\mathrm{Im}[L^{\ell \gamma}_{pp}] \,.
\end{align}
%----------------------------
This chiral matching is particularly advantageous when evaluating quantum loop corrections from NP, as it directly relates the $CP$-violating phases of high-energy operators to the low-energy observable $d_\ell$.

\vspace{0.5em}
\noindent \textbullet~ \textbf{\textit{Electron EDM $(d_e)$:}} Among the leptons, the electron EDM provides the most stringent constraints on BSM $CP$ violation. While direct measurements of free electrons are practically challenging, modern experiments leverage heavy polar molecules. In these molecules, relativistic effects create an enormous internal effective electric field, often on the order of tens of GV/cm, which heavily amplifies the energy shift induced by a non-zero $d_e$. 

Historically, the ACME collaboration utilised a cryogenic beam of Thorium Monoxide (ThO) molecules to set a rigorous upper bound of,
%----------------------------------
\begin{align}\label{eq:ACME_bound}
    |d_e| < 1.1 \times 10^{-29} \, e \cdot \text{cm} \quad (90\% \text{ C.L.})~\text{\cite{ACME:2018yjb}}.
\end{align}
%---------------------------------
More recently, the JILA collaboration employed trapped Hafnium Fluoride ions ($\text{HfF}^+$) to push this frontier even further, establishing the current world-leading limit of:
%--------------------------
\begin{align}\label{eq:jila_bound}
    |d_e| < 4.1 \times 10^{-30} \, e \cdot \text{cm} \quad (90\% \text{ C.L.})\text{~\cite{Roussy:2022cmp}}.
\end{align}
%-------------------------

Looking ahead, next-generation experimental efforts are expected to probe this observable with significantly improved precision. In particular, the ACME~III experiment aims to reach a sensitivity at the level of
\begin{align}
|d_e| &< 10^{-31}\,e\cdot\mathrm{cm} \text{~\cite{ACMEIII:DAMOP2026}},
\end{align}
corresponding to an order-of-magnitude improvement over the current JILA Collaboration limit given in Eq.~\eqref{eq:jila_bound}.

\vspace{0.5em}
\noindent \underline{\textbf{\textit{Molecular Precession Frequencies $(\omega)$:}}} In heavy paramagnetic polar molecule experiments ($\text{ThO}$, $\text{HfF}^+$, $\text{YbF}$), the $P,T$-violating frequency shift $\omega$ measured in these molecular experiments is not sensitive to $d_e$ alone, but to a linear combination
\begin{equation}
\omega = -E_{\rm eff}\,d_e + W_S\,C_S,
\end{equation}
where $C_S$ is the dimensionless coupling of a $CP$-violating scalar-pseudoscalar electron-nucleon contact interaction, and $E_{\rm eff}$, $W_S$ denote molecule-specific enhancement factors obtained from atomic and molecular structure calculations. Explicitly, the parameterised precession frequencies for the leading molecular systems are given by \cite{Chupp:2017rkp, Dekens:2018bci, Degenkolb:2024eve}:
\begin{widetext}
\begin{subequations}
\begin{align}
\omega_{\text{HfF}} &= \left[ (34.9 \pm 1.4) \left( \frac{d_e}{10^{-27} \, e\cdot\text{cm}} \right) + (32_{-2}^{+1}) \left( \frac{C_S}{10^{-7}} \right) \right] \text{mrad/s}, \label{eq:omega_HfF} \\
\omega_{\text{ThO}} &= -\left[ (121_{-39}^{+5}) \left( \frac{d_e}{10^{-27} \, e\cdot\text{cm}} \right) + (182_{-27}^{+42}) \left( \frac{C_S}{10^{-7}} \right) \right] \text{mrad/s}, \label{eq:omega_ThO} \\
\omega_{\text{YbF}} &= -\left[ (19.6 \pm 1.5) \left( \frac{d_e}{10^{-27} \, e\cdot\text{cm}} \right) + (17.6 \pm 2.0) \left( \frac{C_S}{10^{-7}} \right) \right] \text{mrad/s}. \label{eq:omega_YbF}
\end{align}
\end{subequations}
\end{widetext}
Since the ratio $E_{\rm eff}/W_S$ differs across species, combining measurements from ThO, YbF, and HfF$^+$ allows $d_e$ and $C_S$ to be extracted independently, making $C_S$ a distinct and complementary CP-violating observable alongside the electron EDM. The current most stringent limits on  $d_e$ and $C_S$ are set by the JILA collaboration using $\mathrm{HfF}^+$ and ThO measurements \cite{Roussy:2022cmp}. Assuming single-source dominance, the $90\%$ C.L. bounds are
\begin{align}
|d_e| &< 4.1\times10^{-30}\,e\cdot\text{cm} \quad (C_S = 0), \nn\\
|C_S| &< 4.5\times10^{-10} \quad (d_e = 0).
\end{align}
When $d_e$ and $C_S$ are instead varied simultaneously in a combined fit to resolve their mutual degeneracy, the corresponding $90\%$ C.L. limits weaken to
\begin{align}
|d_e| &< 2.1\times10^{-29}\,e\cdot\text{cm}, \nn\\
|C_S| &< 1.9\times10^{-9}.
\end{align}

The nucleon-level scalar coupling $C_S$ can be related to the coefficients of appropriate four-fermion operators in the WET through the corresponding matching relations, as discussed in Ref.~\cite{Kumar:2024yuu}. It is to be noted that in our analysis, we have not got any significant contributions in $C_S$. 
%------------------------------
%- Maybe I will write the expression of CS.
%------------------------------
% In this work, we investigate the contributions of anomalous top-quark interactions to these WET four-fermion operators through virtual corrections. These contributions provide an additional source of $CP$ violation that can be probed through $C_S$, and their impact on the EDM constraints is discussed in the subsequent sections.

%-------------------------
\vspace{0.5em}
\noindent \textbullet~ \textbf{\textit{Muon EDM $(d_{\mu})$:}}
%--------------------------
The current direct constraint on the muon EDM was obtained by the
Fermilab Muon $(g-2)$ Collaboration, which recently reported
\begin{align}
    |d_{\mu}| < 1.10 \times 10^{-19}\,e\cdot\mathrm{cm}
\end{align}
at 95\% C.L. \cite{Muong-2:2026zgd}. This result improves upon the
previous limit from the BNL-E-$0821$ experiment,
$|d_{\mu}| < 1.8\times10^{-19}\,e\cdot\mathrm{cm}$
\cite{Muong-2:2008ebm}. A new dedicated muon EDM experiment has recently been proposed at PSI, with projected sensitivities of
\begin{align}
    \sigma(d_{\mu}) &< 4 \times 10^{-21}\,e\cdot\mathrm{cm}
    &&\text{(Phase I)}, \nonumber\\
    \sigma(d_{\mu}) &< 6 \times 10^{-23}\,e\cdot\mathrm{cm}
    &&\text{(Phase II)}.
\end{align}
In addition to direct searches, the muon EDM can also be constrained indirectly through EDM measurements of other systems. In particular, analyses based on the Hg and ThO EDM experiments yield
\begin{align}
    |d_{\mu}(\mathrm{Hg})| &< 6 \times 10^{-20}\,e\cdot\mathrm{cm}, \nn\\
    |d_{\mu}(\mathrm{ThO})| &< 2 \times 10^{-20}\,e\cdot\mathrm{cm}.
\end{align}
respectively \cite{Ema:2021jds}. These indirect constraints are complementary to the direct measurement and provide additional sensitivity to possible CP-violating interactions involving the muon. 

%-------------------------
\vspace{0.5em}
\noindent \textbullet~ \textbf{\textit{Tau EDM $(d_{\tau})$:}}
%--------------------------
Tau, being the heaviest lepton, has an extremely short lifetime, tau lepton EDM cannot be constrained using static electromagnetic fields or a storage ring. Instead, bounds are extracted from $e^+ e^- \to \gamma^*\to  \tau^+\tau^-$ collision, such as those at the Belle collaboration $(\sqrt{s}\approx 10.58~\text{GeV})$. At these energies, the dipole interaction is governed by a momentum-dependent form factor, which devolop both the real and imaginary component. The real part, $\mathrm{Re}(d_{\tau})$ corresponds to the $CP$- odd static EDM. Current limits on the CP-violating electric dipole moment, $ d_{\tau} $ come from 833 $ \mathrm{fb}^{-1} $ of $ e^{+}e^{-} $ collision at Belle \cite{Belle:2021ybo}.
%---------------------------------
\begin{align}
    \mathrm{Re}(d_{\tau}) &= (-0.62 \pm 0.63) \times 10^{-17} \mathrm{e.cm} \nn\\
    \mathrm{Im}(d_{\tau}) &= (-0.40 \pm 0.32) \times 10^{-17} \mathrm{e.cm}
\end{align}
%-------------------------------
The electron EDM also provides an indirect constraint on the tau EDM
through the three-loop light-by-light contribution~ \cite{Grozin:2008nw},
\begin{align}
    |d_\tau| < 5.94\times10^{-19}\,e\cdot\mathrm{cm}
\end{align}
The above bound is obtained using the most recent experimental constraint on the electron EDM, $d_e$ \cite{Roussy:2022cmp}.

%----------------------------
\subsection{Hadronic and Atomic EDMs and Experimental Status}
%----------------------------
Unlike leptons, quarks can not be isolated due to colour confinement at low energies. Consequently the $CP$- violating phases of high-energy interactions in the hadronic sector must be probed indirectly through the EDMs of composite systems, primarily the nucleon and heavy diamagnetic atoms. Generally, the low-energy effective Lagrangian for the hadronic EDM can be expressed as \cite{Gorbahn:2014sha}
%--------------------------------------
\begin{align}\label{eq:lag_quark_EDM}
   \mathcal{L}_{\rm had}\supset& +d_q(\mu) \frac{i}{2}\bar{q}\sigma^{\mu\nu}\gamma_5 qF_{\mu\nu}+\tilde{d}_q(\mu)\frac{i}{2}g_s(\mu) \bar{q}\sigma^{\mu\nu}T^a\gamma_5 q G^a_{\mu\nu}\nn\\
   &+w(\mu)\frac{1}{3}G^a_{\mu\sigma}G^{b,\sigma}_{\nu}\tilde{G}^{c,\mu\nu}
\end{align}
%---------------------------------------
Here, $q$ runs over all quark flavors except the top quark. $F_{\mu\nu}$ and $G_{\mu\nu}^a$ denote the field strength tensors of QED and QCD, respectively, while $\tilde{G}^{a\,\mu\nu}=\tfrac{1}{2}\,\epsilon^{\mu\nu\alpha\beta}G^a_{\alpha\beta}$ is the dual QCD field strength tensor. The quantities $d_q(\mu)$, $\tilde{d}_q(\mu)$ and $\omega(\mu)$ represent the EDM, chromo-EDM (CEDM) of quarks and Weinberg operator, respectively. In terms of WET coefficient, $d_q$, $\tilde{d}_q$, and $w$ can be written as,
%---------------------------------
\begin{align}\label{eq:quark_EDM_matching}
    d_u&= \frac{2}{\Lambda}~\mathrm{Im}[L^{u\gamma}_{pp}] \,, &\quad \tilde{d}_u&=\frac{2}{g_s(\mu) \Lambda}\mathrm{Im}[L^{uG}_{pp}] \,, \nn\\
    d_d&= \frac{2}{\Lambda}~\mathrm{Im}[L^{d\gamma}_{pp}] \,, &\quad \tilde{d}_d&=\frac{2}{g_s(\mu) \Lambda}\mathrm{Im}[L^{dG}_{pp}] \,,\\
    w&=\frac{3}{\Lambda} L^{\tilde{G}}\,.\nn
\end{align}
%---------------------------------
The effective parameterization in terms of the WET coefficients are essential for utilizing the appropriate anomalous dimension matrices (ADMs) to perform the renormalization group (RG) running. This crucial step bridges the gap between the electroweak scale $(\mu_{\rm EW})$ and the hadronic scale $(\mu_{\rm had})$, where the physical EDM observables are evaluated. A detailed discussion of this matching and running procedure is provided in the subsequent sections.

%-------------------------
\vspace{0.5em}
\noindent \textbullet~ \textbf{\textit{Nucleon EDM $(d_{n}, d_p)$:}}
%--------------------------
Nucleon, being a composite state built from quarks and gluons, receives contributions from constituents' EDMs and CEDM. 
%----------------------------------------
\begin{subequations}
\begin{align}\label{eq:neutron_EDM}
    \frac{d_n}{e}&=\left(g_T^d\frac{d_u}{e}+g_T^u\frac{d_d}{e}+g_T^s\frac{d_s}{e}\right)+(1.1 \pm 0.55) \left(\tilde{d}_d+0.5 \tilde{d}_u\right)\,,\nonumber\\
    &+(22\pm10)\times 10^{-3}\mathrm{GeV}.w\,.
\end{align}

\begin{align}\label{eq:proton_EDM}
    \frac{d_p}{e}&=\left(g_T^u\frac{d_u}{e}+g_T^d\frac{d_d}{e}+g_T^s\frac{d_s}{e}\right)-(0.6 \pm 0.3) \left(\tilde{d}_u+0.5 \tilde{d}_d\right)\,,\nonumber\\
    &-(18 \pm 9)\times 10^{-3}\mathrm{GeV}\,w\,.
\end{align}
\end{subequations}
Here, the values of the hadronic charge currents at the renormalisation scale (Hadronic scale) $\mu_{\rm had}=2\,\mathrm{GeV}$ are~\cite{FlavourLatticeAveragingGroupFLAG:2024oxs}
\begin{align}
    g_T^u&=0.784(28)(10)\,,g_T^d=-0.204(11)(10)\,,\nn\\
    g_T^s&=-0.0027(16)\,.
\end{align}
The corresponding expression for the proton EDM, $d_p$, can be obtained from Eq.~(\ref{eq:neutron_EDM}) by applying isospin symmetry ($u \leftrightarrow d$). This exchange swaps the corresponding intrinsic quark EDMs and CEDMs. It is important to note, however, that standard QCD sum rule evaluations introduce an overall relative minus sign for the effective proton CEDM contribution, and Weinberg operator \cite{Pospelov:2005pr}.

The free neutron EDM is most tightly constrained by a measurement done using ultracold neutrons, which determined 
\begin{align}
    |d_n| <1.8\times 10^{-26} e\,\text{cm}~~(90\%~\text{CL})~\text{\cite{Abel:2020pzs}}
\end{align}

To push this frontier further, several experimental collaborations aim to improve this bound by a full order of magnitude. 
\begin{align}
     |d_n| < 10^{-27} e\,\text{cm}
\end{align}
These dedicated next-generation efforts include the n2EDM experiment at the PSI \cite{n2EDM:2021yah}, PanEDM at the Institut Laue-Langevin (ILL) \cite{Wurm:2019yfj}, TUCAN EDM at TRIUMF \cite{TUCAN:2025rjm}, and the LANL nEDM experiment at Los Alamos National Laboratory \cite{Ito:2017ywc}.

Similarly, direct bounds on the proton EDM are expected to see revolutionary advancements in the near future. The proposed pEDM collaboration plans to employ the frozen-spin technique in a highly symmetric storage ring, potentially utilising existing infrastructure at Brookhaven National Laboratory (BNL) \cite{pEDM:2025nlu}. Building on the technological developments and experimental expertise gained from the successful Muon $(g-2)$ program, the Phase-I pEDM experiment aims to reach a sensitivity of
\begin{align}
    |d_p| &\sim 10^{-29} \, e\cdot\text{cm}.
\end{align}
This achievement would improve upon current indirect limits derived from the Mercury EDM ($d_{\text{Hg}}$) by approximately four orders of magnitude.

\vspace{0.5em}
\noindent \textbullet~ \textbf{\textit{Diamagnetic Atomic EDMs ($d_{\text{A}}$):}}
%---------------------------------
The EDMs of heavy diamagnetic atoms, such as $^{199}\text{Hg}$ , $^{129}\text{Xe}$ ,  $^{171}\text{Yb}$, and $^{225}\text{Ra}$ provide stringent complementary constraints on hadronic $CP$ violation. In such atoms, the electron EDM contributions to the atomic EDM is heavily screened by Schiff's theorem ~\cite{Schiff:1963zz}, so the dominant source of a nonzero atomic EDM is instead the nuclear Schiff moment. This moment is primarily induced by $CP$-odd pion-nucleon interactions ($\bar{g}_{\pi NN}$) and the intrinsic nucleon EDMs ($d_n, d_p$). The experimental upper bounds from different $d_A$ are,
\begin{align}
d_\text{Hg} &<7.4 \times 10^{-30} e\,\text{cm}~\text{\cite{Graner_2016}}\nn\\
    d_{\rm Xe}&< 4.8 \times 10^{-28} e\,\text{cm}~\text{\cite{Sachdeva:2019rkt, PhysRevA.100.022505}}\nn\,,\\
    d_{\rm Yb}&< 1.5 \times 10^{-26} e\,\text{cm}\,~\text{\cite{Zheng:2022jgr}},\\
    d_{\rm Ra}&<  1.4 \times 10^{-23} e\,\text{cm}~\text{\cite{Bishof:2016uqx}}\,.\nn
\end{align}

The expressions for the EDMs of various diamagnetic atoms $(d_A)$ in terms of the hadronic EDMs $(d_n,d_p)$, CP-odd pion--nucleon couplings $(\bar g_0,\bar g_1)$, four-nucleon coupings $(C_1,C_2)$ and semileptonic operator coefficients $(C_S^{(0)},C_S^{(1)})$ can be found in Refs.~\cite{Hubert:2022pnl, Engel:2013lsa, Dmitriev:2004fk, Dzuba:2009kn, Flambaum:2019kbn, Dzuba_2007, Dobaczewski:2018nim, Degenkolb:2024eve}. The resulting EDMs are,

% can be found in Ref.~\cite{Choi:2026tun}, based on the results compiled in Refs.~\cite{Hubert:2022pnl, Engel:2013lsa, Dmitriev:2004fk, Dzuba:2009kn, Flambaum:2019kbn, Dzuba_2007, Dobaczewski:2018nim, Degenkolb:2024eve}.
%-------------------------------
\begin{subequations}
\label{eq:nuclear_EDM}
\begin{align}
d_{\mathrm{Hg}}
={}&-2.26(23)\times10^{-4}
\Bigg[
0.6^{+1.33}_{-0.12}\,d_n
+0.06^{+0.20}_{-0.01}\,d_p
\nonumber\\
&\qquad
+\frac{g_A m_N}{f_\pi}
\left(
0.01^{+0.04}_{-0.005}\,\bar g_0
+0.02^{+0.07}_{-0.05}\,\bar g_1
\right)e\,\mathrm{fm}
\Bigg],
\end{align}

\begin{align}
d_{\mathrm{Xe}}
={}&3.62(25)\times10^{-5}
\Bigg[
0.63^{+0.16}_{-0.12}\,d_n
+0.14(3)\,d_p
\nonumber\\
&\qquad
+\frac{g_A m_N}{f_\pi}
\left(
-0.008^{+0.003}_{-0.042}\,\bar g_0
+0.006^{+0.044}_{-0.003}\,\bar g_1
\right)e\,\mathrm{fm}
\Bigg],
\end{align}

\begin{align}
d_{\mathrm{Yb}}
={}&-2.10^{+0.22}_{-0.00}\times10^{-4}
\Bigg[
0.54^{+0.13}_{-0.11}\,d_n
+0.054^{+0.016}_{-0.014}\,d_p
\nonumber\\
&\qquad
+\frac{g_A m_N}{f_\pi}
\left(
0.01^{+0.02}_{-0.00}\,\bar g_0
+0.02^{+0.034}_{-0.027}\,\bar g_1
\right)e\,\mathrm{fm}
\Bigg],
\end{align}

\begin{align}
d_{\mathrm{Ra}}
={}&-8.5^{+0.25}_{-0.30}\times10^{-4}
\Bigg[
0.63^{+0.16}_{-0.12}\,d_n
+0.14^{+0.04}_{-0.03}\,d_p
\nonumber\\
&\qquad
+\frac{g_A m_N}{f_\pi}
\left(
-0.2(6)\,\bar g_0
+5(3)\,\bar g_1
\right)e\,\mathrm{fm}
\nonumber\\
&\qquad
+m_N^3
\left(
-0.01(3)\,C_1
+0.03(2)\,C_2
\right)e\,\mathrm{fm}
\Bigg].
\end{align}
\end{subequations}
%-------------------------------
where $g_A=1.27$ is the nucleon axial-vector coupling, $f_\pi=92.2~\mathrm{MeV}$ is the pion decay constant, and $m_N \simeq 0.939~\mathrm{GeV}$ denotes the nucleon mass.
The CP-odd pion-nucleon couplings $\bar{g}_0$ and $ \bar{g}_1 $ can receive sizable contributions from purely hadronic UV sources $ \tilde{d}_q $ and $ w $ with $ q=u,d $ ~\cite{deVries:2021sxz}. The resulting expressions in Eq.~\ref{eq:pion_nucleon} are evaluated at the matching scale $ \mu = 1 \text{GeV} $ ~\cite{Choi:2026tun}.
%--------------------------
\begin{subequations}\label{eq:pion_nucleon}
\begin{align}
    \bar{g}_0 (\tilde{d}_q) &\simeq 2.2 (0.7)(\tilde{d}_u+\tilde{d}_d) \text{GeV}, \\
    \bar{g}_1 (\tilde{d}_q) &\simeq 38 (13)(\tilde{d}_u-\tilde{d}_d) \text{GeV}, \\
    \bar{g}_0 (w) &\simeq   9 \times 10^{-3} w . \text{GeV}^2, \\
    \bar{g}_1 (w) &\simeq  \pm (2.6 \pm 1.5) \times 10^{-3} w .\text{GeV}^2 .
 \end{align}
 \end{subequations}
%--------------------------

Currently, the strongest bound is obtained from the $d_\text{Hg} ~\text{\cite{Graner_2016}}$, which predicts
%-----------------------
\begin{align}
    |d_n| &< 1.6 \times 10^{-26}e\,\text{cm}\,, & |d_p|&<2.0 \times 10^{-25}e\,\text{cm}\,\nn\\
    \bar{g}_0& < 2.3 \times 10^{-12}\,, & \bar{g}_1& < 1.1 \times 10^{-12}\,.
\end{align}
%-----------------------

\vspace{0.5em}
\noindent \textbullet~ \textbf{\textit{Light Nuclei EDMs $(d_N^{\rm light})$:}}
%---------------------------------
The EDMs of charged light nuclei such as the proton, deuteron $(D)$, and helion $(^3\text{He}^{++})$ can be directly measured in storage ring experiments ~\cite{pEDM:2025nlu, CPEDM:2019nwp}. These can be parameterised in terms of hadronic EDM as ~\cite{CPEDM:2019nwp},
%-----------------------------------------
\begin{subequations}
    \begin{align}
        d_D &= 0.94(1) (d_n+d_p) +0.18(2) \bar{g}_1 e~\mathrm{fm}, \\
        d_{\mathrm{He}} &= 0.90(1) d_n -0.03(1) d_p +[ 0.11(1) \bar{g}_0  \\ \nn 
        & +0.14(2) \bar{g}_1 - (0.04(2)C_1 -0.09(2)C_2) \mathrm{fm}^{-3}] e~ \mathrm{fm}
    \end{align}
\end{subequations}

%------------------------------
\section{Effective Field Theory Framework}\label{sec:Effective_framework}
%------------------------------
Building upon the experimental landscape established in the previous section \ref{sec:electric_dipole_moments}, we now construct the theoretical framework necessary to evaluate the $CP$- violating phases of the top quark. Assuming that the NP resides at a scale $\Lambda$ significantly higher than the electroweak scale $(\Lambda \gg v)$, its low-energy phenomenological effects can be systematically parametrised using SMEFT operators.

By integrating out the heavy BSM degrees of freedom, the SM Lagrangian is extended by a tower of higher-dimensional operators constructed from the standard SM field content that strictly respect its local $\text{SU}(3)_C \times \text{SU}(2)_L \times \text{U}(1)_Y$ gauge symmetry. In this study, we truncate this expansion at dimension-six, yielding the effective Lagrangian:
\begin{equation}
    \mathcal{L}_{\text{SMEFT}} = \mathcal{L}_{\text{SM}} + \sum_{i} \frac{\mathcal{C}_i}{\Lambda^2} \mathcal{O}_i^{(6)},
\end{equation}
where $\mathcal{L}_{\rm SM}$ denotes the renormalisable dimension $4$ SM Lagrangian, $\mathcal{O}_i^{(6)}$ represent the dimension-six operators and $C_i$ are their corresponding dimensionless Wilson coefficients. We redefine the WCs as
\begin{align}
    C_i = \frac{\mathcal{C}_i}{\Lambda^2},
\end{align}
to absorb the explicit dependence on the scale $\Lambda$ into the coefficients. These rescaled coefficients has mass dimension $[\mathrm{TeV}^{-2}]$.

%--------------------------------
\subsection{The SMEFT operator basis}\label{subsec:SMEFT_op_basis}
%------------------------------------
To comprehensively explore the $CP$-violating landscape of the top sector, we systematically categorise all dimension-six SMEFT operators~\cite{Grzadkowski:2010es} that generate distinct top quark interactions and contribute to low-energy EDM observables. These include flavor-conserving neutral currents, flavor-violating charged and neutral currents, alongside relevant four-fermion and dipole interactions. The complete set of SMEFT operators of interest for our analysis is tabulated in Table~\ref{tab:SMEFT_Ops}.

%-----------------------------------------
\begin{table}[h!]
    \centering
    \renewcommand{\arraystretch}{1.3}
    \setlength{\tabcolsep}{3pt}
    \begin{tabular}{c c c}
        \toprule
        
        \textbf{Class} & \textbf{Operator} & \textbf{Definition} \\
        
        \midrule

        Dipole
        & $\mathcal{O}^{uG}_{pr}$
        & $(\bar q_p\sigma^{\mu\nu}T^A u_r)\tilde{\phi}G_{\mu\nu}^A$ \\

        &
        $\mathcal{O}^{uB}_{pr}$
        & $(\bar q_p\sigma^{\mu\nu}u_r)\tilde{\phi}B_{\mu\nu}$ \\

        &
        $\mathcal{O}^{uW}_{pr}$
        & $(\bar q_p\sigma^{\mu\nu}u_r)\tau^I\tilde{\phi}W_{\mu\nu}^I$ \\

        &
        $\mathcal{O}^{dW}_{pr}$
        & $(\bar q_p\sigma^{\mu\nu}d_r)\tau^I\phi W_{\mu\nu}^I$ \\

        \midrule

        Current
        & $\mathcal{O}^{\phi q(1)}_{pr}$
        & $(\phi^\dagger i\overleftrightarrow{D}_{\mu}\phi)
        (\bar q_p\gamma^\mu q_r)$ \\

        &
        $\mathcal{O}^{\phi q(3)}_{pr}$
        & $(\phi^\dagger i\overleftrightarrow{D}_{\mu}^{I}\phi)
        (\bar q_p\tau^I\gamma^\mu q_r)$ \\

        &
        $\mathcal{O}^{\phi u}_{pr}$
        & $(\phi^\dagger i\overleftrightarrow{D}_{\mu}\phi)
        (\bar u_p\gamma^\mu u_r)$ \\

        &
        $\mathcal{O}^{\phi ud}_{pr}$
        & $i(\tilde{\phi}^{\dagger}D_{\mu}\phi)
        (\bar u_p\gamma^\mu d_r)$ \\

        \midrule

        Yukawa
        & $\mathcal{O}^{u\phi}_{pr}$
        & $(\phi^\dagger\phi)(\bar q_pu_r\tilde{\phi})$ \\

        \midrule

       Four-fermion
        & $\mathcal{O}^{lequ~(1)}_{prst}$
        & $(\bar l_p^je_r)
        \epsilon_{jk}(\bar q_s^ku_t)$ \\
        & $\mathcal{O}^{lequ~(3)}_{prst}$
        & $(\bar l_p^j\sigma_{\mu\nu}e_r)
        \epsilon_{jk}(\bar q_s^k\sigma^{\mu\nu}u_t)$ \\

        &
        $\mathcal{O}^{quqd~(1)}_{prst}$
        & $(\bar q_p^j u_r)
        \epsilon_{jk}(\bar q_s^k d_t)$ \\

        &
        $\mathcal{O}^{quqd~(8)}_{prst}$
        & $(\bar q_p^j T^A u_r)
        \epsilon_{jk}(\bar q_s^k T^A d_t)$ \\

        &
        $\mathcal{O}^{qu~(1)}_{prst}$
        & $(\bar q_p\gamma_\mu q_r)
        (\bar u_s\gamma^\mu u_t)$ \\

        &
        $\mathcal{O}^{qu~(8)}_{prst}$
        & $(\bar q_p\gamma_\mu T^A q_r)
        (\bar u_s\gamma^\mu T^A u_t)$ \\

        \bottomrule
    \end{tabular}
    \caption{Dimension-six SMEFT operators considered in this work,
    classified according to the type of top-quark interaction they
    generate. Here, $\tau^I$ and $T^A$ denote the generators of
    $\mathrm{SU}(2)_L$ and $\mathrm{SU}(3)_C$, respectively. The indices $p,r$ represent different flavor generations.}
    \label{tab:SMEFT_Ops}
\end{table}
%--------------------------------------
Since our primary focus lies in NP interactions involving the top quark, we systematically restrict the fermion generation indices, denoted by $p$ and $r$, within our operator basis. For flavor-conserving neutral current interactions, these indices are strictly fixed to the third generation ($p = r = 3$). To describe FCNC processes, the indices are set to $p, r = (3, i)$ or $(i, 3)$, where $i \in \{1, 2\}$, capturing transitions between the top quark and the lighter up or charm quarks. Finally, for charged-current processes, the top quark can couple to any down-type quark, allowing the relevant generation index to span all three generations ($i = 1, 2, 3$).

%--------------------------------
\vspace{0.5em}
\noindent \underline{\textbf{\textit{Transition from the flavor to the mass basis:}}}
%------------------------------------
All calculations in this work are performed in the phase of spontaneously broken electroweak symmetry, and it is necessary to go from the flavor/ gauge basis to mass basis of the fermions, in order to deal with propagating degrees of freedom.  The fermion mass matrices are diagonalized through bi-unitary transformation acting in generation space,
%-----------------------------
\begin{align}
    f_{L(R)}^{\prime\,i}=U_{L(R)} ^i f_{L(R)}^i\,,
\end{align}
%-----------------------------
with (un)primed fields in the (mass) gauge basis and $i$ denotes any of the fermion flavors and $U_{L(R)}^i$ are unitary matrices for each fermion species. At the order of precision considered in this work, it is sufficient to retain only the SM contributions to the fermion rotation matrix and neglect their higher order corrections (dimension six corrections). The corresponding unitary transformations can be absorbed into redefinitions of the SMEFT WCs as shown in Table~\ref{tab:WC_basis_conversion}.
%-----------------------------------------
\begin{table}[h!]
    \centering
    \renewcommand{\arraystretch}{1.3}
    \setlength{\tabcolsep}{6pt}
    \begin{tabular}{c c}
        \toprule
        \textbf{Mass-basis WC} & \textbf{Flavor-basis WC} \\
        \midrule

        $C^{uG}$
        & $(U_L^u)^\dagger C^{uG\prime} U_R^u$ \\

        $C^{uB}$
        & $(U_L^u)^\dagger C^{uB\prime} U_R^u$ \\[3pt]

        \multirow{2}{*}{$C^{uW}$}
        & $(U_L^u)^\dagger C^{uW\prime} U_R^u$ \\
        & $(U_L^d)^{\dagger}C^{uW\prime} U_R^u$\\[3pt]

        % $C^{dW}$
        % & $(U_L^d)^\dagger C^{dW\prime} U_R^d$ \\
        $C^{dW}$ & $(U_L^u)^\dagger C^{dW\prime} U_R^d$\\

        \midrule

        $C^{\phi q(1)}$
        & $(U_L^u)^\dagger C^{\phi q(1)\prime} U_L^u$ \\[3pt]

        \multirow{2}{*}{$C^{\phi q(3)}$}
        & $(U_L^u)^\dagger C^{\phi q(3)\prime} U_L^u$ \\
        & $(U_L^u)^\dagger C^{\phi q(3)\prime} U_L^d$ \\[3pt]

        $C^{\phi u}$
        & $(U_R^u)^\dagger C^{\phi u~\prime} U_R^u$ \\

        $C^{\phi ud}$
        & $(U_R^u)^\dagger C^{\phi ud ~\prime} U_R^d$ \\

        \midrule

        $C^{u\phi}$
        & $(U_L^u)^\dagger C^{u\phi~\prime} U_R^u$ \\

        \midrule

        $C^{lequ(1,3)}_{abcd}$
&
$ \delta_{ia} \delta_{jb} (U_L^u)^\dagger_{ck} (U_R^u)_{ld} ~C^{lequ(1,3)\prime}_{ijkl} $\\

$C^{quqd(1,8)}_{abcd}$ & $(U^d_L)^{\dagger}_{ai} (U_R^u)_{jb} (U_L^u)^{\dagger}_{ck} (U_R^d)_{ld} ~C^{quqd(1,8)~\prime}_{ijkl}$\\

$C^{qu~(1,8)}_{abcd}$ & $(U_L^u)^{\dagger}_{ai} (U_L^u)_{jb} (U_R^u)^{\dagger}_{ck} (U_R^u)_{ld} ~C_{ijkl}^{qu(1,8)~\prime}$ \\

        \bottomrule
    \end{tabular}
    \caption{Transformation of the dimension-six SMEFT WCs from the flavor basis to the quark mass basis. Here $a,b,c,d$ and $i,j,k,l$ represent different matrix components.}
    \label{tab:WC_basis_conversion}
\end{table}
%--------------------------------------------

%---------------------------------
Because the up- and down-type quark Yukawa matrices require different
unitary transformations to be diagonalized, there is no unique choice
of flavor basis in which all quark rotation matrices can be absorbed
simultaneously into the SMEFT WCs. This is a consequence
of the fact that the SMEFT is formulated in the unbroken phase, where
the left-handed quarks form complete $\mathrm{SU}(2)_L$ doublets and a
$\mathrm{U}(3)_q$ flavor transformation acts on the doublet as a whole.
Consequently, one may choose either the up-type or down-type rotation
to be absorbed into the definition of the SMEFT WCs.
Since we assume that the NP is predominantly coupled to the
top quark, we adopt the \textit{up-aligned} basis. In this basis, the
up-type quark fields are aligned with their mass eigenstates, while the
CKM matrix appears in the down-type sector, thereby minimizing the
explicit CKM dependence in the relevant top-quark interactions. The corresponding transformations between the flavor and mass
eigenstate bases are
\begin{align}
    u_{L(R)}' &= U_{L(R)}^u\,u_{L(R)}, \qquad
    d_{L(R)}' = U_{L(R)}^d\,d_{L(R)},
    \nn\\
    V_{\rm CKM}&=\left(U_{L}^u\right)^{\dagger}U_{L}^d.
\end{align}
In the \textit{up-aligned} basis, we choose
$U_{L/R}^u=\mathbf{1}$ and $U_{L}^d=V_{\rm CKM}$, such that
\begin{align}
    q_L' =
    \begin{pmatrix}
        u_L\\
        V_{\rm CKM}d_L
    \end{pmatrix}.
\end{align}
%----------------------------
% For the operators $\mathcal{O}_{\phi q}^{(1,3)}$, the left-handed quark
% doublet contains both up- and down-type quarks. Hence, the corresponding
% WCs in the two sectors are related to the flavor-basis
% coefficients as
% \begin{align}
%     C_{\phi q}^{u}
%     &= (U_L^u)^\dagger C_{\phi q}^{\prime} U_L^u
%      = C_{\phi q}^{\prime},
%     \nn\\
%     C_{\phi q}^{d}
%     &= (U_L^d)^\dagger C_{\phi q}^{\prime} U_L^d
%      = V_{\rm CKM}^{\dagger} C_{\phi q}^{\prime} V_{\rm CKM},
% \end{align}
% as indicated in Table~\ref{tab:WC_basis_conversion}.

It is important to note that operators involving the $\mathrm{SU}(2)_L$ quark doublets, can generate both up- and down-type interactions through the same flavor-basis WCs. In Table~\ref{tab:WC_basis_conversion}, we therefore display only the corresponding flavor rotations relevant for the interactions that contribute to the EDM observables considered in our analysis.

%---------------------------------------
\subsection{Renormalization group equations}\label{subsec:RGE}
\begin{figure*}[t]
    \centering
    \includegraphics[width=\textwidth]{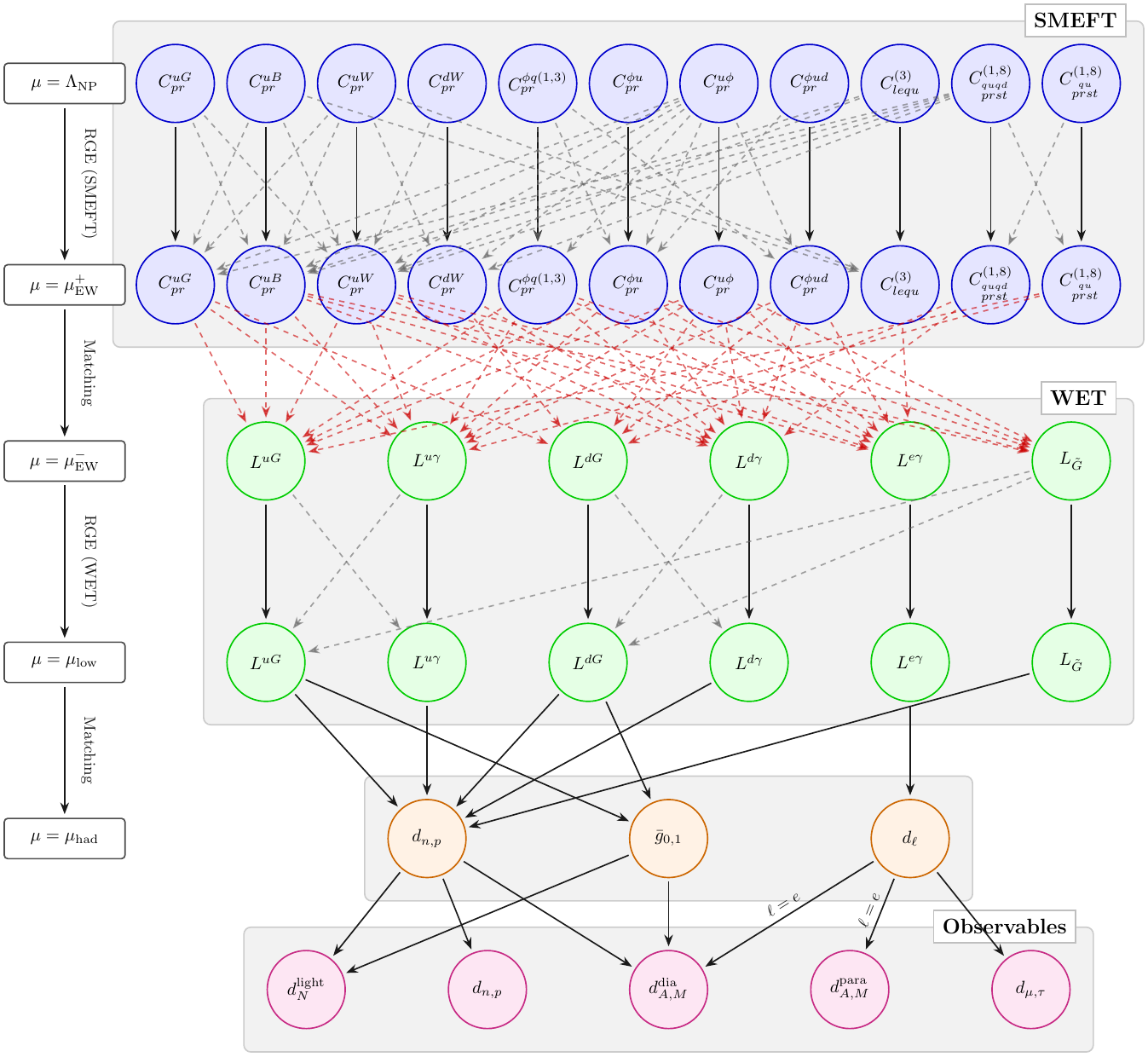}
    \caption{Flowchart illustrating the RGE evolution and matching steps that
connect the SMEFT WCs at the NP scale $(\Lambda)$ to the
low-energy EDM observables at the hadronic scale $(\mu_{\rm had})$.}
    \label{fig:RGE_flowchart}
\end{figure*}

%---------------------------------------
With a knowledge of the effective operators at the new-physics scale $\mu=\Lambda_{\rm NP}$, we now discuss the renormalization-group evolution and matching procedures required to connect them to low-energy EDM observables. The complete multi-scale framework, summarized in Fig.~\ref{fig:RGE_flowchart}, involves evolving the SMEFT operators from $\mu_\Lambda$ to the electroweak scale, matching onto the appropriate low-energy effective theory after integrating out the heavy SM fields, and subsequently evolving the resulting operators to hadronic scales. Throughout this process, operator mixing and threshold corrections are consistently incorporated, ultimately generating the leptonic and hadronic dipole operators relevant for EDM phenomenology.

% \begin{widetext}
%--------------------------------
%--------------------------------------- 
% \end{widetext}

The complete evolution from the NP scale down to the observable scale proceeds through the following stages:

%--------------------------------
\vspace{0.5em}
\noindent \textbullet ~\textit{\textbf{SMEFT Running ($\mu = \Lambda_{\text{NP}}$ to $\mu = \mu_{\text{EW}}^+$):}}
%--------------------------------
We evolve the SMEFT WCs $C_i$ from the NP scale $\Lambda_{\text{NP}}$ down to the electroweak scale $\mu_{\text{EW}}=M_Z$ at leading-logarithmic (LL) accuracy. The running is governed by the anomalous dimension matrix of the SMEFT operators, which we take from the literature ~\cite{Jenkins:2013zja,Jenkins:2013wua,Alonso:2013hga}. The renormalization-group equation governing the running of the WCs is given by
%--------------------------------
\begin{align}
    \frac{d}{d\ln\mu}\,\vec{C}(\mu)
    =
    \frac{1}{16\pi^2}\,\gamma(\mu)\,\vec{C}(\mu)\,,
\end{align}
%--------------------------------
where $\gamma$ denotes the anomalous dimension matrix. At LL accuracy, the WCs at two different scales are related by
%---------------------------------
   \begin{align}\label{eq:RGE}
   C_i(\mu)=&\left(1 +\frac{\gamma_{ii}}{16\pi^2}\log\left(\frac{\mu}{\Lambda}\right)\right)C_i(\Lambda)\,,\nn\\
   &+ \sum_{i\neq j}\frac{\gamma_{ij}}{16\pi^2}\log\left(\frac{\mu}{\Lambda}\right) C_j(\Lambda)\,.
   \end{align} 
%---------------------------------
In our analysis, we choose the reference NP scale to be $\Lambda_{\rm NP}=1~\text{TeV}$. The RGE evolution of the SMEFT WCs
from $\mu_{\rm NP}$ to the electroweak scale is presented for the flavor-conserving $\vec{C}_{33}$ sector in Tables ~\ref{tab:rge_matrix_bosonic} and \ref{tab:rge_matrix_4f}. The corresponding RGE evolution matrices for the top-FCNC WCs $\vec{C}_{3i}$ can be obtained in the same manner using Eq.~\ref{eq:RGE},
employing the SMEFT anomalous dimension matrices
given in the references cited earlier.
%--------------------------------
\begin{table}[t!]
    \renewcommand{\arraystretch}{1.5} 
    \centering
    
    \resizebox{\columnwidth}{!}{%
    \begin{tabular}{|c|c|c|c|c|c|c|c|c|c|}
        \hline
        \diagbox[width=5em]{$C(\mu_{\text{EW}})$}{$C(\Lambda)$} 
        & $C^{uG}_{33}$ 
        & $C^{uB}_{33}$ 
        & $C^{uW}_{33}$ 
        & $C^{dW}_{33}$ 
        & $C^{\phi q(1)}_{33}$ 
        & $C^{\phi q(3)}_{33}$ 
        & $C^{\phi u}_{33}$ 
        & $C^{u\phi}_{33}$ 
        & $C^{\phi ud}_{33}$ \\ \hline
        
        $C^{uG}_{33}$          
        & 1.036 & -0.019 & -0.062 & 0.0 
        & 0.0 & 0.0 & 0.0 & 0.0 & 0.0 \\ \hline
        
        $C^{uB}_{33}$          
        & -0.025 & 0.870 & 0.003 & 0.0 
        & 0.0 & 0.0 & 0.0 & 0.0 & 0.0 \\ \hline
        
        $C^{uW}_{33}$          
        & -0.028 & 0.001 & 0.899 & 0.0 
        & 0.0 & 0.0 & 0.0 & 0.0 & 0.0 \\ \hline
        
        $C^{dW}_{33}$          
        & 0.0 & 0.0 & 0.0 & 0.899 
        & 0.0 & 0.0 & 0.0 & 0.0 & 0.0 \\ \hline
        
        $C^{\phi q(1)}_{33}$   
        & 0.0 & 0.0 & 0.0 & 0.0 
        & 0.894 & 0.094 & 0.007 & 0.0 & 0.0 \\ \hline
        
        $C^{\phi q(3)}_{33}$   
        & 0.0 & 0.0 & 0.0 & 0.0 
        & 0.031 & 0.934 & 0.0 & 0.0 & 0.0 \\ \hline
        
        $C^{\phi u}_{33}$      
        & 0.0 & 0.0 & 0.0 & 0.0 
        & 0.015 & 0.003 & 0.851 & 0.0 & 0.0 \\ \hline
        
        $C^{u\phi}_{33}$       
        & 0.371 & 0.027 & 0.006 & 0.001 
        & 0.017 & 0.089 & -0.008 & 0.865 & -0.002 \\ \hline
        
        $C^{\phi ud}_{33}$     
        & 0.0 & 0.0 & 0.0 & 0.0 
        & 0.0 & 0.0 & 0.0 & 0.0 & 0.904 \\ \hline
        
    \end{tabular}%
    }
    \vspace{0.5em}
    \caption{RGE sub-block for the SMEFT dipole and Higgs-fermion WCs $\vec{C}_{33}$. The columns represent the initial high-energy scale coefficients at $\Lambda$, and the rows represent the evolved coefficients at the electroweak scale $\mu_{\text{EW}}$.}
    \label{tab:rge_matrix_bosonic}
\end{table}
%--------------------------------

\begin{table}[t!]
    \renewcommand{\arraystretch}{1.5} 
    \centering
    
    \resizebox{\columnwidth}{!}{%
    \begin{tabular}{|c|c|c|c|c|c|c|}
        \hline
        \diagbox[width=5em]{$C(\mu_{\text{EW}})$}{$C(\Lambda)$} 
        & $C^{lequ(1)}_{1133}$ & $C^{lequ(3)}_{1133}$ & $C^{quqd(1)}_{1331}$ & $C^{quqd(8)}_{1331}$ 
        & $C^{qu(1)}_{1331}$ & $C^{qu(8)}_{1331}$ \\ \hline
        
        $C^{lequ(1)}_{1133}$   & 1.07 & 0.0 & 0.0 & 0.0 & 0.0 & 0.0 \\ \hline
        $C^{lequ(3)}_{1133}$   & -0.004 & 0.953 & 0.0 & 0.0 & 0.0 & 0.0 \\ \hline
        $C^{quqd(1)}_{1331}$   & 0.0 & 0.0 & 1.364 & -0.029 & 0.0 & 0.0 \\ \hline
        $C^{quqd(8)}_{1331}$   & 0.0 & 0.0 & -0.192 & 0.904 & 0.0 & 0.0 \\ \hline
        $C^{qu(1)}_{1331}$     & 0.0 & 0.0 & 0.0 & 0.0 & 0.982 & 0.044 \\ \hline
        $C^{qu(8)}_{1331}$     & 0.0 & 0.0 & 0.0 & 0.0 & 0.198 & 1.215 \\ \hline
    \end{tabular}%
    }
    \vspace{0.5em}
    \caption{RGE sub-block for the SMEFT four-fermion WCs.}
    \label{tab:rge_matrix_4f}
\end{table}
%-----------------------------

%--------------------------------
\vspace{0.5em}
\noindent \textbullet ~\textit{\textbf{Electroweak Matching ($\mu = \mu_{\text{EW}}^+$ to $\mu = \mu_{\text{EW}}^-$):}}
%---------------------------------
At the electroweak scale $(\mu_{\rm EW})$, the heavy degrees of freedom
of the SM, namely the top quark, Higgs boson, and $W^\pm$ and $Z$ bosons,
are integrated out. At this threshold, the SMEFT operators considered
in our analysis are matched onto the corresponding WET/LEFT operators 
relevant for the EDM observables. We perform the matching, including loop-level contributions up to two
loops. The loop orders at which the different SMEFT WCs contribute to
the relevant EDM observables are partially summarised in
Table~\ref{tab:loop_order} and illustrated schematically in
Fig.~\ref{fig:RGE_flowchart}. We also investigate several
multi-operator matching scenarios, which are discussed in the
corresponding sections. The explicit matching relations are presented
in detail in the subsequent sections.

%--------------------------------
\vspace{0.5em}
\noindent \textbullet ~\textit{\textbf{WET Running ($\mu = \mu_{\text{EW}}^-$ to $\mu = \mu_{\text{low}}$):}}
%------------------------------
Below the electroweak threshold $(\mu<\mu_{\rm EW})$, the RGE is performed within the WET. The relevant gauge symmetry is reduced to $\mathrm{SU}(3)_C\times\mathrm{U}(1)_{\rm em}$. In this low-energy regime, the RGE running is predominantly driven by QCD. As the energy scale decreases towards the hadronic scale $\mu_{\rm had}\sim 2~\text{GeV}$, the strong coupling constant increase significantly, making the QCD corrections increasingly severe. The observables considered in this work are probed at the hadronic scale, where WET dipole operators involving the light fermions $(u,d,s,e)$ contribute directly to the corresponding EDMs. However, WET operators involving heavier fermions, namely the charm, bottom, muon, and tau, can also contribute indirectly to $d_n$, $d_e$, and diamagnetic atomic EDMs through renormalisation-group evolution and threshold corrections generated when these heavy degrees of freedom are integrated out. In the following, we discuss the RGE of these different classes of WET operators separately.

%----------------------------
\vspace{0.5em}
\textit{\textbf{Running of Light-Fermionic Dipole operators:}}
\vspace{0.5em}

%----------------------------
The complete one-loop ADMs of WET operators have been calculated in Ref~\cite{Jenkins:2017dyc}. The running of dipole operators $L^{q\gamma}_{pp}$, $L^{qG}_{pp}$, $L^{e\gamma}$ and Weinberg operator $L^{\tilde{G}}$ are tabulated in the Table~\ref{tab:rge_matrix_light_WET}.
% %------------------------
% \begin{table}[htb!]
%     \renewcommand{\arraystretch}{1.3} 
%     \centering
%     \begin{tabular}{|c|c|c|c|c|c|c|}
%         \hline
%         \diagbox[width=6em]{$C(\mu_{\text{had}})$}{$C(\mu_\text{EW})$} 
%         & $L^{u\gamma}$ & $L^{uG}$ & $L^{d\gamma}$ & $L^{dG}$ & $L^{\tilde{G}}$& $L^{e\gamma}$ \\ \hline

%         $L^{u\gamma}$  & $0.882$ & $-0.062$ & $0.0$ & $0.0$ & $0.0$ & $0.0$ \\ \hline

%         $L^{uG}$       & $-0.046$ & $1.208$ & $0.0$ & $0.0$ & $0.0$ & $0.0$\\ \hline

%         $L^{d\gamma}$  & $0.0$ & $0.0$ & $0.889$ & $0.031$ & $0.0$ & $0.0$\\ \hline

%         $L^{dG}$       & $0.0$ & $0.0$ & $0.023$ & $1.209$ & $0.0$ & $0.0$\\ \hline

%         $L^{\tilde{G}}$ & $0.0$ & $0.0$ & $0.0$ & $0.0$ & $0.570$ & $0.0$\\ \hline

%         $L^{e\gamma}$  & $0.0$ &  $0.0$ & $0.0$ & $0.0$ & $0.0$ & $0.957$ \\
%         \hline
        
%     \end{tabular}
%     \vspace{0.5em}
%     \caption{RGE sub-block describing the evolution of the light-fermion dipole WCs between the electroweak $(\mu_{\rm EW})$ and hadronic scales ($2$~GeV). The dipole operators considered are $L^{u\gamma(G)}_{11}$, $L^{d\gamma(G)}_{11}$, $L^{d\gamma(G)}_{22}$, and $L^{e\gamma}_{11}$.}
%     \label{tab:rge_matrix_light_WET}
% \end{table}
% %----------------------

%------------------------
\begin{table}[htb!]
    \renewcommand{\arraystretch}{1.3} 
    \centering
    \begin{tabular}{|c|c|c|c|c|c|c|}
        \hline
        \diagbox[width=7em]{$C(\mu_{\text{had}})$}{$C(\mu_\text{EW})$} 
        & $L^{u\gamma}$ & $L^{uG}$ & $L^{d\gamma}$ & $L^{dG}$ & $L^{\tilde{G}}$& $L^{e\gamma}$ \\ \hline

        $L^{u\gamma}$  & $0.852$ & $-0.08$ & $0.0$ & $0.0$ & $0.0$ & $0.0$ \\ \hline

        $L^{uG}$       & $-0.061$ & $1.371$ & $0.0$ & $0.0$ & $0.0$ & $0.0$\\ \hline

        $L^{d\gamma}$  & $0.0$ & $0.0$ & $0.857$ & $0.004$ & $0.0$ & $0.0$\\ \hline

        $L^{dG}$       & $0.0$ & $0.0$ & $0.031$ & $1.371$ & $0.0$ & $0.0$\\ \hline

        $L^{\tilde{G}}$ & $0.0$ & $0.0$ & $0.0$ & $0.0$ & $0.523$ & $0.0$\\ \hline

        $L^{e\gamma}$  & $0.0$ &  $0.0$ & $0.0$ & $0.0$ & $0.0$ & $0.957$ \\
        \hline
        
    \end{tabular}
    \vspace{0.5em}
    \caption{RGE sub-block describing the evolution of the light-fermion dipole WCs between the electroweak $(\mu_{\rm EW})$ and hadronic scales ($2$~GeV). The dipole operators considered are $L^{u\gamma(G)}_{11}$, $L^{d\gamma(G)}_{11}$, $L^{d\gamma(G)}_{22}$, and $L^{e\gamma}_{11}$.}
    \label{tab:rge_matrix_light_WET}
\end{table}
%----------------------

% %------------------------
% \begin{table}[htb!]
%     \renewcommand{\arraystretch}{1.3} 
%     \centering
%     \begin{tabular}{|c|c|c|c|c|c|c|}
%         \hline
%         \diagbox[width=6em]{$C(\mu_{\text{had}})$}{$C(\mu_\text{EW})$} 
%         & $L^{u\gamma}$ & $L^{uG}$ & $L^{d\gamma}$ & $L^{dG}$ & $L^{\tilde{G}}$& $L^{e\gamma}$ \\ \hline

%         $L^{u\gamma}$  & $0.817$ & $-0.101$ & $0.0$ & $0.0$ & $0.0$ & $0.0$ \\ \hline

%         $L^{uG}$       & $-0.079$ & $1.503$ & $0.0$ & $0.0$ & $0.0$ & $0.0$\\ \hline

%         $L^{d\gamma}$  & $0.0$ & $0.0$ & $0.821$ & $0.05$ & $0.0$ & $0.0$\\ \hline

%         $L^{dG}$       & $0.0$ & $0.0$ & $0.04$ & $1.502$ & $0.0$ & $0.0$\\ \hline

%         $L^{\tilde{G}}$ & $0.0$ & $0.0$ & $0.0$ & $0.0$ & $0.434$ & $0.0$\\ \hline

%         $L^{e\gamma}$  & $0.0$ &  $0.0$ & $0.0$ & $0.0$ & $0.0$ & $0.957$ \\
%         \hline
        
%     \end{tabular}
%     \vspace{0.5em}
%     \caption{RGE sub-block describing the evolution of the light-fermion dipole WCs between the electroweak $(\mu_{\rm EW})$ and hadronic scales ($1$~GeV). The dipole operators considered are $L^{u\gamma(G)}_{11}$, $L^{d\gamma(G)}_{11}$, $L^{d\gamma(G)}_{22}$, and $L^{e\gamma}_{11}$.}
%     \label{tab:rge_matrix_light_WET}
% \end{table}
% %----------------------
It is important to note that the WET dipole operators can mix at one loop into four-quark operators of the form
$\mathcal{O}^{S(1,8)RR}_{uu}$,
$\mathcal{O}^{S(1,8)RR}_{dd}$, and
$\mathcal{O}^{S(1,8)RR}_{uddu}$,
as well as into the semileptonic tensor operators
$\mathcal{O}^{T,RR}_{eu}$ and $\mathcal{O}^{T,RR}_{ed}$,
and the leptonic operator $\mathcal{O}^{S,RR}_{ee}$.
These operators can subsequently contribute to the neutron EDMs through the CP-odd pion--nucleon couplings $(\bar g_0,\bar g_1)$. 

% In our analysis, however, most of the top-quark SMEFT operators considered here match onto the WET dipole operators only at the two-loop level. Including their one-loop mixing into the aforementioned four-fermion operators would therefore correspond to contributions of higher loop order. So, we neglect this additional operator mixing (four-fermion with light dipole operators) in the present analysis.

%----------------------------
\vspace{0.5em}
\textit{\textbf{Running and Threshold effects of Heavy-Fermionic Operators:}}
\vspace{0.5em}

%----------------------------
Since the ADMs are calculated in the unbroken gauge theory, they are mass independent and generation universal. We can employ the same anomalous dimension matrices to calculate the running of the heavy fermion from $\mu_{\rm EW}$ to the respective fermion mass threshold. At each threshold, the corresponding heavy fermion is integrated out, and the resulting threshold corrections are incorporated into the effective theory below that scale. 

 In particular, the charm- and bottom-quark chromomagnetic dipole couplings, $L^{uG}_{22}$ and $L^{dG}_{33}$, respectively, mixes with the Weinberg coupling $L^{\tilde{G}}$ after integrating out the charm and bottom quarks at their respective mass thresholds \cite{PhysRevLett.64.1709}. The corresponding relations are,
%---------------------------

\begin{subequations}\label{eq:threshold_corr}
    \begin{align}
     [L^{\tilde{G}}](\mu_b^-)&=[L^{\tilde{G}}](\mu_b^+)+\frac{1}{3 m_b}\frac{\alpha_s}{8\pi}\mathrm{Im}[L^{dG}_{33}](\mu_b^+)\nn\\ 
     &= [L^{\tilde{G}}](\mu_b^+)+7.1\times 10^{-4}~\mathrm{Im}[L^{dG}_{33}](\mu_b^+)\\
     [L^{\tilde{G}}](\mu_c^-)&=[L^{\tilde{G}}](\mu_c^+)+\frac{1}{3 m_c}\frac{\alpha_s}{8\pi}\mathrm{Im}[L^{uG}_{22}](\mu_c^+)\nn\\
     &=[L^{\tilde{G}}](\mu_c^+)+5.2 \times 10^{-3}~\mathrm{Im}[L^{uG}_{22}](\mu_c^+)
 \end{align}
\end{subequations}
%-----------------------------
It is worth noting that, in addition to the threshold correction at the charm-quark mass scale, the chromo-EDM of the charm quark, $\tilde d_c$, also induces contributions to the light-quark EDMs and CEDMs through LL RGE. In Ref.~\cite{Gorbahn:2014sha}, this effect was explicitly included in the evolution from the top-quark threshold to the hadronic scale.
%--------------------------------
\begin{align}\label{eq:charm_RGE_EDM}
	\frac{d_d(\mu_{\rm had})}{e} &= 2.3 \times 10^{-8}\,\tilde{d}_c(\mu_t) 
	+ 1.0 \times 10^{-4}~\mathrm{GeV}\cdot w(\mu_t)\,, \nonumber\\[4pt]
	\frac{d_u(\mu_{\rm had})}{e} &= -2.1 \times 10^{-8}\,\tilde{d}_c(\mu_t) 
	- 9.1 \times 10^{-5}~\mathrm{GeV}\cdot w(\mu_t)\,, \nonumber\\[4pt]
	\tilde{d}_d(\mu_{\rm had}) &= 1.8 \times 10^{-6}\,\tilde{d}_c(\mu_t) 
	+ 7.0 \times 10^{-4}~\mathrm{GeV}\cdot w(\mu_t)\,, \nn\\[4pt]
	\tilde{d}_u(\mu_{\rm had}) &= 8.2 \times 10^{-7}\,\tilde{d}_c(\mu_t) 
	+ 3.1 \times 10^{-4}~\mathrm{GeV}\cdot w(\mu_t)\,, \nonumber\\[4pt]
	w(\mu_{\rm had}) &= 1.7 \times 10^{-2}~\mathrm{GeV}^{-1}\,\tilde{d}_c(\mu_t) 
	+ 0.41\,w(\mu_t)\,.
 %    w(\mu_{\rm had}) &= 1.13 \times 10^{-2}~\mathrm{GeV}^{-1}\,\tilde{d}_c(\mu_t) 
	% + 0.60\,w(\mu_t)\,\nonumber\\[4pt]
 %    w(\mu_{\rm had}) &= 1.37 \times 10^{-3}~\mathrm{GeV}^{-1}\,\tilde{d}_b(\mu_t) 
	% + 0.62\,w(\mu_t)\,.
\end{align}
%--------------------------------

\vspace{0.5em}
\noindent \textbullet ~\textit{\textbf{Hadronic Matching and Observables ($\mu = \mu_{\text{had}}$):}}
\vspace{0.5em}
%----------------------------    

Finally, at the hadronic scale, the relevant WET WCs are matched onto
the low-energy hadronic observables. In particular, the light-quark
dipole coefficients $(L^{q\gamma},L^{qG})$, with
$q\in\{u,d,s\}$, and the lepton dipole coefficient $L^{e\gamma}$ are
matched onto the neutron and proton EDMs, $d_n$ and $d_p$, and the
lepton EDMs, $d_\ell$, respectively. The corresponding $CP$-violating
quark chromo-dipole and gluonic operators, $L^{qG}$ and
$L^{\tilde G}$, are matched onto the $CP$-odd pion--nucleon couplings
$\bar g_0$ and $\bar g_1$. The tree-level matching relations used in
our analysis are given in Eqs.~\eqref{eq:quark_EDM_matching},
\eqref{eq:lepton_EDM_matching}, and \eqref{eq:pion_nucleon}. In
addition to the light-fermion dipole operators, heavy-fermion dipole
operators $(L^{dG}_{33}, L^{uG}_{22})$ can also contribute to the low-energy hadronic EDMs through
threshold corrections and subsequent RGE, as discussed in
Eqs.~\eqref{eq:threshold_corr} and \eqref{eq:charm_RGE_EDM}. These low-energy quantities are then used to construct the EDM observables
considered in this work. 

The electron EDMs, $d_e$, are primarily probed
through paramagnetic atoms and molecules, while their contributions to
diamagnetic atomic EDMs are comparatively subdominant. For the heavier
leptons, such as the muon and tau, the EDMs are instead constrained
primarily through direct experimental searches, as discussed earlier in section~\ref{sec:electric_dipole_moments}.
The neutron and proton EDMs, $d_n$ and $d_p$, can be probed through direct
measurements, while diamagnetic atomic and light-nuclear EDMs provide
complementary indirect sensitivity to these hadronic EDMs. In addition,
the $CP$-odd pion--nucleon couplings $\bar g_0$ and $\bar g_1$ contribute
to both light-nuclear and diamagnetic atomic EDMs, making these systems
particularly sensitive to $CP$-violating interactions beyond the
individual nucleon EDMs. Consequently, combining direct nucleon EDM
measurements with diamagnetic atomic and light-nuclear EDMs provides
complementary sensitivity to $d_n$, $d_p$, $\bar g_0$, and $\bar g_1$.

%------------------------------
\section{Analysis and Results}\label{sec:results}
%------------------------------
In this section, we present the constraints on the $CP$-violating components
of the top SMEFT WCs from the current EDM measurements
discussed in section.~\ref{sec:electric_dipole_moments}. Depending on the flavor structure
and operator class, the operators defined in Table~\ref{tab:SMEFT_Ops} contribute to the
leptonic and hadronic EDMs at different loop orders, including one-loop, finite two-loop
Barr-Zee, and sequential one-loop RGE-induced contributions.

%----------------------------------
\begin{table}[t]
	\renewcommand{\arraystretch}{1.5}
	\footnotesize
	\centering
	\begin{tabular}{@{} c | c | c  @{\qquad} c | c | c  @{}}
		\toprule
		\multicolumn{1}{c}{\makecell{\textbf{WCs}\\ \textbf{$(p=r)$}}} &
		\multicolumn{1}{c}{\textbf{$d_\ell$}} &
		\multicolumn{1}{c}{\textbf{$d_{n,p}$}} &
		\multicolumn{1}{c}{\makecell{\textbf{WCs}\\ \textbf{$(p\neq r)$}}} &
		\multicolumn{1}{c}{\textbf{$d_\ell$}} &
		\multicolumn{1}{c}{\textbf{$d_{n,p}$}} \\
		\midrule
		
		$C^{uG}_{33}$ & \xmark & \textbf{1-loop}
		& $C^{uG}_{i3}, C^{uG}_{3i}$ & \xmark & 1-loop \\
		
		$C^{uB}_{33}$ & \textbf{2-loop} & 2-loop
		& $C^{uB}_{i3},C^{uB}_{3i}$ & 2-loop & 1-loop\\
		
		$C^{uW}_{33}$ & \textbf{2-loop} & 2-loop
		& $C^{uW}_{i3}, C^{uW}_{3i}$ & 2-loop & 1-loop\\
		
		$C^{dW}_{33}$ & \textbf{2-loop}  & 1-loop
		& $C^{dW}_{i3}$ & 2-loop & 1-loop \\
		
		$C^{\phi q(1)}_{33}$ & 2-loop & 2-loop
		& $C^{\phi q(1)}_{i3}, C^{\phi q(1)}_{3i}$ & 2-loop & 1-loop \\
		
		$C^{\phi q(3)}_{33}$ & 2-loop & 2-loop
		& $C^{\phi q(3)}_{i3},C^{\phi q(3)}_{3i}$ & 2-loop & 1-loop \\
		
		$C^{\phi u}_{33}$ & 2-loop & 2-loop
		& $C^{\phi u}_{i3},C^{\phi u}_{3i}$ & 2-loop & 1-loop \\
		
		$C^{u\phi}_{33}$ & \textbf{2-loop} & 2-loop
		& $C^{u\phi}_{i3},C^{u\phi}_{3i}$ & 2-loop & \textbf{1-loop}\\
		
		$C^{\phi ud}_{33}$ & 2-loop & 2-loop
		& $C^{\phi ud}_{i3}$ & 2-loop & \textbf{1-loop} \\
		
		$C^{lequ(1)}_{jj33}$ & \textbf{2-loop} & \xmark
		&  $C^{uB(W)}_{i3}, C^{\phi q(-)}_{3i}$& 2-loop & 1-loop  \\
		
		$C^{lequ(3)}_{jj33}$ & \textbf{1-loop} & \xmark
		& $C^{uB(W)}_{i3}, C^{\phi u}_{3i}$ & 2-loop & 1-loop \\
		
		$C^{quqd(1,8)}_{i33i}$ & \xmark & \textbf{1-loop}
		& $C^{uB}_{i3},C^{uW}_{3i}$&  2-loop & 1-loop  \\
		
		$C^{qu(1,8)}_{i33i}$ & \xmark & \textbf{1-loop} & $C^{u\phi}_{i3},C^{uG}_{3i}$ & \xmark & 2-loop \\
		& & & $C^{u\phi}_{i3} ,C^{uB(W)}_{3i}$ & 2-loop &2-loop\\
		& & & $C^{u\phi}_{i3} ,C^{\phi q(-)}_{3i}$ & 2-loop &2-loop\\
		& & & $C^{u\phi}_{i3}, C^{\phi u}_{3i}$ & 2-loop &2-loop\\
			\bottomrule
	\end{tabular}
	\caption{Loop-level contributions of the $CP$-violating top-quark WCs
		to leptonic and hadronic EDMs for flavor-conserving $(p=r=3)$ and
		flavor-violating $(p\neq r)$ interactions.}
	\label{tab:loop_order}
\end{table}
%----------------------------------
%----------------------------------
For each operator class, we identify the leading contributions to EDM observables, determine the loop order at which they first appear, and evaluate their numerical impact on the electron, neutron, proton, and other relevant EDM measurements.
We summarize the various scenarios and their corresponding contributions to leptonic and hadronic EDM observables in Table~\ref{tab:loop_order}. The table systematically classifies the leading contributions according to their perturbative order, distinguishing between one- and two-loop effects for both flavor-conserving and flavor-violating interactions. It includes contributions arising from both single and double operator insertions. Scenarios involving a single WC correspond to single-insertion contributions, whereas entries involving two WCs represent double-insertion effects generated through operator mixing and sequential matching.

Table~\ref{tab:loop_order} also highlights the novelty of the present work. The boldfaced entries denote operator contributions for which explicit EDM calculations already exist in the literature, namely the one-loop contribution of $C^{uG}_{33}$ to hadronic EDMs, the two-loop leptonic-EDM contributions induced by $C^{uB}_{33}$, $C^{uW}_{33}$, $C^{dW}_{33}$, and $C^{u\phi}_{33}$, the one-loop leptonic-EDM contribution from $C^{lequ(3)}_{jj33}$, the two-loop leptonic-EDM contribution from $C^{lequ(1)}_{jj33}$, the one-loop hadronic-EDM contributions generated by $C^{quqd(1,8)}_{i33i}$ and $C^{qu(1,8)}_{i33i}$, as well as the one-loop hadronic-EDM effects of the flavor-violating operators $C^{u\phi}_{i3},C^{\phi ud}_{i3}$. Apart from these isolated cases, all remaining entries in table~\ref{tab:loop_order} represent new calculations performed in this work. In particular, we determine for the first time the leading EDM contributions associated with a broad class of flavor-violating top-quark operators, including dipole, Higgs-current, and Yukawa interactions, and establish the loop order at which they enter leptonic and hadronic EDM observables. Our results therefore provide the first systematic and essentially complete mapping between the full set of CP-violating top-quark SMEFT interactions and low-energy EDM measurements, substantially extending the existing EDM program beyond the limited set of operator contributions previously available in the literature. 

It is important to emphasize that some operators contribute to EDM observables through sequential one-loop effects induced by renormalization-group evolution, as discussed in Sec.~\ref{subsec:sequential_two_one_loop}. In such cases, the EDM contribution is not generated directly through matching onto the dipole operators, but rather through a sequence of operator mixings followed by subsequent threshold corrections. A notable example is the semileptonic tensor operator $\mathcal{O}_{lequ}^{(1)}$. This operator does not generate the leptonic dipole operator through direct one-loop matching. However, under RGE evolution it mixes into the tensor operator $\mathcal{O}_{lequ}^{(3)}$, which subsequently induces the lepton dipole operator through one-loop matching at a lower scale. As a result, $\mathcal{O}_{lequ}^{(1)}$ contributes to the leptonic EDM through a sequential two-step mechanism corresponding effectively to a two-loop effect. Similar RGE-induced pathways can play an important role for several operator classes and must therefore be accounted for in a complete analysis of EDM constraints within the SMEFT framework.

In the following subsections, we analyze the contributions of various single- and multi-operator scenarios to the leptonic and hadronic dipole moments introduced above. Particular emphasis is placed on identifying the dominant mechanisms, the interplay between direct matching and RGE-induced effects, and the resulting constraints on CP-violating top-quark interactions.

We first discuss the contributions to leptonic EDMs and the corresponding constraints, followed by those arising from hadronic EDMs. The results are organized according to the flavor structure of the underlying $CP$-violating top-quark interactions, since the dominant matching contributions, operator-mixing patterns, and EDM sensitivities can differ significantly between flavor-conserving and flavor-violating operators. Using the EFT framework developed in Sec.~\ref{sec:Effective_framework}, we evolve the effective interactions from the new-physics scale $\mu_\Lambda$ to the hadronic scale $\mu_{\rm had}$, consistently incorporating renormalization-group evolution, matching effects, threshold corrections, and sequential operator-mixing contributions, which can compete with or even dominate over direct matching effects. This procedure establishes the connection between high-scale $CP$-violating top-quark interactions and low-energy EDM observables. The analytical matching expressions employed in our analysis are collected in Appendix~\ref{Append:loop_matching}, while representative Feynman diagrams are presented in the relevant sections.

Using the current experimental limits on leptonic and hadronic EDMs, we derive constraints on the relevant WCs at the new-physics scale. We further translate the projected sensitivities of future EDM experiments into expected bounds on the corresponding SMEFT interactions, thereby quantifying the potential of next-generation EDM searches to probe previously unexplored regions of parameter space. The resulting constraints provide a comprehensive assessment of the indirect sensitivity of EDM observables to both flavor-conserving and flavor-violating sources of $CP$ violation in the top-quark sector and allow us to identify the operator structures that can be most effectively tested by future precision measurements.

%---------------------------
\subsection{Contributions to and Constraints from Leptonic EDMs}\label{subsec:Leptonic_EDMs}
%--------------------------
We first consider the constraints on the $CP$-violating top-quark
interactions obtained from leptonic EDM measurements. The electron EDM
provides the strongest sensitivity due to its stringent experimental
bound,
$|d_e|<4.1\times10^{-30}\,e\cdot\mathrm{cm}$,
with the projected sensitivity of the ACME~III experiment expected to
improve upon the current limit by approximately one order of magnitude.
The muon and tau EDMs provide complementary probes of the corresponding
flavor structures. We organize the analysis according to the flavor
structure of the underlying top-quark interactions, beginning with
flavor-conserving couplings, followed by flavor-violating charged- and
neutral-current interactions. The matching relations are presented in a
flavor-general form in Appendix~\ref{Append:loop_matching}, such that the corresponding bounds from the muon
and tau EDMs can be readily reproduced by the appropriate replacements
of the lepton flavor and the associated light-fermion Yukawas.

%---------------------------------
\subsubsection{Flavor-conserving top-quark interactions}\label{subsec:leptonic_FC}
\begin{figure}[t!]
    \centering
    %--------------Diag-1----------------
    \subfloat[]{\label{dl_FCa}% 
        \resizebox{0.48\linewidth}{!}{%
            \begin{tikzpicture}
                \begin{feynman}
                \def\r{0.6 cm}
                \coordinate (O) at (0,0);
                \draw[very thick] (O) circle (\r);
                \vertex (A) at (90:\r);  
                \draw[fermion, very thick] (360:\r) arc (360:180:\r);
                \vertex [](B) at (210:\r);   
                \vertex [red, square dot](C) at (330:\r){};   
                \vertex[above=1.0 of A](F){\(\gamma\)};
                \vertex[below left=1.0cm and 0.5cm of B](G);
                \vertex[below right=1.0cm and 0.5cm of C](H);
                \vertex[left=0.7 cm of G](b){\(\ell\)};
                \vertex[right=0.7 cm of H](d){\(\ell\)};
                \vertex[below=0.6 cm of O](a1);
                \vertex[above right=0.5 cm of O](a2) {\(t\)};
                \vertex[above left=0.6 cm of O](a3);
                \diagram*{
                (A) --[very thick, boson](F),
                (G) --[very thick, scalar, edge label=\(H\)](B),
                (C) --[red, very thick, boson, edge label=\(V_i\)](H),
                (b) --[very thick, fermion](G) --[very thick, fermion, edge label=\(\ell\)](H) --[very thick, fermion](d),
                };
                \end{feynman}
            \end{tikzpicture}%
        }%
    }\hfill% 
    %------------Diag-2--------------------
    \subfloat[]{\label{dl_FCb}%
        \resizebox{0.48\linewidth}{!}{%
            \begin{tikzpicture}
                \begin{feynman}
                \def\r{0.6 cm}
                \coordinate (O) at (0,0);
                \draw[very thick] (O) circle (\r);
                \vertex [red, square dot] (A) at (90:\r) {};  
                \draw[fermion, very thick] (360:\r) arc (360:180:\r);
                \vertex [](B) at (210:\r);   
                \vertex [](C) at (330:\r);   
                \vertex[above=1.0 of A](F){\(\gamma\)};
                \vertex[below left=1.0cm and 0.5cm of B](G);
                \vertex[below right=1.0cm and 0.5cm of C](H);
                \vertex[left=0.7 cm of G](b){\(\ell\)};
                \vertex[right=0.7 cm of H](d){\(\ell\)};
                \vertex[below=0.6 cm of O](a1);
                \vertex[above right=0.5 cm of O](a2) {\(t\)};
                \vertex[above left=0.6 cm of O](a3);
                \diagram*{
                (A) --[red, very thick, boson](F),
                (G) --[very thick, scalar, edge label=\(H\)](B),
                (C) --[very thick, boson, edge label=\(V_i\)](H),
                (b) --[very thick, fermion](G) --[very thick, fermion, edge label=\(\ell\)](H) --[very thick, fermion](d),
                };
                \end{feynman}
            \end{tikzpicture}%
        }%
    }\\ % <-- New Row
    %------------Diag-3--------------------
    \subfloat[]{\label{dl_FCc}%
        \resizebox{0.48\linewidth}{!}{%
            \begin{tikzpicture}
                \begin{feynman}
                \def\r{0.6 cm}
                \coordinate (O) at (0,0);
                \draw[very thick] (O) circle (\r);
                \vertex (A) at (90:\r);  
                \draw[fermion, very thick] (360:\r) arc (360:180:\r);
                \vertex [violet, square dot](B) at (210:\r) {};   
                \vertex [](C) at (330:\r);   
                \vertex[above=1.0 of A](F){\(\gamma\)};
                \vertex[below left=1.0cm and 0.5cm of B](G);
                \vertex[below right=1.0cm and 0.5cm of C](H);
                \vertex[left=0.7 cm of G](b){\(\ell\)};
                \vertex[right=0.7 cm of H](d){\(\ell\)};
                \vertex[below=0.6 cm of O](a1);
                \vertex[above right=0.5 cm of O](a2){\(t\)};
                \vertex[above left=0.6 cm of O](a3);
                \diagram*{
                (A) --[very thick, boson](F),
                (G) --[violet, very thick, scalar, edge label=\(H\)](B),
                (C) --[very thick, boson, edge label=\(V_i\)](H),
                (b) --[very thick, fermion](G) --[very thick, fermion, edge label=\(\ell\)](H) --[very thick, fermion](d),
                };
                \end{feynman}
            \end{tikzpicture}%
        }%
    }\hfill%
    %------------------------------
    \subfloat[]{\label{dl_FCd}%
        \resizebox{0.48\linewidth}{!}{%
            \begin{tikzpicture}
                \begin{feynman}
                \def\r{0.6 cm}
                \coordinate (O) at (0,0);
                \draw[very thick] (O) circle (\r);
                \vertex [red, square dot] (A) at (90:\r) {};  
                \draw[fermion, very thick] (360:\r) arc (360:180:\r);
                \vertex [](B) at (210:\r);   
                \vertex [](C) at (330:\r);   
                \vertex[above=1.2 of A](F){\(\gamma\)};
                \vertex[below left=1.0cm and 0.5cm of B](G);
                \vertex[below right=1.0cm and 0.5cm of C](H);
                \vertex[left=0.7 cm of G](b){\(\ell\)};
                \vertex[right=0.7 cm of H](d){\(\ell\)};
                \vertex[below=0.6 cm of O](a1);
                 \vertex[below=0.6 cm of O](a1) {\(d_i\)};
                \vertex[above right=0.5 cm of O](a2) {\(t\)};
                \vertex[above left=0.6 cm of O](a3) {\(t\)};
                \vertex[above left=0.6 cm of O](a3);
                \diagram*{
                (A) --[red, very thick, boson](F),
                (G) --[very thick, boson, edge label=\(W\)](B),
                (C) --[very thick, boson, edge label=\(W\)](H),
                (b) --[very thick, fermion](G) --[very thick, fermion, edge label'=\(\ell^{\prime}\)](H) --[very thick, fermion](d),
                };
                \end{feynman}
            \end{tikzpicture}%
        }%
    }\\
    %------------------Diag-Contact-EDM---------------
\subfloat[]{\label{fig:lepton_edm_contact}%
    \resizebox{0.48\linewidth}{!}{%
        \begin{tikzpicture}
            \begin{feynman}
                % External lepton lines
                \vertex (a1) {\(\ell\)};
                \vertex[square dot, black, right=1.2cm of a1] (v1) {};
                \vertex[right=1.2cm of v1] (a2) {\(\ell\)};
                
                % Lower vertex of the loop
                \vertex[below=0.8cm of v1] (v2);
                
                % External photon
                \vertex[below=0.5cm of v2] (gamma) {\(\gamma\)};
                
                \diagram*{
                    % Horizontal lepton line
                    (a1) -- [ thick, fermion] (v1) -- [ thick, fermion] (a2),
                    
                    % Fermion loop (clockwise flow)
                    % Right side arc (v1 down to v2)
                    (v1) -- [ thick, fermion, half left, looseness=1.5, edge label=\(t\)] (v2),
                    % Left side arc (v2 up to v1)
                    (v2) -- [ thick, fermion, half left, looseness=1.5, edge label=\(t\)] (v1),
                    
                    % Photon emission
                    (v2) -- [ thick, boson] (gamma)
                };
            \end{feynman}
        \end{tikzpicture}%
    }%
}
%----------------------------------------
    \caption{Barr-Zee topologies contributing to the leptonic EDMs through
flavor-conserving top SMEFT operators.}
    \label{fig:BZ_lep_EDM_FC}
    \end{figure}
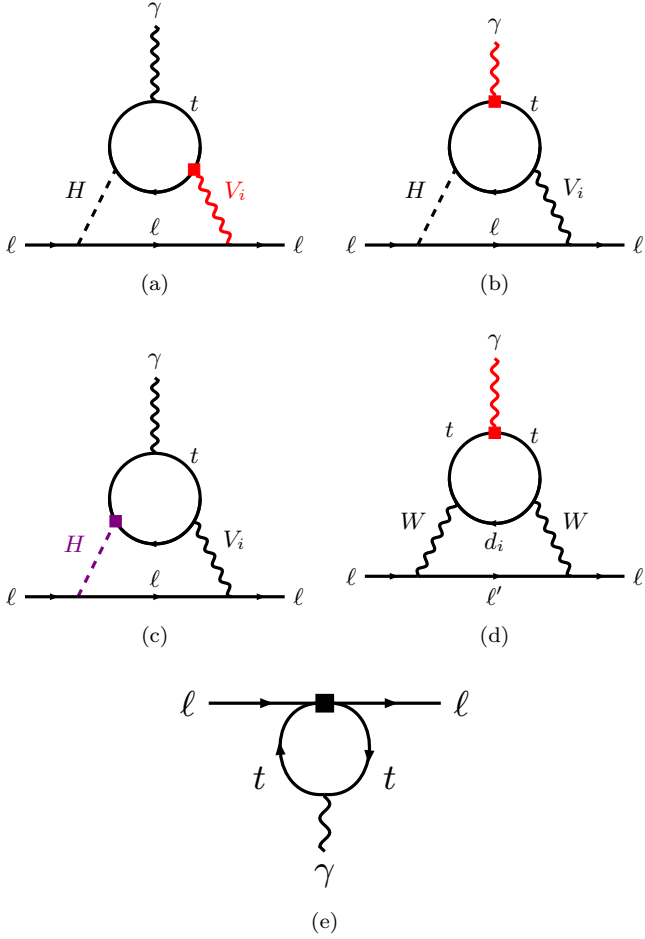

%---------------------------------
For flavor-conserving top-quark interactions, the relevant operators induce anomalous $t\bar{t}\gamma$, $t\bar{t}Z$, $t\bar{t}H$, and semileptonic $t\bar{t}\ell\bar{\ell}$ couplings. The first three classes contribute to leptonic EDMs via two-loop Barr-Zee type diagrams in Fig.~\ref{fig:BZ_lep_EDM_FC}, while the semileptonic four-fermion ($2q2\ell$) operators generate leptonic EDMs through one-loop diagram shown in Fig.~\ref{fig:lepton_edm_contact}. The contributions from these different diagrams can be matched to the coefficients of eq.~\ref{eq:chiral_dipole} and the explicit matching relation for the leptonic dipole coefficient
$L^{e\gamma}_{pp}$ at the electroweak scale, $\mu_{\rm EW}$, is given in
Appendix~\ref{Append:leptonic EDM_FC}. The calculation strategy for the
Barr-Zee-type diagrams entering these matching relations is discussed
in detail in Appendix~\ref{appendix:barr_zee}. In particular, the $t\bar t\gamma$
contribution depends on the dipole WCs $C^{uB}_{33}$ and
$C^{uW}_{33}$, while the $t\bar t Z$ contribution receives contributions
from $C^{uB}_{33}$, $C^{uW}_{33}$, $C^{\phi q(1)}_{33}$,
$C^{\phi q(3)}_{33}$, and $C^{\phi u}_{33}$. The $t\bar t H$
contribution is controlled by the top-Higgs Yukawa operator
$C^{u\phi}_{33}$. The $t\bar t\ell\bar\ell$ interaction is described by
the tensor four-fermion operator $C^{lequ(3)}_{1133}$, which contributes
to the lepton EDM already at one loop. All these contributions are
evaluated at the electroweak scale and subsequently evolved to the
low-energy scale.
%-----------------------------
%-----------------------------
Depending on the internal gauge boson, $V_i=\gamma$ or $Z$, six distinct
Barr-Zee topologies arise, with three corresponding to each choice of
$V_i$. The sums of the three diagrams for $V_i=\gamma$ and $V_i=Z$ are
incorporated into the matching relations given in
Eqs.~\eqref{eq:BarrZee_top_cons_ga} and
\eqref{eq:BarrZee_top_cons_Z}, respectively.

It is important to note that $C^{uW}_{33}$ simultaneously modifies both
the $t\bar t\gamma$ and $tbW$ interactions after electroweak symmetry
breaking. In the diagram shown in Fig.~\ref{dl_FCd},
we associate the $C^{uW}_{33}$ term with the $t\bar t\gamma$ interaction,
while its contribution to the charged-current $tbW$ interaction is
considered separately in the corresponding charged-current analysis.

%-----------------------------------
\begin{table}[htb!]
    \centering
    \renewcommand{\arraystretch}{1.3}
    \setlength{\tabcolsep}{8pt}
    \begin{tabular}{@{} c c c @{}}
        \toprule
        \textbf{WCs~$\,[\mathrm{TeV}^{-2}]$}
        & \textbf{$|d_e|$ (Current)}
        & \textbf{$|d_e|$ (Future)} \\
        \midrule

        $\mathrm{Im}(C^{uB}_{33})$
        & $1.97\times10^{-3}$
        & $4.81\times10^{-5}$ \\

        $\mathrm{Im}(C^{uW}_{33})$
        & $3.64\times10^{-3}$
        & $8.87\times10^{-5}$ \\

        $\mathrm{Im}(C^{u\phi}_{33})$
        & $1.16\times10^{-2}$
        & $2.83\times10^{-4}$ \\

        $\mathrm{Im}(C^{\phi q(1)}_{33})$
        & $3.16\times10^{-1}$
        & $7.70\times10^{-3}$ \\

        $\mathrm{Im}(C^{\phi q(3)}_{33})$
        & $1.35\times10^{-1}$
        & $3.30\times10^{-3}$ \\

        $\mathrm{Im}(C^{\phi u}_{33})$
        & $4.41\times10^{-1}$
        & $1.08\times10^{-2}$ \\

        $\mathrm{Im}(C^{lequ(3)}_{1133})$
        & $ 3.71 \times 10^{-12} $
        & $9.04 \times 10^{-14}$\\
        
        \midrule
        
        \textcolor{DarkEmerald}{$\mathrm{Im}(C^{\phi ud}_{33})$}
        & \textcolor{DarkEmerald}{$5.02$}
        & \textcolor{DarkEmerald}{$1.22\times10^{-1}$} \\ 

       \textcolor{DarkEmerald}{$\mathrm{Im}(C^{dW}_{33})$}
        & \textcolor{DarkEmerald}{$10$}
        & \textcolor{DarkEmerald}{$2.45\times10^{-1}$} \\

      \textcolor{DarkEmerald}{$\mathrm{Im}(C^{uG}_{33})$}
        & \textcolor{DarkEmerald}{$1.66\times10^{-2}$}
        & \textcolor{DarkEmerald}{$4.06\times10^{-4}$} \\  
        \bottomrule
    \end{tabular}
    \caption{Bounds on the flavor-conserving top-quark SMEFT WCs from current and projected electron EDM sensitivities. The WCs are given at $\mu=1~\mathrm{TeV}$ in units of $\mathrm{TeV}^{-2}$; \textcolor{DarkEmerald}{colored entries} denote RGE-induced contributions.}
    \label{tab:de_bound_top_FC}
\end{table}
%-----------------------------------

After we extract the contributions in $L^{e\gamma}_{pp}$ and hence the $d_{e}$ (one could also extract $d_{\mu,\tau}$), using the data given in Sec.~\ref{sec:electric_dipole_moments} we have extracted the relevant WCs. 
The resulting constraints are summarized in Table~\ref{tab:de_bound_top_FC}.
Comparing the two classes of Barr-Zee contributions, we find that the
dipole-type interactions, governed by $C^{uB}_{33}$ and $C^{uW}_{33}$,
together with the top-Higgs Yukawa interaction $C^{u\phi}_{33}$, are more
tightly constrained by the $V_i=\gamma$ contribution. The corresponding
bounds are of $\mathcal{O}(10^{-3}-10^{-2})$ with the current electron EDM limit, and the projected sensitivity can reach $\mathcal{O}(10^{-5}-10^{-4})$. In contrast, the vector-current interactions, which are sensitive to either of
$(C^{\phi q(1)}_{33}$, $C^{\phi q(3)}_{33}$ or $C^{\phi u}_{33})$, receive large contributions from the Barr-Zee type diagrams with an intermediate $Z$
boson. The resulting bounds are typically of $\mathcal{O}(10^{-1})$,
with projected sensitivities reaching $\mathcal{O}(10^{-2})$. For the $2q2\ell$ operators, the bounds are particularly stringent, reaching the level of $\mathcal{O}(10^{-12})$ for $\mathrm{Im}(C^{lequ(3)}_{1133})$, as these contributions arise already at one loop and are dominated by the top-quark mass through a factor of
$m_t$. Similarly, the muon and tau EDMs provide constraints on
the corresponding couplings of the second and third lepton generations,
$\mathrm{Im}(C^{lequ(3)}_{2233})$ and
$\mathrm{Im}(C^{lequ(3)}_{3333})$, respectively. These bounds are given
by
%-------------------------
\begin{align}
    \mathrm{Im}(C^{lequ(3)}_{2233})=&0.99\times 10^{-1}~\mathrm{(Current)}\,,\nn\\
    &5.42\times10^{-5} ~\mathrm{(PSI~Phase~II)}\,,\nn\\
    \mathrm{Im}(C^{lequ(3)}_{3333})=&5.37\times 10^{-1}~\mathrm{(Current)}\,.
\end{align}
%--------------------------
The table also contains several RGE-induced contributions, whose
corresponding entries are highlighted in \textcolor{DarkEmerald}{green}.
These coefficients contribute to the electron EDM through one-loop RGE
mixing into operators associated with the corresponding $t\bar t X$
interactions, with $X=\gamma,Z,H$, followed by the subsequent low-energy
matching. Since these contributions arise through a sequential
RGE-induced mechanism, the resulting bounds are generally weaker than
those obtained from the direct two-loop Barr-Zee contributions, as
reflected in Table~\ref{tab:de_bound_top_FC}. An exception is $\mathrm{Im}(C^{uG}_{33})$, whose constraint remains
comparatively strong due to the sizeable RGE-induced contribution
originating from $C^{u\phi}_{33}$, as shown in
Table~\ref{tab:rge_matrix_bosonic}.

%---------------------------------------------
% \begin{table}[htb!]
%     \centering
%     \renewcommand{\arraystretch}{1.3}
%     \setlength{\tabcolsep}{8pt}
%     \begin{tabular}{@{} c c c @{}}
%         \toprule
%         \textbf{WCs}
%         & \textbf{$|d_e|$ (Current)}
%         & \textbf{$|d_e|$ (Future)} \\
%         & $\left(<4.1\times10^{-30}\,\mathrm{e\!-\!cm}\right)$
%         & $\left(< 10^{-31}\,\mathrm{e\!-\!cm}\right)$ \\
%         \midrule

%         $\mathrm{Im}\,(C^{uB}_{33})$
%         & $3.27\times10^{-3}$
%         & $7.98\times10^{-5}$ \\

%         $\mathrm{Im}\,(C^{uW}_{33})$
%         & $6.29\times10^{-3}$
%         & $1.53\times10^{-4}$ \\

%         $\mathrm{Im}\,(C^{u\phi}_{33})$
%         & $9.61\times10^{-3}$
%         & $2.34\times10^{-4}$ \\

%         $\mathrm{Im}\,(C^{\phi q(1)}_{33})$
%         & $5.50\times10^{-1}$
%         & $1.34\times10^{-2}$ \\

%         $\mathrm{Im}\,(C^{\phi q(3)}_{33})$
%         & $5.50\times10^{-1}$
%         & $1.34\times10^{-2}$ \\

%         $\mathrm{Im}\,(C^{\phi u}_{33})$
%         & $5.50\times10^{-1}$
%         & $1.34\times10^{-2}$ \\

%         \bottomrule
%     \end{tabular}
%     \caption{Bounds on the WCs from the current and future experimental bounds on the electron EDM without considering RGE effects.}
%     \label{tab:bounds-WC-FCNC-new}% \end{table}
%---------------------------------------------

%---------------------------------
\subsubsection{Flavor-violating charged current interactions}\label{subsec:leptonic_FCCC}
%---------------------------------

%----------------------------
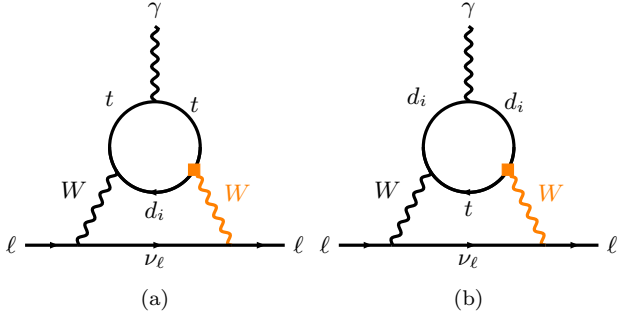
\begin{figure}[t]
    \centering
    %------------Diag-1---------------
    \subfloat[]{\label{dl_FCCCa}%
        \resizebox{0.48\linewidth}{!}{%
            \begin{tikzpicture}
                \begin{feynman}
                \def\r{0.6 cm}
                \coordinate (O) at (0,0);
                \draw[very thick] (O) circle (\r);
                \vertex (A) at (90:\r);  
                \draw[fermion, very thick] (360:\r) arc (360:180:\r);
                \vertex [](B) at (210:\r);   
                \vertex [orange, square dot](C) at (330:\r) {};   
                \vertex[above=1.0 of A](F){\(\gamma\)};
                \vertex[below left=1.0cm and 0.5cm of B](G);
                \vertex[below right=1.0cm and 0.5cm of C](H);
                \vertex[left=0.7 cm of G](b){\(\ell\)};
                \vertex[right=0.7 cm of H](d){\(\ell\)};
                \vertex[below=0.6 cm of O](a1) {\(d_i\)};
                \vertex[above right=0.5 cm of O](a2) {\(t\)};
                \vertex[above left=0.6 cm of O](a3) {\(t\)};
                \diagram*{
                (A) --[very thick, boson](F),
                (G) --[very thick, boson, edge label=\(W\)](B),
                (C) --[very thick, orange, boson, edge label=\(W\)](H),
                (b) --[very thick,fermion](G) --[very thick, fermion, edge label'=\(\nu_\ell\)](H) --[very thick, fermion](d),
                };
                \end{feynman}
            \end{tikzpicture}%
        }%
    }
  %------------Diag-2---------------
    \subfloat[]{\label{dl_FCCCa}%
        \resizebox{0.48\linewidth}{!}{%
            \begin{tikzpicture}
                \begin{feynman}
                \def\r{0.6 cm}
                \coordinate (O) at (0,0);
                \draw[very thick] (O) circle (\r);
                \vertex (A) at (90:\r);  
                \draw[fermion, very thick] (360:\r) arc (360:180:\r);
                \vertex [](B) at (210:\r);   
                \vertex [orange, square dot](C) at (330:\r) {};   
                \vertex[above=1.0 of A](F){\(\gamma\)};
                \vertex[below left=1.0cm and 0.5cm of B](G);
                \vertex[below right=1.0cm and 0.5cm of C](H);
                \vertex[left=0.7 cm of G](b){\(\ell\)};
                \vertex[right=0.7 cm of H](d){\(\ell\)};
                \vertex[below=0.6 cm of O](a1) {\(t\)};
                \vertex[above right=0.5 cm of O](a2) {\(d_i\)};
                \vertex[above left=0.6 cm of O](a3) {\(d_i\)};
                \diagram*{
                (A) --[very thick, boson](F),
                (G) --[very thick, boson, edge label=\(W\)](B),
                (C) --[very thick, orange, boson, edge label=\(W\)](H),
                (b) --[very thick,fermion](G) --[very thick, fermion, edge label'=\(\nu_\ell \)](H) --[very thick, fermion](d),
                };
                \end{feynman}
            \end{tikzpicture}%
        }%
    }
   \caption{Barr-Zee topologies contributing to the leptonic EDMs through
flavor-violating charged-current SMEFT operators.}
    \label{fig:BZ_lep_EDM_FCCC}
\end{figure}
%---------------------------------

For flavor-violating charged-current top-quark interactions, we consider
the effective coupling of the form $t\bar d_i W^-$, where $d_i$ denotes
a down-type quark of any generation. Four types of dimension-six SMEFT
operators can generate the corresponding charged-current interactions.
The operators $\mathcal{O}^{\phi q(3)}_{i3}$ and
$\mathcal{O}^{\phi ud}_{i3}$ induce left- and right-handed vector
charged-current couplings, respectively, while
$\mathcal{O}^{dW}_{i3}$ and $\mathcal{O}^{uW}_{i3}$ generate
left- and right-chiral tensor couplings, respectively. 
The relevant Barr-Zee diagrams are shown in Fig.~\ref{fig:BZ_lep_EDM_FCCC}. Among the two topologies, the contribution shown in Fig.~\ref{dl_FCCCa} is found to be dominant, due to the enhancement associated with the top-quark mass in the loop. 
After taking into account these loop-level contributions, we find that
the dominant contributions to the leptonic EDM arise from dipole operators
$C^{dW}_{i3}$ as evident from the corresponding
matching relations in Eq.~\eqref{eq:BarrZee_top_vio_W},whereas $C^{uW}_{i3}$ is light quark mass dependent.

%---------------------------------------------
\begin{table}[t!]
    \centering
    \renewcommand{\arraystretch}{1.3}
    \setlength{\tabcolsep}{8pt}
    \begin{tabular}{@{} c c c @{}}
        \toprule
        \textbf{WCs~$[\text{TeV}^{-2}]$}
        & \textbf{$|d_e|$ (Current)}
        & \textbf{$|d_e|$ (Future)} \\
        \midrule

        $\mathrm{Im}\,(\mathcal{C}^{dW}_{13})$
        & $1.61$
        & $3.92\times10^{-2}$ \\

        $\mathrm{Im}\,(\mathcal{C}^{dW}_{23})$
        & $3.38\times10^{-1}$
        & $8.24\times10^{-3}$ \\

        $\mathrm{Im}\,(\mathcal{C}^{dW}_{33})$
        & $1.43\times10^{-2}$
        & $3.48\times10^{-4}$ \\

        %  $\mathrm{Im}\,(\mathcal{C}^{uB}_{33})$
        % & $5.53\times10^{-2}$
        % & $1.35\times10^{-3}$ \\

        $\mathrm{Im}\,(\mathcal{C}^{uW}_{33})$
        & $1.91\times10^{-1}$
        & $4.67\times10^{-3}$ \\

        % \midrule

        % \textcolor{rgegray}{$\mathrm{Im}\,(\mathcal{C}^{uG}_{33})$}
        % & \textcolor{rgegray}{$1.20$}
        % & \textcolor{rgegray}{$2.94\times10^{-2}$} \\

        \bottomrule
    \end{tabular}
    \caption{Bounds on the top-quark FCCC SMEFT WCs.}
    \label{tab:bound_top_FCCC}
\end{table}
%---------------------------------------------
We summarize our results in Table~\ref{tab:bound_top_FCCC}. It is
interesting to note that the WC $C^{uW}_{33}$ contributes
to both charged- and neutral-current top-quark interactions. However, as
evident from Table~\ref{tab:de_bound_top_FC}, the neutral-current analysis
provides a particularly strong constraint on $C^{uW}_{33}$.

% %---------------------------------------------
% \begin{table}[htb!]
%     \centering
%     \renewcommand{\arraystretch}{1.3}
%     \setlength{\tabcolsep}{8pt}
%     \begin{tabular}{@{} c c c @{}}
%         \toprule
%         \textbf{WCs}
%         & \textbf{$|d_e|$ (Current)}
%         & \textbf{$|d_e|$ (Future)} \\
%         & $\left(<4.1\times10^{-30}\,\mathrm{e\!-\!cm}\right)$
%         & $\left(< 10^{-31}\,\mathrm{e\!-\!cm}\right)$ \\
%         \midrule

%         $\mathrm{Im}\,(\mathcal{C}^{uW}_{13})$
%         & $2.89\times10^{-3}$
%         & $7.05\times10^{-5}$ \\

%         $\mathrm{Im}\,(\mathcal{C}^{uW}_{23})$
%         & $4.24\times10^{-5}$
%         & $1.03\times10^{-6}$ \\

%         $\mathrm{Im}\,(\mathcal{C}^{uW}_{33})$
%         & $7.06\times10^{-8}$
%         & $1.72\times10^{-9}$ \\

%         $\mathrm{Im}\,(\mathcal{C}^{\phi ud}_{31})$
%         & $1.10\times10^{-2}$
%         & $2.68\times10^{-4}$ \\

%         $\mathrm{Im}\,(\mathcal{C}^{\phi ud}_{32})$
%         & $1.61\times10^{-4}$
%         & $3.94\times10^{-6}$ \\

%         $\mathrm{Im}\,(\mathcal{C}^{\phi ud}_{33})$
%         & $2.69\times10^{-7}$
%         & $6.55\times10^{-9}$ \\

%         \bottomrule
%     \end{tabular}
%     \caption{Bounds on the WCs from the current and future experimental bounds on the electron EDM for flavor violating charged current processes without considering RGE effects.}
%     \label{tab:bounds-WC-FCNC}
% \end{table}
% %---------------------------------------------

%---------------------------------
\subsubsection{Flavor-changing neutral current interactions}\label{subsec:leptonic_FCNC}
\begin{figure}[t!]
    \centering
    %--------------Diag-1-----------------
    \subfloat[]{\label{dl_FCNCa}%
        \resizebox{0.48\linewidth}{!}{%
            \begin{tikzpicture}
                \begin{feynman}
                \def\r{0.6 cm}
                \coordinate (O) at (0,0);
                \draw[very thick] (O) circle (\r);
                \vertex (A) at (90:\r);  
                \draw[fermion, very thick] (360:\r) arc (360:180:\r);
                \vertex [violet, square dot](B) at (210:\r) {};   
                \vertex [red, square dot](C) at (330:\r) {};   
                \vertex[above=1.0 of A](F){\(\gamma\)};
                \vertex[below left=1.0cm and 0.5cm of B](G);
                \vertex[below right=1.0cm and 0.5cm of C](H);
                \vertex[left=0.7 cm of G](b){\(\ell\)};
                \vertex[right=0.7 cm of H](d){\(\ell\)};
                \vertex[below=0.6 cm of O](a1) {\(u_i\)};
                \vertex[above right=0.5 cm of O](a2) {\(t\)};
                \vertex[above left=0.6 cm of O](a3) {\(t\)};
                \diagram*{
                (A) --[very thick, boson](F),
                (G) --[violet, very thick, scalar, edge label=\(H\)](B),
                (C) --[red, very thick, boson, edge label=\(V_i\)](H),
                (b) --[very thick, fermion](G) --[very thick, fermion](H) --[very thick, fermion](d),
                };
                \end{feynman}
            \end{tikzpicture}%
        }%
    }\hfill%
    %------------Diag-2--------------------
    \subfloat[]{\label{dl_FCNCb}%
        \resizebox{0.48\linewidth}{!}{%
            \begin{tikzpicture}
                \begin{feynman}
                \def\r{0.6 cm}
                \coordinate (O) at (0,0);
                \draw[very thick] (O) circle (\r);
                \vertex [blue, square dot] (A) at (90:\r) {};  
                \draw[fermion, very thick] (360:\r) arc (360:180:\r);
                \vertex [violet, square dot](B) at (210:\r) {};   
                \vertex [](C) at (330:\r) ;   
                \vertex[above=1.0 of A](F){\(\gamma\)};
                \vertex[below left=1.0cm and 0.5cm of B](G);
                \vertex[below right=1.0cm and 0.5cm of C](H);
                \vertex[left=0.7 cm of G](b){\(\ell\)};
                \vertex[right=0.7 cm of H](d){\(\ell\)};
                \vertex[below=0.6 cm of O](a1) {\(t\)};
                \vertex[above right=0.5 cm of O](a2) {\(t\)};
                \vertex[above left=0.6 cm of O](a3) {\(u_i\)};
                \diagram*{
                (A) --[blue, very thick, boson](F),
                (G) --[violet, very thick, scalar, edge label=\(H\)](B),
                (C) --[very thick, boson, edge label=\(V_i\)](H),
                (b) --[very thick, fermion](G) --[very thick, fermion](H) --[very thick, fermion](d),
                };
                \end{feynman}
            \end{tikzpicture}%
        }%
    }\\ % <-- This double slash forces the 3rd diagram to the next row
    %------------Diag-3--------------------
    \subfloat[]{\label{dl_FCNCc}%
        \resizebox{0.48\linewidth}{!}{%
            \begin{tikzpicture}
                \begin{feynman}
                \def\r{0.6 cm}
                \coordinate (O) at (0,0);
                \draw[very thick] (O) circle (\r);
                \vertex [blue, square dot] (A) at (90:\r) {};  
                \draw[fermion, very thick] (360:\r) arc (360:180:\r);
                \vertex [](B) at (210:\r) ;   
                \vertex [red, square dot](C) at (330:\r) {};   
                \vertex[above=1.0 of A](F){\(\gamma\)};
                \vertex[below left=1.0cm and 0.5cm of B](G);
                \vertex[below right=1.0cm and 0.5cm of C](H);
                \vertex[left=0.7 cm of G](b){\(\ell\)};
                \vertex[right=0.7 cm of H](d){\(\ell\)};
                \vertex[below=0.6 cm of O](a1) {\(t\)};
                \vertex[above right=0.5 cm of O](a2) {\(u_i\)};
                \vertex[above left=0.6 cm of O](a3) {\(t\)};
                \diagram*{
                (A) --[blue, very thick, boson](F),
                (G) --[very thick, scalar, edge label=\(H\)](B),
                (C) --[red, very thick, boson, edge label=\(V_i\)](H),
                (b) --[very thick, fermion](G) --[very thick, fermion](H) --[very thick, fermion](d),
                };
                \end{feynman}
            \end{tikzpicture}%
        }%
    }
    %--------------------------
    \caption{Barr-zee type diagrams contributing to lepton dipole moments via top-FCNC couplings. Here $ u_i = u,c$ , $ V_i = \gamma , Z $.}
    \label{fig:EDM_flavor_vio_neutral}
\end{figure}
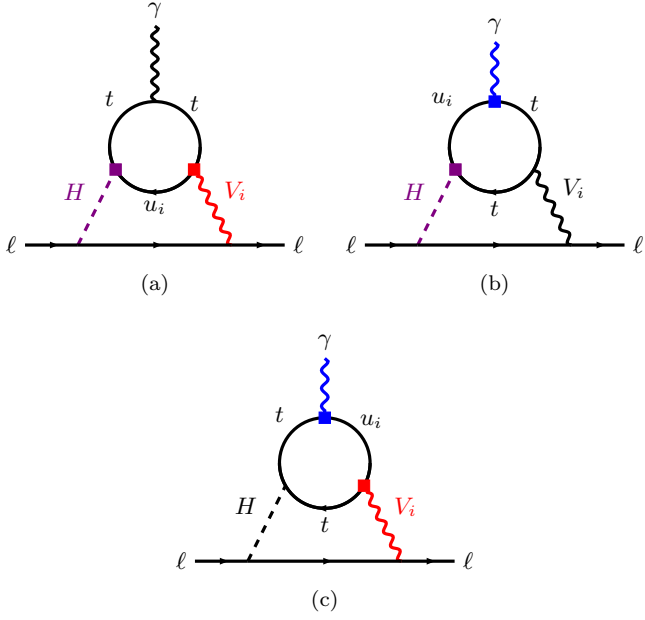

%---------------------------------
We next consider top-FCNC
interactions of the form $t u_j X$, where $u_j=u,c$ and
$X=\gamma,Z,H$. The corresponding contributions to the leptonic EDMs
involve several distinct finite Barr-Zee topologies. Since these
contributions require two insertions of the top-FCNC
interaction, the relevant diagrams for the different choices of
$X=\gamma,Z,H$ are shown in Fig.~\ref{fig:EDM_flavor_vio_neutral}. To simplify the
matching calculation, we reparametrize the relevant SMEFT WCs in terms of dimensionless effective couplings, as shown in Eqs.~\eqref{eq:top_effective_matching}. 
%---------------------------------
\begin{widetext}
\begin{subequations}\label{eq:top_effective_matching}
\begin{align}
(\xi_L)_{pr}
&= \sqrt{2}\,v\frac{m_t}{g_s}\left(C^{uG}_{rp}\right)^*,
&
(\xi_R)_{pr}
&= \sqrt{2}\,v\frac{m_t}{g_s}C^{uG}_{pr},
\end{align}

\begin{align}
(\lambda_L)_{pr}
&= \sqrt{2}\,v\frac{m_t}{e}
\left[
s_W\left(C^{uW}_{rp}\right)^*
+c_W\left(C^{uB}_{rp}\right)^*
\right],
&
(\lambda_R)_{pr}
&= \sqrt{2}\,v\frac{m_t}{e}
\left(
s_W C^{uW}_{pr}
+c_W C^{uB}_{pr}
\right),
\\
(\kappa_L)_{pr}
&= \sqrt{2}\,v\frac{c_W m_t}{g_W}
\left[
c_W\left(C^{uW}_{rp}\right)^*
-s_W\left(C^{uB}_{rp}\right)^*
\right],
&
(\kappa_R)_{pr}
&= \sqrt{2}\,v\frac{c_W m_t}{g_W}
\left(
c_W C^{uW}_{pr}
-s_W C^{uB}_{pr}
\right),
\end{align}

\begin{align}
(X_L)_{pr}
&= v^2
\left(
C^{\phi q(1)}_{pr}
-C^{\phi q(3)}_{pr}
\right)
\equiv v^2 C^{\phi q(-)}_{pr},
&
(X_R)_{pr}
&= v^2 C^{\phi u}_{pr},
\\
(\eta_L)_{pr}
&= \frac{3}{2}v^2\left(C^{u\phi}_{rp}\right)^*,
&
(\eta_R)_{pr}
&= \frac{3}{2}v^2 C^{u\phi}_{pr}.
\end{align}
\end{subequations}
\end{widetext}
%---------------------------------

The explicit
matching relations at the electroweak scale, $\mu_{\rm EW}$, are given
in Eqs.~\eqref{eq:matching_lepEDM_topFCNC}. The resulting constraints from the leptonic EDM
measurements are presented in Table~\ref{tab:bounds-de-FCNC-effective} in terms of these
dimensionless effective couplings. The corresponding bounds can be
straightforwardly translated into constraints on the associated pairs of
SMEFT WCs using the definitions given in
Eqs.~\eqref{eq:top_effective_matching}.

%-----------------------------------------
%------------------------------------------
%------------------------------------------
\begin{table}[t!]
    \centering
    \footnotesize
    \renewcommand{\arraystretch}{1.3}
    \setlength{\tabcolsep}{6pt}
    \begin{tabular}{@{} c c c @{}}
        \toprule
        \textbf{Effective Couplings}
        & \textbf{$|d_e|$ (Current)}
        & \textbf{$|d_e|$ (Future)} \\
        \midrule

        $\displaystyle
        \mathrm{Im}\left(
        \lambda_L^{ct}\eta_R^{tc^*}
        -\lambda_R^{ct}\eta_L^{tc*}
        \right)$
        & $8.32\times10^{-1}$
        & $2.02\times10^{-2}$ \\

        $\displaystyle
        \mathrm{Im}\left(
        \eta_L^{u_jt}\lambda_L^{tu_j*}
        -\eta_R^{u_jt}\lambda_R^{tu_j*}
        \right)$
        & $2.13\times10^{-2}$
        & $5.21\times10^{-4}$ \\

        $\displaystyle
        \mathrm{Im}\left(
        X_R^{u_jt}\eta_R^{tu_j*}
        -X_L^{u_jt}\eta_L^{*tu_j}
        \right)$
        & $1.95\times10^{-2}$
        & $4.76\times10^{-4}$ \\

       $\displaystyle
        \mathrm{Im}\left(
        \eta_L^{tu_j*}\kappa_L^{u_jt}
        -\eta_R^{tu_j*}\kappa_R^{u_jt}
        \right)$
        & $9.23\times10^{-2}$
        & $2.25\times10^{-3}$ \\

        % \midrule

        % \textcolor{rgegray}{$\displaystyle
        % \mathrm{Im}\left(
        % X_R^{ut}\eta_L^{tu*}
        % -X_L^{ut}\eta_R^{tu*}
        % \right)$}
        % & \textcolor{rgegray}{$6.15\times10^{3}$}
        % & \textcolor{rgegray}{$1.50\times10^{2}$} \\

        % \textcolor{rgegray}{$\displaystyle
        % \mathrm{Im}\left(
        % X_R^{ct}\eta_L^{tc*}
        % -X_L^{ct}\eta_R^{tc*}
        % \right)$}
        % & \textcolor{rgegray}{$8.86$}
        % & \textcolor{rgegray}{$2.16\times10^{-1}$} \\

        % \textcolor{rgegray}{$\displaystyle
        % \mathrm{Im}\left(
        % \eta_R^{tu*}\kappa_L^{ut}
        % -\eta_L^{tu*}\kappa_R^{ut}
        % \right)$}
        % & \textcolor{rgegray}{$2.49\times10^{3}$}
        % & \textcolor{rgegray}{$6.09\times10^{1}$} \\

        % \textcolor{rgegray}{$\displaystyle
        % \mathrm{Im}\left(
        % \eta_R^{tc*}\kappa_L^{ct}
        % -\eta_L^{tc*}\kappa_R^{ct}
        % \right)$}
        % & \textcolor{rgegray}{$3.60$}
        % & \textcolor{rgegray}{$8.77\times10^{-2}$} \\

        \bottomrule
    \end{tabular}
    \caption{Bounds on the dimensionless top-quark FCNC couplings at the $\mu_{\rm EW}$ scale. Here $u_j=u,c$ quarks respectively.}
    \label{tab:bounds-de-FCNC-effective}
\end{table}
%---------------------------------------------
In Table~\ref{tab:bounds-de-FCNC-effective}, we present only those
effective-coupling combinations that receive sizeable constraints from
the leptonic EDM. The strongest bounds correspond to combinations whose
matching contributions are enhanced by the top-quark mass and are
therefore largely insensitive to the light-quark flavor, yielding
similar constraints for the $tu_j$ interactions with $u_j=u,c$. Other
coupling combinations exhibit an explicit dependence on the light-quark
mass and consequently lead to much weaker constraints; these have been
removed from the Table~\ref{tab:bounds-de-FCNC-effective} for brevity.

%--------------------------
\subsubsection{Logarithmic enhancement contributions to EDMs}\label{subsec:sequential_two_one_loop}
%--------------------------

%------------------------------
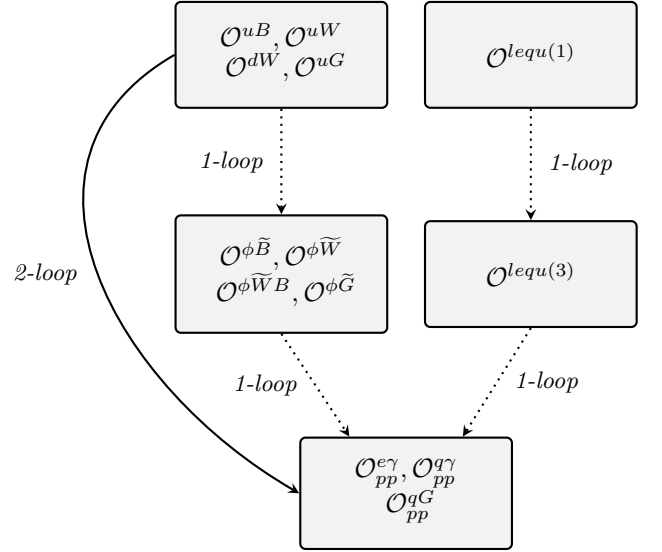
\begin{figure}[h!]
	\centering
	\begin{tikzpicture}[
		>=stealth, % Standard, clean arrow tips
		% Set a strict minimum width/height so the gaps can be calculated precisely
		box/.style={draw, thick, fill=gray!10, minimum height=1.4cm, minimum width=2.8cm, inner sep=8pt, align=center, rounded corners=2pt},
		every node/.style={font=\normalsize}
		]
		% -------------------------------
		% 1. Define the Nodes (Exact coordinates for 0.5cm H-gap and 1.5cm V-gap)
		% -------------------------------
		
		% Top Row (y = 5.8)
		\node[box] (TopL) at (-1.65, 5.8) {$\mathcal{O}^{uB}, \mathcal{O}^{uW}$ \\ \vspace{0.2em} $\mathcal{O}^{dW}, \mathcal{O}^{uG}$};
		\node[box] (TopR) at (1.65, 5.8) {$\mathcal{O}^{lequ(1)}$};
		
		% Middle Row (y = 2.9)
		\node[box] (MidL) at (-1.65, 2.9) {$\mathcal{O}^{\phi \widetilde{B}}, \mathcal{O}^{\phi \widetilde{W}}$ \\ \vspace{0.2em} $\mathcal{O}^{\phi \widetilde{W}B}, \mathcal{O}^{\phi \widetilde{G}}$};
		\node[box] (MidR) at (1.65, 2.9) {$\mathcal{O}^{lequ(3)}$};
		
		% Bottom Box (y = 0)
		\node[box] (Bot) at (0, 0) {$\mathcal{O}^{e\gamma}_{pp}, \mathcal{O}^{q\gamma}_{pp}$ \\ \vspace{0.2em} $\mathcal{O}^{qG}_{pp}$};
		
		% -------------------------------
		% 2. Draw the Arrows
		% -------------------------------
		
		% Left Column Arrows (Original Path)
		\draw[->, thick, dotted] (TopL.south) -- (MidL.north) node[midway, left=3pt, font=\small] {\textit{1-loop}};
		\draw[->, thick, dotted] (MidL.south) -- (Bot.135) node[midway, left=3pt, font=\small] {\textit{1-loop}};
		
		% 2-loop curvy arrow (Left side only)
		\draw[->, thick] (TopL.west) to[out=210, in=150, looseness=1.2] node[midway, left=4pt, font=\small] {\textit{2-loop}} (Bot.west);
		
		% Right Column Arrows (New lequ path)
		\draw[->, thick, dotted] (TopR.south) -- (MidR.north) node[midway, right=3pt, font=\small] {\textit{1-loop}};
		\draw[->, thick, dotted] (MidR.south) -- (Bot.45) node[midway, right=3pt, font=\small] {\textit{1-loop}};
		
	\end{tikzpicture}
	\caption{Schematic illustration of the direct two-loop and sequential one-loop contributions to the light-fermion dipole operators, highlighting the logarithmically enhanced path.}
	
	\label{fig:RGE _enhanced_top_dip}
\end{figure}

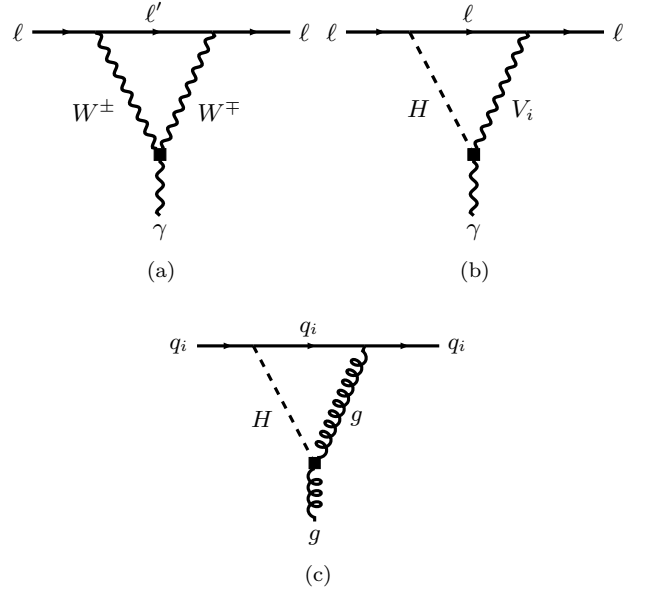
\begin{figure}[t!]
	\centering
	%------------------Diag-1---------------
	\subfloat[]{\label{fig:lepton_edm_triangle}%
		\resizebox{0.48\linewidth}{!}{%
			\begin{tikzpicture}
				\begin{feynman}
					\vertex (a1) {\(\ell\)};
					\vertex[right=1.0cm of a1] (v1);
					\vertex[right=1.5cm of v1] (v2);
					\vertex[right=1.0cm of v2] (a2) {\(\ell\)};
					\vertex[square dot, black, below right=1.5cm and 0.75cm of v1] (v3) {};
					
					% External photon
					\vertex[below=1.cm of v3] (gamma) {\(\gamma\)};
					
					\diagram*{
						(a1) -- [very thick, fermion] (v1) -- [very thick, fermion, edge label=\(\ell^{\prime}\)] (v2) -- [very thick, fermion] (a2),
						(v1) -- [very thick, boson, edge label'=\(W^\pm\)] (v3),
						(v3) -- [very thick, boson, edge label'=\(W^\mp\)] (v2),
						(v3) -- [very thick, boson] (gamma)
					};
				\end{feynman}
			\end{tikzpicture}%
		}%
	}
	%------------------Diag-2---------------
	\subfloat[]{\label{fig:lepton_edm_triangle}%
		\resizebox{0.48\linewidth}{!}{%
			\begin{tikzpicture}
				\begin{feynman}
					\vertex (a1) {\(\ell\)};
					\vertex[right=1.0cm of a1] (v1);
					\vertex[right=1.5cm of v1] (v2);
					\vertex[right=1.0cm of v2] (a2) {\(\ell\)};
					\vertex[square dot, black, below right=1.5cm and 0.75cm of v1] (v3) {};
					
					% External photon
					\vertex[below=1.cm of v3] (gamma) {\(\gamma\)};
					
					\diagram*{
						(a1) -- [very thick, fermion] (v1) -- [very thick, fermion, edge label=\(\ell\)] (v2) -- [very thick, fermion] (a2),
						(v1) -- [very thick, scalar, edge label'=\(H\)] (v3),
						(v3) -- [very thick, boson, edge label'=\(V_i\)] (v2),
						(v3) -- [very thick, boson] (gamma)
					};
				\end{feynman}
			\end{tikzpicture}%
		}%
	}\\
	%------------------Diag-3---------------
	\subfloat[]{\label{fig:lepton_edm_triangle}%
		\resizebox{0.48\linewidth}{!}{%
			\begin{tikzpicture}
				\begin{feynman}
					\vertex (a1) {\(q_i\)};
					\vertex[right=1.0cm of a1] (v1);
					\vertex[right=1.5cm of v1] (v2);
					\vertex[right=1.0cm of v2] (a2) {\(q_i\)};
					\vertex[square dot, black, below right=1.5cm and 0.75cm of v1] (v3) {};
					
					% External photon
					\vertex[below=1.cm of v3] (gamma) {\(g\)};
					
					\diagram*{
						(a1) -- [very thick, fermion] (v1) -- [very thick, fermion, edge label=\(q_i\)] (v2) -- [very thick, fermion] (a2),
						(v1) -- [very thick, scalar, edge label'=\(H\)] (v3),
						(v3) -- [very thick, gluon, edge label'=\(g\)] (v2),
						(v3) -- [very thick, gluon] (gamma)
					};
				\end{feynman}
			\end{tikzpicture}%
		}%
	}
	%------------------------------------
	\caption{Possible one-loop $\mathrm{CP}$-violating topologies arising from the bosonic $X^2\phi^2$ operator class.}
	\label{fig:RGE_induced}
\end{figure}

An interesting feature of our analysis is the significantly enhanced
sensitivity of EDM observables to the flavor-conserving top-quark dipole
operators. The direct contribution of these operators to the electron
and neutron EDMs arises through finite two-loop Barr-Zee diagrams, as
shown in Figs.~\ref{fig:BZ_lep_EDM_FC},\ref{fig:BZ_lep_EDM_FCCC},\ref{fig:qEDM_Barr_Zee}.
However, the same top-dipole operators can also mix at one loop into
$CP$-violating bosonic operators of the $X^2\phi^2$ type, as shown in Fig.~\ref{fig:RGE_induced}. These
intermediate operators subsequently contribute to the EDMs through
one-loop matching and RGE, as shown in the
left panel of Fig.~\ref{fig:RGE _enhanced_top_dip}. Although both
contributions are of the same nominal order in the loop expansion, the
sequential one-loop contribution receives a double-logarithmic
enhancement from the RG evolution. Similar, sequential one-loop contributions arise
for other operators as well. The scalar operator, $\mathcal{O}_{lequ}^{(1)}$
does not contribute directly to the leptonic EDMs at one loop or two loop, but it
mixes into $\mathcal{O}_{lequ}^{(3)}$ through RGE evolution. The latter
then contributes to the leptonic EDMs through one-loop matching,
providing a two-loop contribution to the EDM through this sequential
RGE-induced mechanism. This logarithmic enhancement can therefore
substantially increase the sensitivity of EDM observables to the
top-dipole and top-scalar WCs. The corresponding running equations can be found in Refs.~\cite{Jenkins:2013wua, Alonso:2013hga}.
%--------------------------------------
\begin{subequations}\label{eq:log_enhanced_ADM}
	\begin{align}
		\frac{d ~C^{\phi \tilde{B}}}{d \ln \mu}&=-\frac{10}{3} N_c\frac{g'}{16\pi^2}\mathrm{Im}\left([Y_u]_{pq} C^{uB}_{qr}\right) \,,\\
		\frac{d ~C^{\phi \tilde{W}}}{d \ln \mu}&= -2 N_c\frac{g}{16 \pi^2}\mathrm{Im}\left([Y_u]_{pq} C^{uW}_{qr}+[Y_d]_{pq} C^{dW}_{qr}\right)\,,\\
		\frac{d ~C^{\phi \tilde{W}B}}{d \ln \mu}&= -\frac{1}{3}N_c\frac{g'}{16\pi^2}\mathrm{Im}\left([Y_d]_{pq} C^{dW}_{qr}-5~[Y_u]_{pq} C^{uW}_{qr}\right)\,,\\
		\frac{d ~C^{\phi \tilde{G}}}{d \ln \mu}&= -4 \frac{g_s}{16\pi^2}\mathrm{Im}\left([Y_u]_{pq} C^{uG}_{qr} \right)\,,\\
		\frac{d~C^{lequ(3)}_{prst}}{d\ln \mu}&= \frac{1}{8}\frac{(5 g^{\prime ~2}+3 g^2)}{16\pi^2}C^{lequ(1)}_{prst}\,.
	\end{align}
\end{subequations}
%----------------------------------
Since we adopt the \textit{up-aligned} basis, the up-type Yukawa matrix is completely diagonalized, $Y_u=\mathrm{diag}(y_u,y_c,y_t)$, while the down-type Yukawa matrix becomes $Y_d=V_{\rm CKM}Y_d^{\rm diag}$. Consequently, for the $C^{dW}_{pr}$ WCs, we also obtain constraints on the flavor-violating components $V_{td_i}C^{dW}_{i3}$. Here, we take $y_t\simeq1$ and $y_b=0.014$ at $\mu=1~\mathrm{TeV}$.
The $CP$-violating bosonic operators of the $X^2\phi^2$ class generate
a variety of anomalous gauge and Higgs interactions. In particular,
$\mathcal{O}^{\phi\tilde{W}B}$ induces a $CP$-violating $WW\gamma$
coupling, while $\mathcal{O}^{\phi\tilde{W}}$,
$\mathcal{O}^{\phi\tilde{B}}$, and $\mathcal{O}^{\phi\tilde{W}B}$
generate $CP$-violating $h\gamma\gamma$ and $hZ\gamma$ interactions.
Similarly, $\mathcal{O}^{\phi\tilde{G}}$ induces a $CP$-violating
$hgg$ interaction. The corresponding Feynman diagrams are shown in
Fig.~\ref{fig:RGE_induced}. The corresponding matching relations of the $X^2\phi^2$ operators to the
leptonic and hadronic EDMs are collected in Eqs.~\eqref{Append_eq:lepEDM_log_enhance}--\eqref{Append_eq:qEDM_log_enhance} in Appendix~\ref{Append:Logarithmic Enhancement Contribution}.
The matching relation for $\mathcal{O}^{lequ(3)}$ has been discussed
previously in Eq.~\eqref{eq:op_lequ3}.
%----------------------------
%--------------------------------------------
%---------------------------------------------
\begin{table}[t!]
	\centering
	\renewcommand{\arraystretch}{1.3}
	\setlength{\tabcolsep}{6pt}
	\begin{tabular}{@{} c c c @{}}
		\toprule
		\textbf{WCs~ [$\mathrm{TeV}^{-2}$]}
		& \textbf{$|d_e|$ (Current)}
		& \textbf{$|d_e|$ (Future)} \\
		\midrule
		
		$\displaystyle
		\mathrm{Im}(C^{uW}_{33}
		)$
		& $3.69\times10^{-5}$
		& $0.89\times10^{-6}$ \\
		
		$\displaystyle
		\mathrm{Im}(
		C^{uB}_{33})$
		& $7.15\times10^{-5}$
		& $1.74\times10^{-6}$ \\
		
		$\displaystyle
		\mathrm{Im}(C^{dW}_{33}
		)$
		& $2.66\times10^{-3}$
		& $6.49\times10^{-5}$ \\
		
		$\displaystyle
		\mathrm{Im}(C^{dW}_{23}
		)$
		& $1.24\times10^{-1}$
		& $3.01\times10^{-3}$ \\
		
		$\displaystyle
		\mathrm{Im}(C^{dW}_{13}
		)$
		& $5.89\times10^{-1}$
		& $1.43\times10^{-2}$ \\
		
		$\displaystyle
		\mathrm{Im}(C^{lequ(1)}_{1133})
		$
		& $ 8.85 \times 10^{-10} $
		& $2.15 \times 10^{-11}$\\
		\midrule
		
%		& \textbf{$|d_n|$ (Current)}
%		& \textbf{$|d_p|$ (Future)} \\
%		\midrule
%		$\displaystyle
%		\mathrm{Im}(C^{uG}_{33}
%		)$
%		& $ 1.19 \times 10^{-1}  $
%		& $1.30\times10^{-4}$\\
%		\bottomrule
	\end{tabular}
	\caption{Bounds on the top-dipole SMEFT WCs from double-logarithmically
		enhanced two-loop RGE contributions at the NP scale,
		$\mu=1~\mathrm{TeV}$.}
	\label{tab:bounds_sequential_loop}
\end{table}
%----------------------------------

We summarize the bounds on these RGE-induced contributions in
Table~\ref{tab:bounds_sequential_loop}. We find that including the
RGE-induced contributions strengthens the bounds on the imaginary parts
of the flavor-conserving top-dipole coefficients by approximately two
orders of magnitude for $C^{uW}_{33}$ and $C^{uB}_{33}$, compared with
those obtained from the finite two-loop Barr-Zee contributions to
$d_e$. For the charged-current operator $C^{dW}_{33}$, the inclusion of
RGE effects improve the constraint by at least one order of magnitude.
The scalar four-fermion coefficient $C^{lequ(1)}_{1133}$ is constrained
at the level of $\mathcal{O}(10^{-10})$. 

% The corresponding one-loop
% matching relations for the bosonic operator WCs are provided in
% Appendix~\ref{Append:Logarithmic Enhancement Contribution}, while the associated RGE mixing contributions are
% collected in Eq.~\eqref{eq:log_enhanced_ADM}.

%-------------------------
\subsection{Constraints from Hadronic EDMs}\label{subsec:Hadronic_EDMs}
%----------------------------
We next investigate the constraints on $CP$-violating top-quark interactions arising from hadronic EDM observables. The relevant SMEFT operators generate, through one- and two-loop matching contributions at the electroweak scale, the low-energy hadronic effective Lagrangian defined in Eq.~\ref{eq:lag_quark_EDM}, which consists of the quark electric dipole moments $d_q$, quark chromoelectric dipole moments $\tilde d_q$, and the Weinberg three-gluon operator $w$. The corresponding WCs are subsequently evolved from the electroweak scale to the hadronic scale using the appropriate renormalization-group equations, including heavy-quark threshold corrections. At low energies, these operators induce contributions to the neutron and proton EDMs, $d_n$ and $d_p$, as well as to the $CP$-odd pion-nucleon couplings $\bar g_0$ and $\bar g_1$. These hadronic quantities, in turn, contribute to a wide variety of observables, including nucleon EDMs, diamagnetic atomic EDMs, and light-nuclear EDMs. In particular, we consider the EDMs of $^{199}$Hg, $^{129}$Xe, $^{171}$Yb, and $^{225}$Ra, together with the deuteron and helion EDMs. As discussed previously, the most stringent current constraint on hadronic $CP$ violation arises from the EDM of $^{199}$Hg. Several ongoing and proposed experiments aim to significantly improve the sensitivities of proton, diamagnetic-atom, and light-nuclear EDM measurements, and we include these projected sensitivities in our analysis. In the following, we first discuss the constraints obtained from the neutron and proton EDMs, and subsequently those arising from diamagnetic atomic and light-nuclear EDM observables.

Unlike the leptonic EDMs, where most of the relevant top-quark
interactions contribute predominantly through finite two-loop Barr-Zee
diagrams. In contrast, the hadronic EDM constraints obtained in this work arise from several distinct matching topologies, each probing different classes of CP-violating top-quark interactions. This classification provides valuable insight into the origin of the resulting bounds and the complementarity among the various hadronic observables. In particular, the same SMEFT operators can contribute
through different loop orders and matching structures, depending on the
low-energy operator they induce. We therefore organize the hadronic
EDM analysis according to the underlying contribution topology rather
than the flavor structure of the SMEFT operators. Specifically, we
separate the contributions arising at one loop, from finite two-loop
Barr-Zee diagrams, and from the Weinberg three-gluon contribution. This classification allows us to identify which topological contributions
provide the dominant sensitivity to the different hadronic EDM
observables.

%---------------------------------
\subsubsection{One-Loop contributions}\label{subsec:Hadronic_oneloop}
%---------------------------------
%---------------------------------
\begin{figure}[t!]
    \centering
    %----------------------------------------
	\subfloat[]{\label{fig:EDM_double_a}%
        \resizebox{0.48\linewidth}{!}{%
            \begin{tikzpicture}
                \begin{feynman}
                    \vertex[](a1){\(u_i\)};
                    \vertex[square dot,blue,right=0.8cm of a1](a2){};
                    \vertex[right=0.8cm of a2](a3);
                    \vertex[square dot,blue,right=0.8cm of a3](a4){};
                    \vertex[right=0.8cm of a4](a5){\(u_i\)};
                    \vertex[below= 0.9cm of a3](a6){\(\gamma,g\)};
                    \diagram*{
                        (a1) --[very thick,fermion](a2) --[very thick, fermion, edge label'=\(t\)](a3) --[very thick, fermion, edge label'=\(t\)](a4) --[very thick, fermion](a5),
                        (a3) --[very thick,boson](a6),
                        (a2) --[blue,gluon,very thick, half left, looseness=1.5, edge label=\(g\)](a4)
                    };
                \end{feynman}
            \end{tikzpicture}%
        }%
	}\hfill%
	%----------------------------------------
	\subfloat[]{\label{fig:EDM_double_b}%
        \resizebox{0.48\linewidth}{!}{%
            \begin{tikzpicture}
                \begin{feynman}
                    \vertex[](a1){\(u_i\)};
                    \vertex[square dot,red,right=0.8cm of a1](a2){};
                    \vertex[right=0.8cm of a2](a3);
                    \vertex[square dot,red,right=0.8cm of a3](a4){};
                    \vertex[right=0.8cm of a4](a5){\(u_i\)};
                    \vertex[below= 0.9cm of a3](a6){\(\gamma,g\)};
                    \diagram*{
                        (a1) --[very thick,fermion](a2) --[very thick,fermion, edge label'=\(t\)](a3) --[very thick,fermion, edge label'=\(t\)](a4) --[very thick,fermion](a5),
                        (a3) --[very thick,boson](a6),
                        (a2) --[red,boson,very thick, half left, looseness=1.5, edge label=\(\gamma/Z\)](a4)
                    };
                \end{feynman}
            \end{tikzpicture}%
        }%
	}\\
	%----------------------------------------
	\subfloat[]{\label{fig:EDM_double_c}%
        \resizebox{0.48\linewidth}{!}{%
            \begin{tikzpicture}
                \begin{feynman}
                    \vertex[](a1){\(u_i\)};
                    \vertex[square dot,violet,right=0.8cm of a1](a2){};
                    \vertex[right=0.8cm of a2](a3);
                    \vertex[square dot,violet,right=0.8cm of a3](a4){};
                    \vertex[right=0.8cm of a4](a5){\(u_i\)};
                    \vertex[below= 0.9cm of a3](a6){\(\gamma,g\)};
                    \diagram*{
                        (a1) --[very thick,fermion](a2) --[very thick,fermion, edge label'=\(t\)](a3) --[very thick,fermion, edge label'=\(t\)](a4) --[very thick,fermion](a5),
                        (a3) --[very thick,boson](a6),
                        (a2) --[violet,scalar,very thick, half left, looseness=1.5, edge label=\(H\)](a4)
                    };
                \end{feynman}
            \end{tikzpicture}%
        }%
	}\hfill%
	%----------------------------------------
    \subfloat[]{\label{fig:CEDM_three_gluon}%
        \resizebox{0.48\linewidth}{!}{%
            \begin{tikzpicture}
                \begin{feynman}
                    \vertex[](a1){\(u_i\)};
                    \vertex[square dot,blue,right=0.8cm of a1](a2){};
                    
                    % Invisible midpoint to center the three-gluon vertex
                    \vertex[right=0.8cm of a2](mid); 
                    \vertex[square dot,blue,right=0.8cm of mid](a4){};
                    \vertex[right=0.8cm of a4](a5){\(u_i\)};
                    
                    % Standard Model 3-gluon interaction node (no NP styling)
                    \vertex[above=1.0cm of mid](a3);
                    
                    % The emitted external gluon
                    \vertex[above=0.8cm of a3](a6){\(g\)};
                    
                    \diagram*{
                        % Continuous internal heavy quark line
                        (a1) --[very thick,fermion](a2) --[very thick, fermion, edge label'=\(t\)](a4) --[very thick, fermion](a5),
                        
                        % Segmented gluon arc meeting at the SM 3-gluon vertex
                        (a2) --[blue, gluon, very thick, quarter left](a3) --[blue, gluon, very thick, quarter left](a4),
                        
                        % External gluon emission
                        (a3) --[blue, gluon, very thick](a6)
                    };
                \end{feynman}
            \end{tikzpicture}%
        }%
    }\\
    %--------------------------
    \subfloat[]{\label{fig:qEDM_FCCC_a}%
        \resizebox{0.48\linewidth}{!}{%
            \begin{tikzpicture}
                \begin{feynman}
                    \vertex[](a1){\(d_i\)};
                    \vertex[square dot,orange,right=0.8cm of a1](a2){};
                    \vertex[right=0.8cm of a2](a3);
                    \vertex[square dot,red,right=0.8cm of a3](a4);
                    \vertex[right=0.8cm of a4](a5){\(d_i\)};
                    \vertex[below= 0.9cm of a3](a6){\(\gamma,g\)};
                    \diagram*{
                        (a1) --[very thick,fermion](a2) --[very thick,fermion, edge label'=\(t\)](a3) --[very thick,fermion, edge label'=\(t\)](a4) --[very thick,fermion](a5),
                        (a3) --[very thick,boson](a6),
                        (a2) --[orange ,boson,very thick, half left, looseness=1.5, edge label=\(W\)](a4)
                    };
                \end{feynman}
            \end{tikzpicture}%
        }%
	}\hfill%
    %--------------------------------------
    \subfloat[]{\label{fig:qEDM_FCCC_b}%
        \resizebox{0.48\linewidth}{!}{%
            \begin{tikzpicture}
                \begin{feynman}
                    \vertex[](a1){\(d_i\)};
                    \vertex[square dot,orange,right=0.8cm of a1](a2){};
                    
                    % Invisible midpoint to center the three-gluon vertex
                    \vertex[right=0.8cm of a2](mid); 
                    \vertex[square dot,red,right=0.8cm of mid](a4);
                    \vertex[right=0.8cm of a4](a5){\(d_i\)};
                    
                    % Standard Model 3-gluon interaction node (no NP styling)
                    \vertex[above=1.0cm of mid](a3);
                    
                    % The emitted external gluon
                    \vertex[above=0.8cm of a3](a6){\(\gamma\)};
                    
                    \diagram*{
                        % Continuous internal heavy quark line
                        (a1) --[very thick,fermion](a2) --[very thick, fermion, edge label'=\(t\)](a4) --[very thick, fermion](a5),
                        
                        % Segmented W boson arc
                        (a2) --[orange, boson, very thick, quarter left, edge label=\(W\)](a3) --[orange, boson, very thick, quarter left, edge label=\(W\)](a4),
                        
                        % External photon emission
                        (a3) --[black, boson, very thick](a6)
                    };
                \end{feynman}
            \end{tikzpicture}%
        }%
    }\hfill
     %-----Diag-Contact-EDM-------------
\subfloat[]{\label{fig:quark_edm_contact1}%
    \resizebox{0.48\linewidth}{!}{%
        \begin{tikzpicture}
            \begin{feynman}
                % External lepton lines
                \vertex (a1) {\(q_i\)};
                \vertex[square dot, black, right=1.2cm of a1] (v1) {};
                \vertex[right=1.2cm of v1] (a2) {\(q_i\)};
                
                % Lower vertex of the loop
                \vertex[below=0.8cm of v1] (v2);
                
                % External photon
                \vertex[below=0.5cm of v2] (gamma) {\(\gamma,g\)};
                
                \diagram*{
                    % Horizontal lepton line
                    (a1) -- [ thick, fermion] (v1) -- [ thick, fermion] (a2),
                    
                    % Fermion loop (clockwise flow)
                    % Right side arc (v1 down to v2)
                    (v1) -- [ thick, fermion, half left, looseness=1.5, edge label=\(t\)] (v2),
                    % Left side arc (v2 up to v1)
                    (v2) -- [ thick, fermion, half left, looseness=1.5, edge label=\(t\)] (v1),
                    
                    % Photon emission
                    (v2) -- [ thick, boson] (gamma)
                };
            \end{feynman}
        \end{tikzpicture}%
    }%
}
 %-----Diag-Contact-EDM-------------
\subfloat[]{\label{fig:quark_edm_contact2}%
    \resizebox{0.48\linewidth}{!}{%
        \begin{tikzpicture}
            \begin{feynman}
                % External lepton lines
                \vertex (a1) {\(q_i\)};
                \vertex[square dot, black, right=1.2cm of a1] (v1) {};
                \vertex[right=1.2cm of v1] (a2) {\(q_i\)};
                
                % Lower vertex of the loop
                \vertex[below=0.8cm of v1] (v2);
                
                % External photon
                \vertex[below=0.5cm of v2] (gamma) {\(g\)};
                
                \diagram*{
                    % Horizontal lepton line
                    (a1) -- [ thick, fermion] (v1) -- [ thick, fermion] (a2),
                    
                    % Fermion loop (clockwise flow)
                    % Right side arc (v1 down to v2)
                    (v1) -- [ thick, fermion, half left, looseness=1.5, edge label=\(t\)] (v2),
                    % Left side arc (v2 up to v1)
                    (v2) -- [ thick, fermion, half left, looseness=1.5, edge label=\(t\)] (v1),
                    
                    % Photon emission
                    (v2) -- [ thick, gluon] (gamma)
                };
            \end{feynman}
        \end{tikzpicture}%
    }%
}\\
%---------------------------------------------
 %---------------------------------------
\subfloat[]{\label{fig:Weinberg_EDM_FC1}
		\begin{tikzpicture}
			\begin{feynman}
				\def\r{0.9 cm}
				\coordinate (O) at (0,0);
				\draw[very thick] (O) circle (\r);
				\draw[fermion, very thick] (360:\r) arc (360:180:\r);
				\vertex[square dot, blue] (A) at (90:\r){} ;   
				\vertex (B) at (210:\r) ;   
				\vertex (C) at (330:\r) ;   
				\vertex (D) at (0:\r)   {}; 
				\vertex (E) at (180:\r) {};
				\vertex[above=0.9 of A](F){\(g\)};
				\vertex[ below left=0.5cm of B](G){\(g\)};
				\vertex[below right=0.5cm of C](H){\(g\)};
				\vertex[below=1.0cm of O](a1) {\(t\)};
				\vertex[above right=1.0 cm of O](a2) {\(t\)};
				\vertex[square dot, violet, at=(D)] ;
				\vertex[square dot, violet, at=(E)] ;
				\diagram*{
					(A) --[very thick,blue, gluon](F),
					(B) --[very thick, gluon](G),
					(C) --[very thick, gluon](H),
				};
			\end{feynman}
		\end{tikzpicture}
	}
%---------------------------------------
\subfloat[]{\label{fig:Weinberg_EDM_FC2}
		\begin{tikzpicture}
			\begin{feynman}
				\def\r{0.9 cm}
				\coordinate (O) at (0,0);
				\draw[very thick] (O) circle (\r);
				\draw[fermion, very thick] (360:\r) arc (360:180:\r);
				\vertex (A) at (90:\r) ;   
				\vertex[square dot, blue] (B) at (210:\r){};   
				\vertex[square dot, blue] (C) at (330:\r){} ;   
				\vertex (D) at (0:\r)   {}; 
				\vertex (E) at (180:\r) {};
				\vertex[above=0.5 of A](F){\(g\)};
				\vertex[ below left=0.9cm of B](G){\(g\)};
				\vertex[below right=0.9cm of C](H){\(g\)};
				\vertex[below=1.1cm of O](a1) {\(u_i\)};
				\vertex[above right=1.0 cm of O](a2) {\(t\)};
				\vertex[square dot, violet, at=(D)] ;
				\vertex[square dot, violet, at=(E)] ;
				\diagram*{
					(A) --[very thick,  gluon](F),
					(B) --[very thick,blue, gluon](G),
					(C) --[very thick,blue, gluon](H),
				};
			\end{feynman}
		\end{tikzpicture}
	}
%----------------------------------------  
    \caption{Representative one-loop topologies contributing to the EDMs and
CEDMs of quarks through anomalous top-quark interactions. Here,
$q_i=u_i,d_i$, with $u_i=u,c$ and $d_i=d,s,b$.}
    \label{fig:Had_EDM_FCNC_oneloop}
\end{figure}
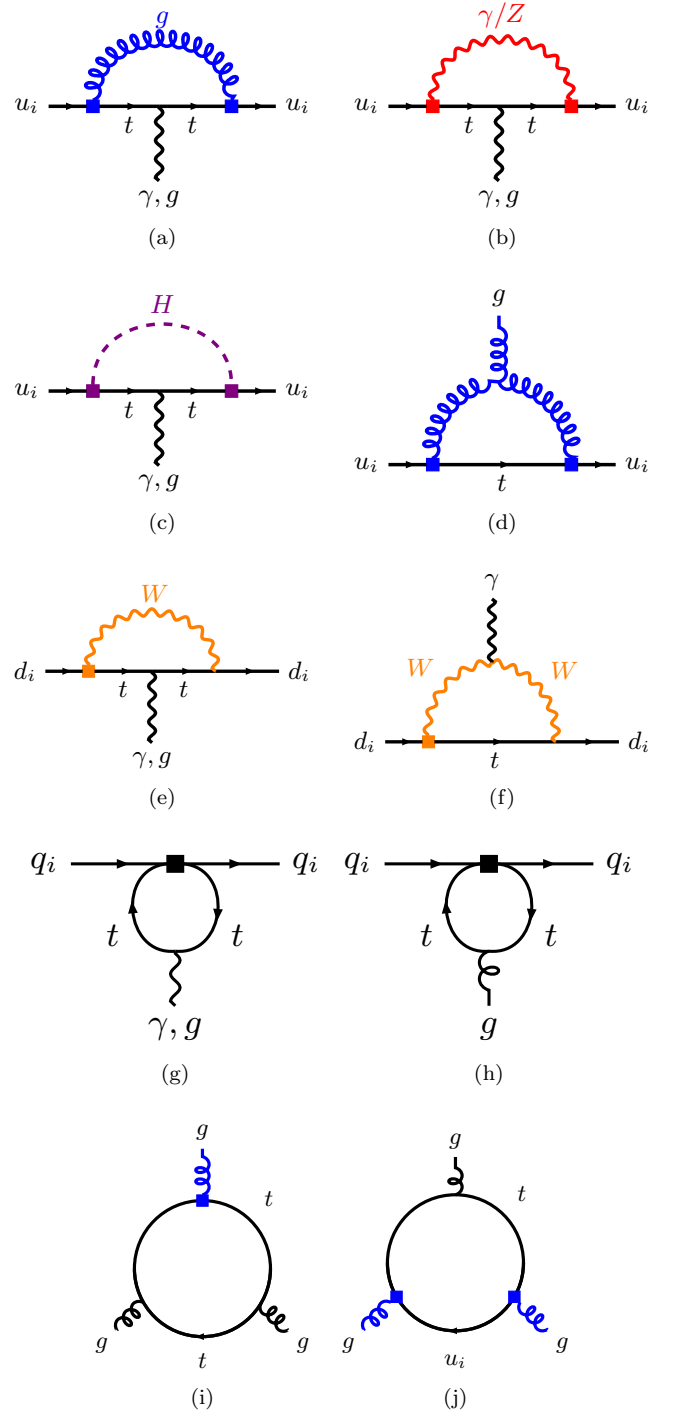
%----------------------------------------

The first class consists of direct one-loop contributions to the light-quark EDMs and CEDMs. Flavor-changing neutral-current top-quark interactions contribute through the topologies shown in Figs.~\ref{fig:EDM_double_a}-\ref{fig:CEDM_three_gluon} and Fig.~\ref{fig:Weinberg_EDM_FC2}, while flavor-violating charged-current interactions contribute through Figs.~\ref{fig:qEDM_FCCC_a} and \ref{fig:qEDM_FCCC_b}. Four-quark interactions generate quark EDMs and CEDMs through the contact topologies shown in Figs.~\ref{fig:quark_edm_contact1} and \ref{fig:quark_edm_contact2}. These diagrams induce the low-energy operators $d_q$ and $\tilde d_q$, which subsequently contribute to $d_n$, $d_p$, $\bar g_0$, and $\bar g_1$. We also include the effects of the charm- and bottom-quark CEDMs, $\tilde d_c$ and $\tilde d_b$, which contribute to hadronic EDM observables through heavy-quark threshold corrections to the Weinberg operator and through their subsequent RGE-induced contributions to the light-quark EDMs and CEDMs. The corresponding matching conditions at the electroweak scale, $\mu_{\rm EW}$, are summarized in Appendix~\ref{Append:Hadronic EDM_oneloop}. The constraints obtained from the one-loop contributions are presented in Table~\ref{tab:top_WC_values_FCCC}, which naturally separates into three qualitatively distinct categories according to the underlying mechanism responsible for the low-energy CP-violating observables.

\begin{table*}[htbp]
	\centering
	\renewcommand{\arraystretch}{1.2}
	\setlength{\tabcolsep}{8pt}
	\begin{tabular}{@{} c c c c c c @{}}
		\toprule
		\textbf{WCs~$\,[\mathrm{TeV}^{-2}]$}
		& \textbf{$|d_n|$ (Current)}
		& \textbf{$|\bar{g}_0|$ (Current)}
		& \textbf{$|\bar{g}_1|$ (Current)}
		& \textbf{$|d_n|$ (Future)}
		& \textbf{$|d_p|$ (Future)} \\
		\midrule
		
		$\mathrm{Im}(C^{uG}_{33})$
		& $\mathbf{4.05\times10^{-3}}$
		& $3.01\times10^{-2}$
		& $4.98\times10^{-2}$
		& $2.25\times10^{-4}$
		& $2.76\times10^{-6}$ \\
		
		$\mathrm{Im}(C^{dW}_{13})$
		& $7.54\times10^{-6}$
		& $1.27\times10^{-5}$
		& $\mathbf{3.52\times10^{-7}}$
		& $4.19\times10^{-7}$
		& $1.56\times10^{-8}$ \\
		
		$\mathrm{Im}(C^{dW}_{23})$
		& $\mathbf{1.45\times10^{-3}}$
		& \xmark 
		&\xmark 
		& $8.04\times10^{-5}$
		& $8.04\times10^{-7}$ \\
		
		$\mathrm{Im}(C^{\phi ud}_{13})$
		& $9.96\times10^{-4}$
		& $4.64\times10^{-4}$
		& $\mathbf{1.29\times10^{-5}}$
		& $5.53\times10^{-5}$
		& $1.90\times10^{-6}$ \\
		
		$\mathrm{Im}(C^{\phi ud}_{23})$
		& $\mathbf{4.14\times10^{-2}}$
		& \xmark 
		& \xmark 
		& $2.30\times10^{-3}$
		& $2.30\times10^{-5}$ \\
		
		$\mathrm{Im}(C^{qu(1)}_{1331})$
		& $9.98\times10^{-7}$
		& $4.80\times10^{-7}$
		& $\mathbf{1.32\times10^{-8}}$
		& $5.54\times10^{-8}$
		& $2.69\times10^{-10}$ \\
		
		$\mathrm{Im}(C^{qu(8)}_{1331})$
		& $1.42\times10^{-6}$
		& $2.60\times10^{-6}$
		& $\mathbf{7.20\times10^{-8}}$
		& $7.88\times10^{-8}$
		& $3.39\times10^{-10}$ \\
		
		$\mathrm{Im}(C^{quqd(1)}_{1331})$
		& $1.09\times10^{-7}$
		& $1.83\times10^{-7}$
		& $\mathbf{5.07\times10^{-9}}$
		& $6.11\times10^{-9}$
		& $5.71\times10^{-10}$ \\
		
		$\mathrm{Im}(C^{quqd(1)}_{2332})$
		& $\mathbf{1.03\times10^{-4}}$
		& \xmark 
		& \xmark 
		& $5.70\times10^{-6}$
		& $5.70\times10^{-8}$ \\
		
		$\mathrm{Im}(C^{quqd(8)}_{1331})$
		& $7.48\times10^{-7}$
		& $1.47\times10^{-6}$
		& $\mathbf{4.07\times10^{-8}}$
		& $4.15\times10^{-8}$
		& $7.25\times10^{-10}$ \\
		
		$\mathrm{Im}(C^{quqd(8)}_{2332})$
		& $\mathbf{1.37\times10^{-4}}$
		& \xmark 
		& \xmark 
		& $7.60\times10^{-6}$
		& $7.60\times10^{-8}$ \\
		
		\midrule
		
		\textcolor{lightblue}{$\mathrm{Im}(C^{dW}_{33})$}
		& \textcolor{lightblue}{$\mathbf{3.47\times10^{-3}}$}
		& \textcolor{lightblue}{$2.13\times10^{-2}$}
		& \textcolor{lightblue}{$3.53\times10^{-2}$}
		& \textcolor{lightblue}{$1.93\times10^{-4}$}
		& \textcolor{lightblue}{$2.35\times10^{-6}$} \\
		
		\textcolor{lightblue}{$\mathrm{Im}(C^{\phi ud}_{33})$}
		& \textcolor{lightblue}{$\mathbf{1.22\times10^{-1}}$}
		& \textcolor{lightblue}{$7.48\times10^{-1}$}
		& \textcolor{lightblue}{$1.24\times10^{-1}$}
		& \textcolor{lightblue}{$6.75\times10^{-3}$}
		& \textcolor{lightblue}{$8.23\times10^{-5}$} \\
		
		\textcolor{lightblue}{$\mathrm{Im}(C^{qu(1)}_{2332})$}
		& \textcolor{lightblue}{$\mathbf{1.19\times10^{-3}}$}
		& \textcolor{lightblue}{$6.86\times10^{-3}$}
		& \textcolor{lightblue}{$1.14\times10^{-2}$}
		& \textcolor{lightblue}{$6.60\times10^{-5}$}
		& \textcolor{lightblue}{$1.59\times10^{-7}$} \\
		
		\textcolor{lightblue}{$\mathrm{Im}(C^{qu(8)}_{2332})$}
		& \textcolor{lightblue}{$\mathbf{7.11\times10^{-3}}$}
		& \textcolor{lightblue}{$4.12\times10^{-2}$}
		& \textcolor{lightblue}{$6.81\times10^{-2}$}
		& \textcolor{lightblue}{$3.95\times10^{-4}$}
		& \textcolor{lightblue}{$4.78\times10^{-6}$} \\
		
		\textcolor{lightblue}{$\mathrm{Im}(C^{quqd(1)}_{3333})$}
		& \textcolor{lightblue}{$\mathbf{5.37\times10^{-3}}$}
		& \textcolor{lightblue}{$4.85\times10^{-2}$}
		& \textcolor{lightblue}{$8.03\times10^{-2}$}
		& \textcolor{lightblue}{$2.68\times10^{-4}$}
		& \textcolor{lightblue}{$3.64\times10^{-6}$} \\
		
		\textcolor{lightblue}{$\mathrm{Im}(C^{quqd(8)}_{3333})$}
		& \textcolor{lightblue}{$\mathbf{5.66\times10^{-2}}$}
		& \textcolor{lightblue}{$2.91\times10^{-1}$}
		& \textcolor{lightblue}{$4.82\times10^{-1}$}
		& \textcolor{lightblue}{$3.14\times10^{-3}$}
		& \textcolor{lightblue}{$3.83\times10^{-5}$} \\
		
		\midrule
		
		\textcolor{DarkEmerald}{$\mathrm{Im}(C^{uB}_{33})$}
		& \textcolor{DarkEmerald}{$\mathbf{2.21\times10^{-1}}$}
		& \textcolor{DarkEmerald}{$1.64$}
		& \textcolor{DarkEmerald}{$2.72$}
		& \textcolor{DarkEmerald}{$1.23\times10^{-2}$}
		& \textcolor{DarkEmerald}{$1.50\times10^{-4}$} \\
		
		\textcolor{DarkEmerald}{$\mathrm{Im}(C^{uW}_{33})$}
		& \textcolor{DarkEmerald}{$\mathbf{6.78\times10^{-2}}$}
		& \textcolor{DarkEmerald}{$5.02\times10^{-1}$}
		& \textcolor{DarkEmerald}{$8.33\times10^{-1}$}
		& \textcolor{DarkEmerald}{$3.77\times10^{-3}$}
		& \textcolor{DarkEmerald}{$4.61\times10^{-5}$} \\
		
		\bottomrule
	\end{tabular}
	\caption{Bounds on the SMEFT WCs from one-loop, single-insertion
		contributions to the hadronic EDMs. The table is divided into three
		groups: WCs directly constrained by light-quark EDMs and CEDMs;
		\textcolor{lightblue}{blue} entries induced predominantly by charm-
		and bottom-quark CEDM threshold corrections; and
		\textcolor{DarkEmerald}{green} entries arising predominantly from RGE-induced
		operator mixing. The WCs are evaluated at $\mu=1~\mathrm{TeV}$ and have
		mass dimension $\mathrm{TeV}^{-2}$.}
	\label{tab:top_WC_values_FCCC}
\end{table*}
%-----------------------------------

%---------------------------------------------
% Table Hadronic EDM at one loop (Top FCNC)
\begin{table*}[t!]
	\centering
	\renewcommand{\arraystretch}{0.6}
	\setlength{\tabcolsep}{5pt}
	\begin{tabular}{@{} c c c c c c @{}}
		\toprule
		\textbf{Effective Couplings}
		& \textbf{$|d_n|$ (Current)}
		& \textbf{$|\bar{g}_0|$ (Current)}
		& \textbf{$|\bar{g}_1|$ (Current)}
		& \textbf{$|d_n|$ (Future)}
		& \textbf{$|d_p|$ (Future)} \\
		\midrule
		
		$\displaystyle
		\mathrm{Im}\left(\xi_L^{ut}\xi_R^{tu*}\right)$
		& $2.20\times10^{-8}$
		& $1.33\times10^{-8}$
		& $\mathbf{3.68\times10^{-9}}$
		& $4.73\times10^{-9}$
		& $1.64\times10^{-11}$ \\
		
		$\displaystyle
		\mathrm{Im}\left(\lambda_L^{ut}\lambda_R^{tu*}\right)$
		& $2.19\times10^{-7}$
		& $3.86\times10^{-7}$
		& $\mathbf{1.07\times10^{-8}}$
		& $4.48\times10^{-8}$
		& $6.82\times10^{-10}$ \\
		
		$\displaystyle
		\mathrm{Im}\left(\eta_L^{ut}\eta_R^{tu*}\right)$
		& $1.61\times10^{-7}$
		& $2.84\times10^{-7}$
		& $\mathbf{7.85\times10^{-9}}$
		& $3.29\times10^{-8}$
		& $5.01\times10^{-10}$ \\
		
		$\displaystyle
		\mathrm{Im}\left(X_L^{ut}X_R^{tu*}\right)$
		& $2.54\times10^{-7}$
		& $3.54\times10^{-7}$
		& $\mathbf{9.81\times10^{-9}}$
		& $4.12\times10^{-8}$
		& $6.26\times10^{-10}$ \\
		
		$\displaystyle
		\mathrm{Im}\left(\kappa_L^{ut}\kappa_R^{tu*}\right)$
		& $3.63\times10^{-8}$
		& $5.06\times10^{-8}$
		& $\mathbf{1.40\times10^{-9}}$
		& $5.88\times10^{-9}$
		& $8.94\times10^{-11}$ \\
		
		\midrule
		
		\textcolor{lightblue}{$\displaystyle
			\mathrm{Im}\left(\xi_L^{ct}\xi_R^{tc*}\right)$}
		& \textcolor{lightblue}{$\mathbf{0.89\times10^{-4}}$}
		& \textcolor{lightblue}{$2.17\times10^{-3}$}
		& \textcolor{lightblue}{$3.58\times10^{-3}$}
		& \textcolor{lightblue}{$1.98\times10^{-5}$}
		& \textcolor{lightblue}{$2.39\times10^{-7}$} \\
		
		\textcolor{lightblue}{$\displaystyle
			\mathrm{Im}\left(\lambda_L^{ct}\lambda_R^{tc*}\right)$}
		& \textcolor{lightblue}{$\mathbf{2.39\times10^{-4}}$}
		& \textcolor{lightblue}{$5.83\times10^{-3}$}
		& \textcolor{lightblue}{$9.66\times10^{-3}$}
		& \textcolor{lightblue}{$5.32\times10^{-5}$}
		& \textcolor{lightblue}{$6.43\times10^{-7}$} \\
		
		\textcolor{lightblue}{$\displaystyle
			\mathrm{Im}\left(\eta_L^{ct}\eta_R^{tc*}\right)$}
		& \textcolor{lightblue}{$\mathbf{1.76\times10^{-4}}$}
		& \textcolor{lightblue}{$4.28\times10^{-3}$}
		& \textcolor{lightblue}{$7.09\times10^{-3}$}
		& \textcolor{lightblue}{$3.91\times10^{-5}$}
		& \textcolor{lightblue}{$4.73\times10^{-7}$} \\
		
		\textcolor{lightblue}{$\displaystyle
			\mathrm{Im}\left(X_L^{ct}X_R^{tc*}\right)$}
		& \textcolor{lightblue}{$\mathbf{2.79\times10^{-4}}$}
		& \textcolor{lightblue}{$5.36\times10^{-3}$}
		& \textcolor{lightblue}{$8.87\times10^{-3}$}
		& \textcolor{lightblue}{$4.89\times10^{-5}$}
		& \textcolor{lightblue}{$5.91\times10^{-7}$} \\
		
		\textcolor{lightblue}{$\displaystyle
			\mathrm{Im}\left(\kappa_L^{ct}\kappa_R^{tc*}\right)$}
		& \textcolor{lightblue}{$\mathbf{3.97\times10^{-5}}$}
		& \textcolor{lightblue}{$5.65\times10^{-4}$}
		& \textcolor{lightblue}{$8.36\times10^{-4}$}
		& \textcolor{lightblue}{$6.98\times10^{-6}$}
		& \textcolor{lightblue}{$8.45\times10^{-8}$} \\
		
		\bottomrule
	\end{tabular}
	\caption{Bounds on products of dimensionless top-quark FCNC couplings
		from one-loop contributions to the hadronic EDMs. The first group
		contains constraints from light-quark EDMs and CEDMs, while the
		\textcolor{lightblue}{blue} entries in the second group arise
		predominantly through charm- and bottom-quark CEDM threshold corrections.}
	\label{tab:qEDM_FCNC_oneloop}
\end{table*}
%---------------------------------------------

The \textbf{\textit{first group}} contains operators that contribute directly to the light-quark EDMs and CEDMs through the one-loop topologies shown in Figs.~\ref{fig:qEDM_FCCC_a}-\ref{fig:Weinberg_EDM_FC1}. In this category, the neutral-current top-gluon dipole coefficient $C^{uG}_{33}$ contributes through the Weinberg operator $w$ shown in Fig.~\ref{fig:Weinberg_EDM_FC1}, while the charged-current coefficients $C^{dW}_{i3}$ and $C^{\phi ud}_{i3}$ induce direct contributions to $d_{d,s}$ and $\tilde d_{d,s}$. The four-fermion coefficients $C^{qu(1)}_{1331}$ and $C^{qu(8)}_{1331}$ contribute to $d_u$ and $\tilde d_u$, respectively, whereas $C^{quqd(1)}_{i33i}$ and $C^{quqd(8)}_{i33i}$ generate $d_{d,s}$ and $\tilde d_{d,s}$. These operators primarily feed into the neutron and proton EDMs through the light-quark dipole operators and, more importantly, into the CP-odd pion-nucleon couplings $\bar g_0$ and $\bar g_1$. Consequently, some of the strongest hadronic constraints arise in this class. In particular, the coefficient $C^{dW}_{13}$ is constrained at the level of ${\cal O}(10^{-7})\,{\rm TeV}^{-2}$, while the four-quark coefficients $C^{quqd(1)}_{1331}$ and $C^{quqd(8)}_{1331}$ are probed at the ${\cal O}(10^{-8})-{\cal O}(10^{-9})\,{\rm TeV}^{-2}$ level level through the pion-nucleon coupling $\bar g_1$. The coefficients $C^{{qu}(1,8)}_{1331}$ are constrained at the ${\cal O}(10^{-8})\,{\rm TeV}^{-2}$ level, while the charged-current current operator $C^{\phi ud}$ is constrained at approximately ${\cal O}(10^{-5})\,{\rm TeV}^{-2}$.

Furthermore, a distinct class of contributions arises from the one-loop generation of the Weinberg three-gluon operator. The flavor-conserving top chromodipole interaction induces the Weinberg operator through the topology shown in Fig.~\ref{fig:Weinberg_EDM_FC1}, while flavor-changing top-gluon interactions contribute through the double-insertion topology depicted in Fig \ref{fig:Weinberg_EDM_FC2}. Unlike the quark EDM and CEDM operators, the Weinberg operator does not suffer from light-quark mass suppression and therefore provides a particularly powerful probe of gluonic sources of $CP$ violation. Consequently, this class yields the strongest sensitivity to the top chromodipole operator $C^{uG}_{33}$, which is currently constrained at the level of ${\cal O}(10^{-3})\,{\rm TeV}^{-2}$ by neutron-EDM measurements. Future proton-EDM experiments are expected to significantly enhance this reach, potentially probing $C^{uG}_{33}$ down to the ${\cal O}(10^{-6})\,{\rm TeV}^{-2}$ level. These results highlight the unique role of hadronic EDM observables in constraining gluonic $CP$-violating interactions that are only indirectly accessible through leptonic observables.

The \textbf{\textit{second group}}, highlighted in \textcolor{lightblue}{blue}, contains operators whose dominant sensitivity arises through threshold corrections induced by the heavy-quark CEDMs $\tilde d_c$ and $\tilde d_b$, 
% The corresponding contributions are generated through the one-loop Weinberg-operator topologies shown in Figs.~\ref{fig:CEDM_three_gluon}, \ref{fig:Weinberg_EDM_FC1}, and \ref{fig:Weinberg_EDM_FC2}, 
followed by the threshold corrections and RGE evolution described in Eqs.~\eqref{eq:threshold_corr} and \eqref{eq:charm_RGE_EDM}. Since these effects involve an additional heavy-quark loop and subsequent RG evolution, they are generally suppressed relative to the direct one-loop contributions discussed above. As a result, the corresponding WCs are typically constrained at the ${\cal O}(10^{-3})$-${\cal O}(10^{-1})\,{\rm TeV}^{-2}$ level, considerably weaker than those entering directly through the light-quark EDMs and CEDMs. Nevertheless, these threshold effects provide a unique sensitivity to several operator structures that would otherwise be only weakly constrained.

The \textbf{\textit{third group}}, highlighted in \textcolor{DarkEmerald}{green}, corresponds to WCs whose leading sensitivity is generated through operator mixing under the SMEFT and WET renormalization-group evolution between the NP scale and the hadronic scale. These coefficients do not induce sizeable hadronic EDM contributions through direct matching, but instead generate low-energy dipole operators through a sequence of RGE-induced mixings. Their constraints therefore reflect the cumulative effects of running, operator mixing, and threshold corrections. Although the resulting bounds are generally weaker than those obtained from direct one-loop contributions, they provide important complementary information and ensure that the full chain of CP-violating effects generated by anomalous top-quark interactions is consistently taken into account.

%The corresponding FCNC coupling products are typically probed at the level of ${\cal O}(10^{-8})$ for $tuX$ interactions and ${\cal O}(10^{-4})$ for the corresponding charm-sector couplings.
The flavor-changing neutral-current top-quark interactions of the form $t u_i X$, with $X=g,\gamma,Z,H$, generate the light-quark dipole operators through the topologies shown in Figs.~\ref{fig:EDM_double_a}-\ref{fig:CEDM_three_gluon} and Fig.~\ref{fig:Weinberg_EDM_FC2}. These contributions depend on products of FCNC couplings with opposite chiralities and induce both quark EDMs and CEDMs, which subsequently contribute to the neutron and proton EDMs as well as to the CP-odd pion-nucleon couplings $\bar g_0$ and $\bar g_1$. The resulting hadronic observables provide some useful probes of top-FCNC interactions. To keep the presentation concise
and transparent, we therefore express the corresponding constraints in
terms of the dimensionless effective couplings defined in
Eq.~\eqref{eq:top_effective_matching}, with the resulting bounds
presented in Table~\ref{tab:qEDM_FCNC_oneloop}. The table is divided
into two parts: \textbf{\textit{the first}} contains the constraints arising from the
light-quark EDMs and CEDMs, while \textbf{\textit{the second}}, highlighted in \textcolor{lightblue}{blue},
shows the contributions induced through the threshold corrections. Although the same types of couplings appear in the topologies shown in
Figs.~\ref{fig:EDM_double_a}-\ref{fig:CEDM_three_gluon}, only the
crossed-chirality combinations of the couplings survive in the
$\mathrm{CP}$-violating amplitudes. The products of FCNC couplings involving the up quark are constrained at the level of ${\cal O}(10^{-9})$-${\cal O}(10^{-8})$, while the corresponding charm-sector combinations are typically constrained at the ${\cal O}(10^{-4})$ level. Also, the coupling combinations contributing directly to
the light-quark EDMs and CEDMs are more tightly constrained than those
entering through the Weinberg-operator threshold corrections, i.e.,
through $\tilde d_b$ and $\tilde d_c$.

Among the current measurements, the most constraining bounds on the WCs are highlighted in \textbf{bold}.
As defined in Eq.~\eqref{eq:pion_nucleon}, the pion--nucleon couplings
$\bar{g}_0$ and $\bar{g}_1$ depend on the light-quark CEDMs,
$\tilde d_q$, and the Weinberg operator coefficient, $w$. We find that,
for most WCs contributing simultaneously to $d_n$ and $\bar{g}_1$,
the constraint from $\bar{g}_1$ is stronger. In contrast, for WCs whose
contributions arise through the Weinberg operator, the corresponding
constraints from $\bar{g}_1$ are comparatively weaker than those from
$d_n$, due to the underlying hadronic form factors.

%-----------------------------------
% Table Hadronic EDM at one loop (Top FCCC and 4 Fermi)

\subsubsection{Barr-Zee diagrams}
%-----------------------------
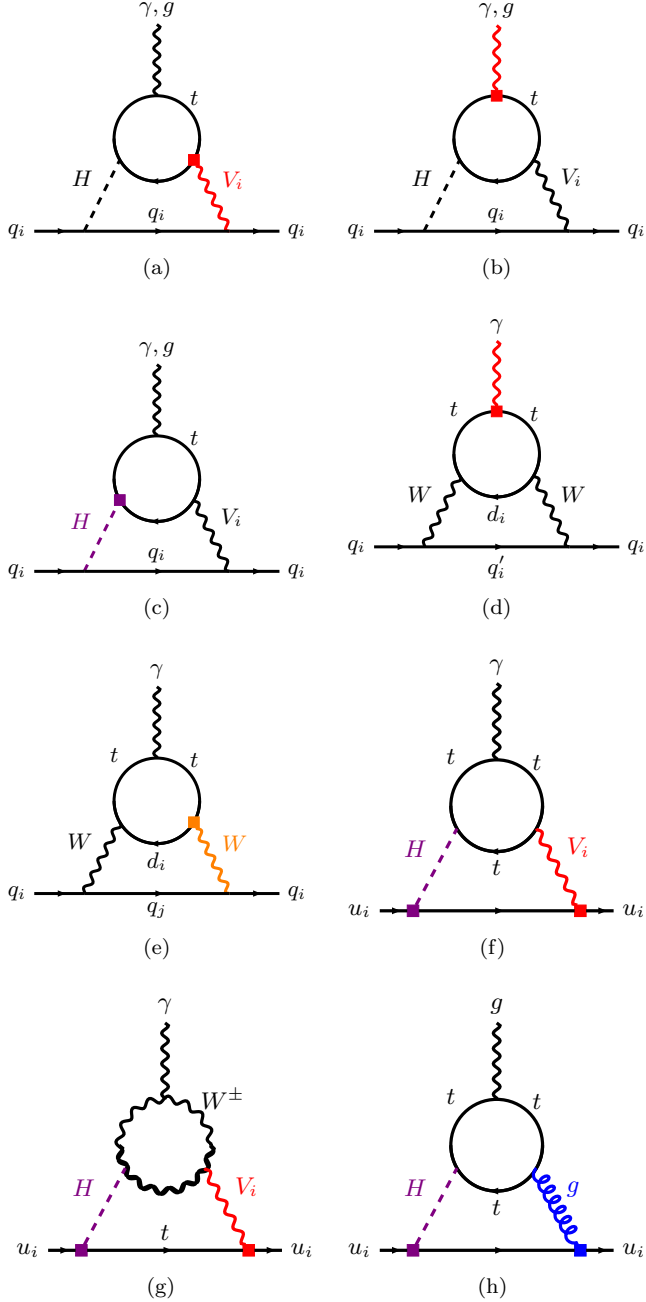
\begin{figure}[h!]
    \centering
    %--------------Diag-1----------------
    \subfloat[]{\label{fig:dq_FCa}% 
        \resizebox{0.48\linewidth}{!}{%
            \begin{tikzpicture}
                \begin{feynman}
                \def\r{0.6 cm}
                \coordinate (O) at (0,0);
                \draw[very thick] (O) circle (\r);
                \vertex (A) at (90:\r);  
                \draw[fermion, very thick] (360:\r) arc (360:180:\r);
                \vertex [](B) at (210:\r);   
                \vertex [red, square dot](C) at (330:\r){};   
                \vertex[above=1.0 of A](F){\(\gamma, g\)};
                \vertex[below left=1.0cm and 0.5cm of B](G);
                \vertex[below right=1.0cm and 0.5cm of C](H);
                \vertex[left=0.7 cm of G](b){\(q_i\)};
                \vertex[right=0.7 cm of H](d){\(q_i\)};
                \vertex[below=0.6 cm of O](a1);
                \vertex[above right=0.5 cm of O](a2) {\(t\)};
                \vertex[above left=0.6 cm of O](a3);
                \diagram*{
                (A) --[very thick, boson](F),
                (G) --[very thick, scalar, edge label=\(H\)](B),
                (C) --[red, very thick, boson, edge label=\(V_i\)](H),
                (b) --[very thick, fermion](G) --[very thick, fermion, edge label=\(q_i\)](H) --[very thick, fermion](d),
                };
                \end{feynman}
            \end{tikzpicture}%
        }%
    }\hfill% 
    %------------Diag-2--------------------
    \subfloat[]{\label{fig:dq_FCb}%
        \resizebox{0.48\linewidth}{!}{%
            \begin{tikzpicture}
                \begin{feynman}
                \def\r{0.6 cm}
                \coordinate (O) at (0,0);
                \draw[very thick] (O) circle (\r);
                \vertex [red, square dot] (A) at (90:\r) {};  
                \draw[fermion, very thick] (360:\r) arc (360:180:\r);
                \vertex [](B) at (210:\r);   
                \vertex [](C) at (330:\r);   
                \vertex[above=1.2 of A](F){\(\gamma,g\)};
                \vertex[below left=1.0cm and 0.5cm of B](G);
                \vertex[below right=1.0cm and 0.5cm of C](H);
                \vertex[left=0.7 cm of G](b){\(q_i\)};
                \vertex[right=0.7 cm of H](d){\(q_i\)};
                \vertex[below=0.6 cm of O](a1);
                \vertex[above right=0.5 cm of O](a2) {\(t\)};
                \vertex[above left=0.6 cm of O](a3);
                \diagram*{
                (A) --[red, very thick, boson](F),
                (G) --[very thick, scalar, edge label=\(H\)](B),
                (C) --[very thick, boson, edge label=\(V_i\)](H),
                (b) --[very thick, fermion](G) --[very thick, fermion, edge label=\(q_i\)](H) --[very thick, fermion](d),
                };
                \end{feynman}
            \end{tikzpicture}%
        }%
    }\\ % <-- New Row
    %------------Diag-3--------------------
    \subfloat[]{\label{fig:dq_FCc}%
        \resizebox{0.48\linewidth}{!}{%
            \begin{tikzpicture}
                \begin{feynman}
                \def\r{0.6 cm}
                \coordinate (O) at (0,0);
                \draw[very thick] (O) circle (\r);
                \vertex (A) at (90:\r);  
                \draw[fermion, very thick] (360:\r) arc (360:180:\r);
                \vertex [violet, square dot](B) at (210:\r) {};   
                \vertex [](C) at (330:\r);   
                \vertex[above=1.0 of A](F){\(\gamma,g\)};
                \vertex[below left=1.0cm and 0.5cm of B](G);
                \vertex[below right=1.0cm and 0.5cm of C](H);
                \vertex[left=0.7 cm of G](b){\(q_i\)};
                \vertex[right=0.7 cm of H](d){\(q_i\)};
                \vertex[below=0.6 cm of O](a1);
                \vertex[above right=0.5 cm of O](a2){\(t\)};
                \vertex[above left=0.6 cm of O](a3);
                \diagram*{
                (A) --[very thick, boson](F),
                (G) --[violet, very thick, scalar, edge label=\(H\)](B),
                (C) --[very thick, boson, edge label=\(V_i\)](H),
                (b) --[very thick, fermion](G) --[very thick, fermion, edge label=\(q_i\)](H) --[very thick, fermion](d),
                };
                \end{feynman}
            \end{tikzpicture}%
        }%
    }\hfill%
    %------------Diag-4---------------
    \subfloat[]{\label{fig:dq_FCd}%
        \resizebox{0.48\linewidth}{!}{%
            \begin{tikzpicture}
                \begin{feynman}
                \def\r{0.6 cm}
                \coordinate (O) at (0,0);
                \draw[very thick] (O) circle (\r);
                \vertex [red, square dot] (A) at (90:\r) {};  
                \draw[fermion, very thick] (360:\r) arc (360:180:\r);
                \vertex [](B) at (210:\r);   
                \vertex [](C) at (330:\r);   
                \vertex[above=1.2 of A](F){\(\gamma\)};
                \vertex[below left=1.0cm and 0.5cm of B](G);
                \vertex[below right=1.0cm and 0.5cm of C](H);
                \vertex[left=0.7 cm of G](b){\(q_i\)};
                \vertex[right=0.7 cm of H](d){\(q_i\)};
                \vertex[below=0.6 cm of O](a1);
                 \vertex[below=0.6 cm of O](a1) {\(d_i\)};
                \vertex[above right=0.5 cm of O](a2) {\(t\)};
                \vertex[above left=0.6 cm of O](a3) {\(t\)};
                \vertex[above left=0.6 cm of O](a3);
                \diagram*{
                (A) --[red, very thick, boson](F),
                (G) --[very thick, boson, edge label=\(W\)](B),
                (C) --[very thick, boson, edge label=\(W\)](H),
                (b) --[very thick, fermion](G) --[very thick, fermion, edge label'=\(q_i^{\prime}\)](H) --[very thick, fermion](d),
                };
                \end{feynman}
            \end{tikzpicture}%
        }%
    }\\
    %------------Diag-5--------------------
    \subfloat[]{\label{fig:dq_FCCCa_BZ}%
        \resizebox{0.48\linewidth}{!}{%
            \begin{tikzpicture}
                \begin{feynman}
                \def\r{0.6 cm}
                \coordinate (O) at (0,0);
                \draw[very thick] (O) circle (\r);
                \vertex (A) at (90:\r);  
                \draw[fermion, very thick] (360:\r) arc (360:180:\r);
                \vertex [](B) at (210:\r);   
                \vertex [orange, square dot](C) at (330:\r) {};   
                \vertex[above=1.0 of A](F){\(\gamma\)};
                \vertex[below left=1.0cm and 0.5cm of B](G);
                \vertex[below right=1.0cm and 0.5cm of C](H);
                \vertex[left=0.7 cm of G](b){\(q_i\)};
                \vertex[right=0.7 cm of H](d){\(q_i\)};
                \vertex[below=0.6 cm of O](a1) {\(d_i\)};
                \vertex[above right=0.5 cm of O](a2) {\(t\)};
                \vertex[above left=0.6 cm of O](a3) {\(t\)};
                \diagram*{
                (A) --[very thick, boson](F),
                (G) --[very thick, boson, edge label=\(W\)](B),
                (C) --[very thick, orange, boson, edge label=\(W\)](H),
                (b) --[very thick,fermion](G) --[very thick, fermion, edge label'=\(q_j\)](H) --[very thick, fermion](d),
                };
                \end{feynman}
            \end{tikzpicture}%
        }%
    }\hfill%% <-- New Row
    %-----------Diag-6--------------------
    \subfloat[]{\label{fig:dq_fcnc_a}
        \resizebox{0.48\linewidth}{!}{%
            \begin{tikzpicture}
                \begin{feynman}
                \def\r{0.6 cm}
                \coordinate (O) at (0,0);
                \draw[very thick] (O) circle (\r);
                \vertex (A) at (90:\r);  
                \draw[fermion, very thick] (360:\r) arc (360:180:\r);
                \vertex [](B) at (210:\r);   
                \vertex [](C) at (330:\r);   
                \vertex[above=1.0 of A](F){\(\gamma\)};
                \vertex[square dot, violet, below left=1.0cm and 0.5cm of B](G){};
                \vertex[square dot, red, below right=1.0cm and 0.5cm of C](H){};
                \vertex[left=0.7 cm of G](b){\(u_i\)};
                \vertex[right=0.7 cm of H](d){\(u_i\)};
                \vertex[below=0.6 cm of O](a1) {\(t\)};
                \vertex[above right=0.5 cm of O](a2) {\(t\)};
                \vertex[above left=0.6 cm of O](a3) {\(t\)};
                \diagram*{
                (A) --[very thick, boson](F),
                (G) --[very thick, violet, scalar, edge label=\(H\)](B),
                (C) --[very thick, red, boson, edge label=\(V_i\)](H),
                (b) --[very thick, fermion](G) --[very thick, fermion](H) --[very thick, fermion](d),
                };
                \end{feynman}
            \end{tikzpicture}%
        }%
    }\\
    %------------Diag-7--------------------
    \subfloat[]{\label{fig:dq_fcnc_b}
        \resizebox{0.48\linewidth}{!}{%
            \begin{tikzpicture}
                \begin{feynman}
                \def\r{0.6 cm}
                \coordinate (O) at (0,0);
                \draw[boson, very thick] (O) circle (\r);
                \vertex (A) at (90:\r);  
                \draw[boson, very thick] (360:\r) arc (360:180:\r);
                \vertex [](B) at (210:\r);   
                \vertex [](C) at (330:\r);   
                \vertex[above=1.0 of A](F){\(\gamma\)};
                \vertex[square dot, violet, below left=1.0cm and 0.5cm of B](G){};
                \vertex[square dot, red, below right=1.0cm and 0.5cm of C](H){};
                \vertex[left=0.7 cm of G](b){\(u_i\)};
                \vertex[right=0.7 cm of H](d){\(u_i\)};
                \vertex[below=0.6 cm of O](a1);
                \vertex[above right=0.5 cm of O](a2) {\(W^{\pm}\)};
                \vertex[above left=0.6 cm of O](a3);
                \diagram*{
                (A) --[very thick, boson](F),
                (G) --[very thick, violet, scalar, edge label=\(H\)](B),
                (C) --[very thick, red, boson, edge label=\(V_i\)](H),
                (b) --[very thick, fermion](G) --[very thick, fermion, edge label=\(t\)](H) --[very thick, fermion](d),
                };
                \end{feynman}
            \end{tikzpicture}%
        }%
    }\hfill% % <-- New Row
    %------------Diag-8--------------------
    \subfloat[]{\label{fig:dq_fcnc_c}
        \resizebox{0.48\linewidth}{!}{%
            \begin{tikzpicture}
                \begin{feynman}
                \def\r{0.6 cm}
                \coordinate (O) at (0,0);
                \draw[very thick] (O) circle (\r);
                \vertex (A) at (90:\r);  
                \draw[fermion, very thick] (360:\r) arc (360:180:\r);
                \vertex [](B) at (210:\r);   
                \vertex [](C) at (330:\r);   
                \vertex[above=1.0 of A](F){\(g\)};
                \vertex[square dot, violet, below left=1.0cm and 0.5cm of B](G){};
                \vertex[square dot, blue, below right=1.0cm and 0.5cm of C](H){};
                \vertex[left=0.7 cm of G](b){\(u_i\)};
                \vertex[right=0.7 cm of H](d){\(u_i\)};
                \vertex[below=0.6 cm of O](a1) {\(t\)};
                \vertex[above right=0.5 cm of O](a2) {\(t\)};
                \vertex[above left=0.6 cm of O](a3) {\(t\)};
                \diagram*{
                (A) --[very thick, boson](F),
                (G) --[very thick, violet, scalar, edge label=\(H\)](B),
                (C) --[very thick, blue, gluon, edge label=\(g\)](H),
                (b) --[very thick, fermion](G) --[very thick, fermion](H) --[very thick, fermion](d),
                };
                \end{feynman}
            \end{tikzpicture}%
        }%
    }
    %--------------------------
    \caption{Barr-Zee-type diagrams contributing to the quark EDMs through
different anomalous top-quark couplings. Here, $V_i=\gamma,Z$.}
    \label{fig:qEDM_Barr_Zee}
\end{figure}
%-----------------------------

For several of the operators discussed above, the leading contributions to hadronic EDM observables arise through the two-loop Barr-Zee topologies shown in Figs.~\ref{fig:dq_FCa}--\ref{fig:dq_fcnc_c}. These diagrams involve anomalous top-quark interactions with electroweak gauge bosons and the Higgs boson and generate light-quark EDMs and CEDMs after integrating out the heavy degrees of freedom. In contrast to the direct one-loop matching contributions discussed previously, the Barr-Zee mechanism provides the dominant finite matching contribution for many electroweak top-quark operators and therefore plays a central role in determining their hadronic EDM sensitivity.

The flavor-conserving electroweak dipole operators $C_{uB}^{33}$ and $C_{uW}^{33}$, the CP-violating top-Higgs Yukawa interaction $C_{u\phi}^{33}$, and the neutral-current operators $C_{\phi q}^{(1,3)}$ and $C_{\phi u}$ contribute predominantly through the single-insertion topologies shown in Figs.~\ref{fig:dq_FCa}--\ref{fig:dq_FCCCa_BZ}. These diagrams generate both light-quark EDMs, $d_{q_i}$, and CEDMs, $\tilde d_{q_i}$, which subsequently induce contributions to the neutron and proton EDMs as well as the CP-odd pion--nucleon couplings $\bar g_0$ and $\bar g_1$. As a result, hadronic EDM observables provide significant constraints on these operators, with the top-Higgs Yukawa coupling $C_{u\phi}^{33}$ and the electroweak dipole operators receiving particularly important contributions through this mechanism.

The multi-operator topologies shown in Figs.~\ref{fig:dq_fcnc_a}-\ref{fig:dq_fcnc_c} exhibit a qualitatively different behavior and lead to some of the strongest constraints on products of FCNC couplings. In these diagrams, the chirality flip is generated by an anomalous top-quark interaction rather than by a light-quark Yukawa insertion. Consequently, the suppression associated with the light-quark Yukawa coupling is replaced by interactions involving the top quark and electroweak-scale particles, substantially enhancing the resulting $CP$-violating amplitude. This enhancement translates into remarkably strong bounds on combinations of FCNC top-Higgs and top-gauge couplings, making hadronic EDM observables highly sensitive probes of correlated FCNC interactions in the top-quark sector.

The complete set of two-loop matching relations for the light-quark EDMs
and CEDMs is collected in Appendix~\ref{Append: Had_EDM_BZ}. We predominantly employ the
standard Barr-Zee loop functions, $f(z)$ and $g(z)$, following
Ref.~\cite{PhysRevLett.65.21}. For topologies requiring additional loop
functions, their explicit definitions are provided alongside the
corresponding matching relations. To facilitate reproducibility, we
also describe the procedure adopted for the two-loop Barr-Zee
calculation and provide an explicit derivation for one representative
topology in Appendix~\ref{appendix:barr_zee}; the remaining topologies
can be obtained following the same procedure. The resulting constraints on the anomalous top-quark couplings are
summarized in Tables~\ref{tab:top_WC_FC_qEDM_BZ} and
\ref{tab:qEDM_Barr_Zee_FCNC}. As in the case of the lepton EDM, the same
Barr-Zee topologies contribute to the electron EDM, $d_e$. Since the
experimental bound on $d_e$ is at least four orders of magnitude more
stringent than the corresponding hadronic EDM constraints, it provides
stronger bounds on these WCs, as shown in Table~\ref{tab:de_bound_top_FC}. However, the projected sensitivity to the proton EDM, $d_p$, can
substantially improve the constraints on these SMEFT WCs.

In Table~\ref{tab:qEDM_Barr_Zee_FCNC}, we present several multi-operator
top-FCNC scenarios in which the top-Higgs WCs are correlated with
top-gauge couplings of the form $tu_jX$, with $X=\gamma,g,Z$. The
resulting bounds on the products of the corresponding dimensionless
couplings are found to be at the level of $\mathcal{O}(10^{-4}-10^{-5})$.
It is important to note that Figs.~\ref{fig:dq_fcnc_a} and
\ref{fig:dq_fcnc_b} involve the same sets of SMEFT couplings; however,
we find numerically that the products of couplings entering the topology
in Fig.~\ref{fig:dq_fcnc_b} are constrained at least two orders of
magnitude more strongly.

%-----------------------------------
\begin{table*}[t!]
    \centering
    \renewcommand{\arraystretch}{1.3}
    \setlength{\tabcolsep}{5pt} % Slightly tightened column separation to fit PRD column width
    \begin{tabular}{@{} c c c c c c @{}}
        \toprule
        \textbf{WCs~$\,[\mathrm{TeV}^{-2}]$}
        & \textbf{$|d_n|$ (Current)}
        & \textbf{$|\bar{g}_0|$ (Current)}
        & \textbf{$|\bar{g}_1|$ (Current)}
        & \textbf{$|d_n|$ (Future)}
        & \textbf{$|d_p|$ (Future)} \\
        \midrule

        $\mathrm{Im}(C^{uB}_{33})$
        & $1.19$
        & $1.90$
        & $\mathbf{1.48\times10^{-1}}$
        & $6.64\times10^{-2}$
        & $9.58\times10^{-4}$ \\

        $\mathrm{Im}(C^{uW}_{33})$
        & $8.28\times10^{-1}$
        & $1.13$
        & $\mathbf{8.69\times10^{-2}}$
        & $4.60\times10^{-2}$
        & $7.96\times10^{-4}$ \\

        $\mathrm{Im}(C^{u\phi}_{33})$
        & $1.08\times10^{-1}$
        & $1.16\times10^{-1}$
        & $\mathbf{8.81\times10^{-3}}$
        & $6.02\times10^{-3}$
        & $1.23\times10^{-4}$ \\

        $\mathrm{Im}(C^{\phi q(1)}_{33})$
        & $6.17$
        & $5.89$
        & $\mathbf{4.49\times10^{-1}}$
        & $3.43\times10^{-1}$
        & $7.15\times10^{-2}$ \\

        $\mathrm{Im}(C^{\phi q(3)}_{33})$
        & $1.03$
        & $1.12$
        & $\mathbf{8.56\times10^{-2}}$
        & $5.7\times10^{-2}$
        & $1.16\times10^{-3}$ \\

        $\mathrm{Im}(C^{\phi u}_{33})$
        & $1.51\times10^{1}$
        & $1.25\times10^{1}$
        & $\mathbf{9.56\times10^{-1}}$
        & $8.40\times10^{-1}$
        & $1.81\times10^{-2}$ \\

        $\mathrm{Im}(C^{uG}_{33})$
        & $9.29\times10^{-2}$
        & $1.0\times10^{-1}$
        & $\mathbf{7.64\times10^{-3}}$
        & $5.16\times10^{-3}$
        & $1.04\times10^{-4}$ \\

        \midrule

        \textcolor{DarkEmerald}{$\mathrm{Im}(C^{dW}_{33})$}
        & \textcolor{DarkEmerald}{$9.37\times10^{1}$}
        & \textcolor{DarkEmerald}{$1.0\times10^{2}$}
        & \textcolor{DarkEmerald}{$\mathbf{7.62}$}
        & \textcolor{DarkEmerald}{$5.20$}
        & \textcolor{DarkEmerald}{$1.06\times10^{-1}$} \\

        \textcolor{DarkEmerald}{$\mathrm{Im}(C^{\phi ud}_{33})$}
        & \textcolor{DarkEmerald}{$4.68\times10^{1}$}
        & \textcolor{DarkEmerald}{$5.0\times10^{1}$}
        & \textcolor{DarkEmerald}{$\mathbf{3.81}$}
        & \textcolor{DarkEmerald}{$2.60$}
        & \textcolor{DarkEmerald}{$5.31\times10^{-2}$} \\

        \bottomrule
    \end{tabular}

    \caption{Bounds on the top-quark couplings through
    Barr-Zee topologies contributing to the quark EDMs. The
    \textcolor{DarkEmerald}{green colored} entries arise predominantly through
    RGE-induced mixing. The WCs are evaluated at $\mu=1~\mathrm{TeV}$ and
    have mass dimension $\mathrm{TeV}^{-2}$.}
    
    \label{tab:top_WC_FC_qEDM_BZ}
\end{table*}
%-----------------------------------

%---------------------------------------------
% Table: Hadronic EDM through Barr-Zee Topology
\begin{table*}[t!]
    \centering
    \renewcommand{\arraystretch}{0.6}
    \setlength{\tabcolsep}{3pt}
    \begin{tabular}{@{} c c c c c c @{}}
        \toprule
        \textbf{Effective Couplings}
        & \textbf{$|d_n|$ (Current)}
        & \textbf{$|\bar{g}_0|$ (Current)}
        & \textbf{$|\bar{g}_1|$ (Current)}
        & \textbf{$|d_n|$ (Future)}
        & \textbf{$|d_p|$ (Future)} \\
        \midrule

        $\displaystyle
        \mathrm{Im}\left(
        \lambda_L^{ut}\eta_R^{tu^*}
        -\lambda_R^{ut}\eta_L^{tu*}
        \right)$
        & $4.29\times10^{-4}$
        & $6.36\times10^{-3}$
        & $\mathbf{1.76\times10^{-4}}$
        & $2.3\times10^{-5}$
        & $6.35 \times10^{-8}$ \\

        $\displaystyle
        \mathrm{Im}\left(
        X_R^{ut}\eta_R^{tu^*}
        -X_L^{ut}\eta_L^{*tu}
        \right)$
        & $8.90\times10^{-5}$
        & $1.32\times10^{-3}$
        & $\mathbf{3.65\times10^{-5}}$
        & $4.94\times10^{-6}$
        & $1.32 \times10^{-8}$ \\

        $\displaystyle
        \mathrm{Im}\left(
        \eta_L^{tu*}\kappa_R^{ut}
        -\eta_R^{tu*}\kappa_L^{ut}
        \right)$
        & $4.45\times10^{-5}$
        & $6.60\times10^{-4}$
        & $\mathbf{1.83\times10^{-5}}$
        & $2.47\times10^{-6}$
        & $6.59 \times10^{-9}$ \\

        $\displaystyle
        \mathrm{Im}\left(
        \eta_R^{ut}\xi_L^{tu*}
        -\eta_L^{ut}\xi_R^{tu*}
        \right)$
        & $1.93\times10^{-5}$
        & $1.25\times10^{-5}$
        & $\mathbf{3.45\times10^{-7}}$
        & $1.07\times10^{-6}$
        & $7.71 \times10^{-9}$ \\

        \midrule

        \textcolor{lightblue}{$\displaystyle
        \mathrm{Im}\left(
        \eta_R^{ct}\xi_L^{tc*}
        -\eta_L^{ct}\xi_R^{tc*}
        \right)$}
        & \textcolor{lightblue}{$\mathbf{3.12\times10^{-2}}$}
        & \textcolor{lightblue}{$2.86\times10^{-1}$}
        & \textcolor{lightblue}{$4.73\times10^{-1}$}
        & \textcolor{lightblue}{$1.73\times10^{-3}$}
        & \textcolor{lightblue}{$2.01\times10^{-5}$} \\

        \bottomrule
    \end{tabular}

    \caption{Bounds on products of the dimensionless top-quark FCNC couplings
    from Barr-Zee contributions to the quark EDMs. The table summarizes
    multi-operator scenarios involving top-Higgs couplings correlated with
    top-gauge couplings of the form $tu_jX$, with $X=\gamma,g,Z$. The
    \textcolor{lightblue}{blue} entry denotes the contribution arising from
    the charm-quark CEDM, $\tilde d_c$.}

    \label{tab:qEDM_Barr_Zee_FCNC}
\end{table*}
%---------------------------------------------

%---------------------------------
\subsubsection{Weinberg contributions}\label{subsec:Hadronic_Weinberg}
%---------------------------------
%-----------------------------
\begin{figure}[h!]
    \centering
    %---------------------------
  
    %----------------Diag-1------------------
	\subfloat[]{\label{fig:weinberg1}
		\begin{tikzpicture}
			\begin{feynman}
				\def\r{1.0 cm}
				\coordinate (O) at (0,0);
				\draw[very thick] (O) circle (\r);
				\draw[fermion, very thick] (360:\r) arc (360:180:\r);
				\vertex (A) at (90:\r)  ;   
				\vertex (B) at (210:\r) ;   
				\vertex (C) at (330:\r) ;   
				\vertex (D) at (0:\r)   {}; 
				\vertex (E) at (180:\r) {};
				\vertex[above=0.5 of A](F){\(g\)};
				\vertex[below left=0.5cm of B](G){\(g\)};
				\vertex[below right=0.5cm of C](H){\(g\)};
				\vertex[below=1.1cm of O](a1) {\(t\)};
				\vertex[above right=1.0 cm of O](a2) {\(t\)};
				\vertex[square dot, red, at=(D)] ;
				\vertex[square dot, red, at=(E)] {};
				\diagram*{
					(A) --[very thick, gluon](F),
					(B) --[very thick, gluon](G),
					(C) --[very thick, gluon](H),
					(D) --[boson, very thick, red, edge label'=\(\gamma/Z\)](E),
				};
			\end{feynman}
		\end{tikzpicture}
	}\hfill
    %------------------Diag-2---------------
	\subfloat[]{\label{fig:weinberg2}
		\begin{tikzpicture}
			\begin{feynman}
				\def\r{1.0 cm}
				\coordinate (O) at (0,0);
				\draw[very thick] (O) circle (\r);
				\draw[fermion, very thick] (360:\r) arc (360:180:\r);
				\vertex (A) at (90:\r)  ;   
				\vertex (B) at (210:\r) ;   
				\vertex (C) at (330:\r) ;   
				\vertex (D) at (0:\r)   {}; 
				\vertex (E) at (180:\r) {};
				\vertex[above=0.5 of A](F){\(g\)};
				\vertex[below left=0.5cm of B](G){\(g\)};
				\vertex[below right=0.5cm of C](H){\(g\)};
				\vertex[below=1.1cm of O](a1) {\(t\)};
				\vertex[above right=1.0 cm of O](a2) {\(t\)};
				\vertex[square dot, violet, at=(D)] ;
				\vertex[square dot, violet, at=(E)] {};
				\diagram*{
					(A) --[very thick, gluon](F),
					(B) --[very thick, gluon](G),
					(C) --[very thick, gluon](H),
					(D) --[scalar, very thick, violet, edge label'=\(H\)](E),
				};
			\end{feynman}
		\end{tikzpicture}
	}\\
	\subfloat[]{\label{fig:weinberg5}
		\begin{tikzpicture}
			\begin{feynman}
				\def\r{1.0 cm}
				\coordinate (O) at (0,0);
				\draw[very thick] (O) circle (\r);
				\draw[fermion, very thick] (360:\r) arc (360:180:\r);
				
				% Gluon vertices (fixed)
				\vertex (A) at (90:\r) ;
				\vertex (B) at (210:\r) ;
				\vertex (C) at (330:\r) ;
				
				% NP insertion vertices
				\vertex[square dot, red] (NP1) at (110:\r) ;
				\vertex[square dot, red] (NP2) at (190:\r) {};
				
				\vertex[above=0.5 of A](F){\(g\)};
				\vertex[below left=0.5cm of B](G){\(g\)};
				\vertex[below right=0.5cm of C](H){\(g\)};
				\vertex[below=1.1cm of O](a1) {\(t\)};
				\vertex[above right=1.0cm of O](a2) {\(t\)};
				\diagram*{
					(A) --[very thick, gluon](F),
					(B) --[very thick, gluon](G),
					(C) --[very thick, gluon](H),
					(NP1) --[boson, very thick, red, bend left=45, edge label=\(\gamma/Z\)](NP2),
				};
			\end{feynman}
		\end{tikzpicture}
	}\hfill
    %------------------Diag-6 (Off-Vertex Topology)---------------
	\subfloat[]{\label{fig:weinberg6}
		\begin{tikzpicture}
			\begin{feynman}
				\def\r{1.0 cm}
				\coordinate (O) at (0,0);
				\draw[very thick] (O) circle (\r);
				\draw[fermion, very thick] (360:\r) arc (360:180:\r);
				
				% Gluon vertices (fixed)
				\vertex (A) at (90:\r) ;
				\vertex (B) at (210:\r) ;
				\vertex (C) at (330:\r) ;
				
				% NP insertion vertices
				\vertex[square dot, violet] (NP1) at (110:\r) ;
				\vertex[square dot, violet] (NP2) at (190:\r) {};
				
				\vertex[above=0.5 of A](F){\(g\)};
				\vertex[below left=0.5cm of B](G){\(g\)};
				\vertex[below right=0.5cm of C](H){\(g\)};
				\vertex[below=1.1cm of O](a1) {\(u_i\)};
				\vertex[above right=1.0cm of O](a2) {\(t\)};
				\diagram*{
					(A) --[very thick, gluon](F),
					(B) --[very thick, gluon](G),
					(C) --[very thick, gluon](H),
					(NP1) --[scalar, very thick, violet, bend left=45, edge label=\(H\)](NP2),
				};
			\end{feynman}
		\end{tikzpicture}
	}
	%------------------------------------
    %---------------------------
    \caption{Representative loop topologies generating the $CP$-violating
Weinberg three-gluon operator from anomalous top-quark interactions.}
    \label{fig:qEDM_weinberg}
\end{figure}
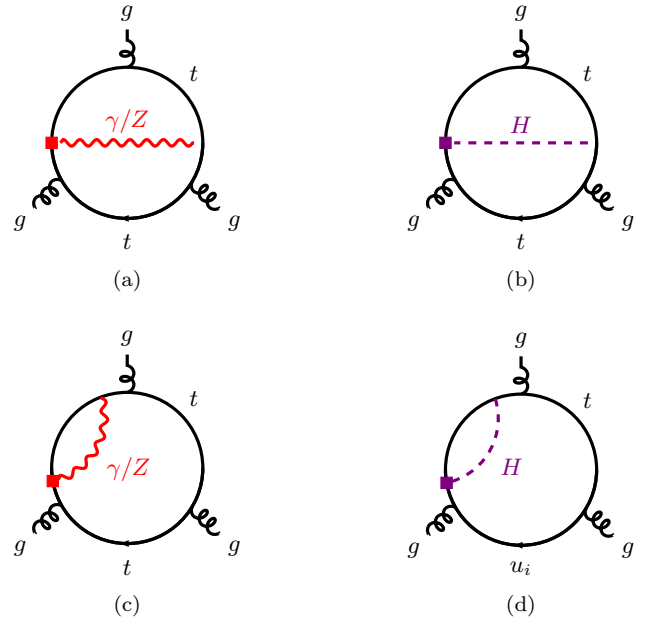
%----------------------------------------

%-----------------------------------
\begin{table*}[t!]
	\centering
	\renewcommand{\arraystretch}{1.3}
	\setlength{\tabcolsep}{5pt} % Slightly tightened column separation to fit PRD column width
	\begin{tabular}{@{} c c c c c c @{}}
		\toprule
		\textbf{WCs~$\,[\mathrm{TeV}^{-2}]$}
		& \textbf{$|d_n|$ (Current)}
		& \textbf{$|\bar{g}_0|$ (Current)}
		& \textbf{$|\bar{g}_1|$ (Current)}
		& \textbf{$|d_n|$ (Future)}
		& \textbf{$|d_p|$ (Future)} \\
		\midrule
		
		$\mathrm{Im}(C^{uB}_{33})$
		& $\mathbf{3.44 \times 10^{-2}}$
		& $2.64\times10^{-1}$
		& $4.38\times10^{-1}$
		& $1.91\times10^{-3}$
		& $2.33\times10^{-5}$ \\
		
		$\mathrm{Im}(C^{uW}_{33})$
		& $\mathbf{1.85 \times 10^{-2}}$
		& $1.37\times10^{-1}$
		& $2.27\times10^{-1}$
		& $1.03\times10^{-3}$
		& $1.25\times10^{-5}$ \\
		
		$\mathrm{Im}(C^{u\phi}_{33})$
		& $\mathbf{5.22\times10^{-2}}$
		& $2.80\times10^{-1}$
		& $4.63\times10^{-1}$
		& $2.90\times10^{-3}$
		& $3.54\times10^{-5}$ \\
		
		$\mathrm{Im}(C^{\phi q(1)}_{33})$
		& $\mathbf{2.77 \times10^{-1}}$
		& $2.06$
		& $3.43$
		& $1.55 \times10^{-2}$
		& $1.89 \times10^{-4}$ \\
		
		$\mathrm{Im}(C^{\phi q(3)}_{33})$
		& $\mathbf{2.77 \times10^{-1}}$
		& $2.06$
		& $3.43$
		& $1.55 \times10^{-2}$
		& $1.89 \times10^{-4}$ \\
		
		$\mathrm{Im}(C^{\phi u}_{33})$
		& $\mathbf{2.77 \times10^{-1}}$
		& $2.06$
		& $3.43$
		& $1.55 \times10^{-2}$
		& $1.89 \times10^{-4}$ \\
		
		\midrule
		
		\textcolor{DarkEmerald}{$\mathrm{Im}(C^{uG}_{33})$}
		& \textcolor{DarkEmerald}{$\mathbf{1.21\times10^{-1}}$}
		& \textcolor{DarkEmerald}{$6.53\times10^{-1}$}
		& \textcolor{DarkEmerald}{$1.08$}
		& \textcolor{DarkEmerald}{$6.76\times10^{-3}$}
		& \textcolor{DarkEmerald}{$8.26\times10^{-5}$} \\
		
		\textcolor{DarkEmerald}{$\mathrm{Im}(C^{dW}_{33})$}
		& \textcolor{DarkEmerald}{$\mathbf{3.93\times10^{1}}$}
		& \textcolor{DarkEmerald}{$2.42\times10^{2}$}
		& \textcolor{DarkEmerald}{$4.01\times10^{2}$}
		& \textcolor{DarkEmerald}{$2.19$}
		& \textcolor{DarkEmerald}{$2.67\times10^{-2}$} \\
		
		\textcolor{DarkEmerald}{$\mathrm{Im}(C^{\phi ud}_{33})$}
		& \textcolor{DarkEmerald}{$\mathbf{1.97\times10^{1}}$}
		& \textcolor{DarkEmerald}{$1.21\times10^{2}$}
		& \textcolor{DarkEmerald}{$2.0\times10^{2}$}
		& \textcolor{DarkEmerald}{$1.09$}
		& \textcolor{DarkEmerald}{$1.34\times10^{-2}$} \\
		
		\bottomrule
	\end{tabular}
	
	\caption{Bounds on WCs from two loop topologies generating Weinberg three-gluon operator. The \textcolor{DarkEmerald}{colored} entries denote WCs whose contributions arise through RGE mixing. The WCs are given at $\mu=1~\mathrm{TeV}$.}
	
	\label{tab:top_WC_values_qEDM_wein}
\end{table*}
%-----------------------------------

The Weinberg three-gluon operator plays an important role in constraining
$CP$-violating top-quark interactions, as it contributes directly to both
the neutron and proton EDMs without suffering from the light-quark mass
suppression associated with the quark dipole operators. In addition, the
Weinberg operator mixes with the light-quark EDM and CEDM operators under
RGE and can receive threshold contributions
when heavy quarks ($b$ and $c$ quarks) are integrated out. The leading contribution to the
Weinberg operator arises at one loop from a single insertion of the
top-quark chromo-dipole operator, as shown in Fig.~\ref{fig:Weinberg_EDM_FC1}.
Flavor-violating top-quark chromo-dipole interactions can also generate
the Weinberg operator through two insertions, as shown in
Fig.~\ref{fig:Weinberg_EDM_FC2}.

Beyond the contributions from the quark chromo-dipole operators, the
Weinberg three-gluon operator can also mediate contributions to the
low-energy EDMs through two-loop corrections. In the seminal work of
Weinberg~\cite{PhysRevLett.63.2333}, the $CP$-violating three-gluon operator
was generated by integrating out a top-quark loop and a neutral Higgs
boson. The corresponding contribution arises from a two-loop topology
in which three gluons couple to the top-quark loop, with the neutral
Higgs boson emitted and reabsorbed within the loop. Weinberg expressed
the resulting loop integral in terms of a dimensionless function
$h(m_t/m_H)$ and showed that this anomalous $t\bar{t}H$ coupling receives a strong constraint from the neutron EDM. Motivated by this result,
we adopt a similar strategy to evaluate the corresponding two-loop
contributions for the more general anomalous top-quark interactions
considered in this work. While the scalar $t\bar t H$ interaction admits
a compact representation in terms of the loop function $h(m,M)$, 
%-------------------------------
\begin{align}
    w^{(b)}=12\sqrt{2}\frac{g_s^3(\mu_t)}{(4\pi)^4}\frac{1}{y_t}\mathrm{Im}(C^{u\phi}_{33})~h(m_t,M_H)\,,
\end{align}
%-------------------------------
Here, $h(m,M)$ is defined by~\cite{PhysRevLett.63.2333},
%--------------------------
\begin{align}
    h(m,M)&=\frac{m^4}{4}\int_0^1dx\int_0^1dy\nn\\
    &\times\frac{y^3 x^3(1-x)}{[m^2x(1-yx)+M^2(1-y)(1-x)]^2}\,.
\end{align}
%--------------------------
The
corresponding loop integrals for the other anomalous top-quark
interactions generally involve more complicated Lorentz and momentum
structures and do not permit a comparably simple analytical form. We therefore employ a hybrid semi-analytical procedure to evaluate these
loop integrals. A detailed description of the methodology can be found
in Appendix~\ref{appendix:weinberg_corrections}.

We summarize our results for the two-loop Weinberg-operator
contributions in Table~\ref{tab:top_WC_values_qEDM_wein}, while the
one-loop Weinberg contributions from the topologies shown in
Figs.~\ref{fig:Weinberg_EDM_FC1} and~\ref{fig:Weinberg_EDM_FC2} are
already collected in Table~\ref{tab:top_WC_values_FCCC} and ~\ref{tab:qEDM_FCNC_oneloop}. It is evident
from Table~\ref{tab:top_WC_values_qEDM_wein} that the top-scalar WC
$C^{u\phi}_{33}$ is more tightly constrained through the two-loop
Weinberg contribution than through the finite Barr-Zee contribution
associated with the topology shown in Fig.~\ref{fig:dq_FCc}. In contrast,
the contribution from the topology shown in Fig.~\ref{fig:weinberg6} vanishes,
as also verified in Ref.~\cite{Dicus:1989va}. A topology analogous to Fig.~\ref{fig:weinberg2} can also arise
for charged-current interactions, with the intermediate $H$ replaced by
a $W$ boson and the internal top quark by a down-type quark, where
appropriate. Since this contribution is proportional to
$m_{d_i}m_t V_{td_i}$ and is therefore suppressed by the light down-type-quark
mass and CKM, we neglect it. In addition, the same charged-current operators
already contribute through one-loop diagrams~(\ref{fig:qEDM_FCCC_a}) and (\ref{fig:qEDM_FCCC_b}) with a single insertion,
which yield substantially stronger constraints.

%--------------------------------------------
A comparison of the results of  Tables~\ref{tab:top_WC_FC_qEDM_BZ} and \ref{tab:top_WC_values_qEDM_wein} reveals several interesting features regarding the relative importance of these mechanisms for different classes of top-quark interactions. For the electroweak dipole operators $C^{uB}_{33}$ and $C^{uW}_{33}$, the Weinberg-operator contributions provide relatively stronger constraints than the Barr-Zee-induced quark dipoles. For the current sensitivities, the Weinberg topologies improve the bounds by approximately one order of magnitude compared to those obtained from the Barr-Zee contributions. The same pattern persists for the projected sensitivities, where future proton-EDM measurements are expected to probe these operators at the level of ${\cal O}(10^{-5})\,\mathrm{TeV}^{-2}$ through the Weinberg operator, representing nearly an order-of-magnitude improvement over the corresponding Barr-Zee limits.

A different pattern emerges for the neutral-current operators $C^{\phi q(1)}_{33}$, $C^{\phi q(3)}_{33}$, $C^{u\phi}_{33}$, and $C^{\phi u}_{33}$. For the current experimental sensitivities, the Barr-Zee contributions yield constraints that are stronger by roughly a factor of two to three compared with those obtained from the Weinberg operator. In the Barr-Zee analysis, the most stringent bounds typically arise from the CP-odd pion-nucleon coupling $\bar g_1$, whereas the dominant sensitivity in the Weinberg-operator analysis originates from the neutron EDM. This difference reflects the distinct low-energy structures generated by the two mechanisms. The Barr-Zee topologies efficiently induce light-quark EDMs and, in particular, CEDMs, which generate sizable contributions to the pion-nucleon coupling $\bar g_1$. In contrast, the Weinberg operator contributes directly to hadronic $CP$ violation and and is less strongly linked to the chiral mechanisms responsible for the generation of $\bar g_1$, therefore primarily manifests itself through the neutron EDM. As a result, the Barr-Zee contributions currently provide the leading hadronic constraints on these operators.

However, this hierarchy is reversed for future EDM measurements. In particular, the projected proton-EDM sensitivities arising from the Weinberg topologies reach the ${\cal O}(10^{-4})\,\mathrm{TeV}^{-2}$ regime, corresponding to an improvement of one to two orders of magnitude relative to the Barr-Zee contributions. This demonstrates that, while Barr-Zee-induced quark dipoles dominate the present hadronic sensitivity to these operators, gluonic contributions mediated by the Weinberg operator are expected to become the leading probe in future hadronic EDM searches.

Overall, the comparison reveals a clear hierarchy. The Barr-Zee topologies dominate the hadronic sensitivity to the CP-violating top-Higgs interaction $C^{u\phi}_{33}$ and generate strong constraints through the pion-nucleon coupling $\bar g_1$. In contrast, for the electroweak dipole and current operators, the two-loop generation of the Weinberg operator provides the dominant hadronic probe and yields substantially stronger limits, particularly for future proton-EDM measurements. These results highlight the complementary role of quark-dipole and gluonic contributions and demonstrate that a complete assessment of hadronic EDM constraints on top-quark interactions requires the simultaneous inclusion of both Barr-Zee and Weinberg-operator effects.

%----------------------------------
\section{Summary of Constraints from EDM Observables}\label{sec:EDM_summary}
%----------------------------------
\begin{figure*}[t!]
    \centering
    \includegraphics[width=1.0\linewidth]{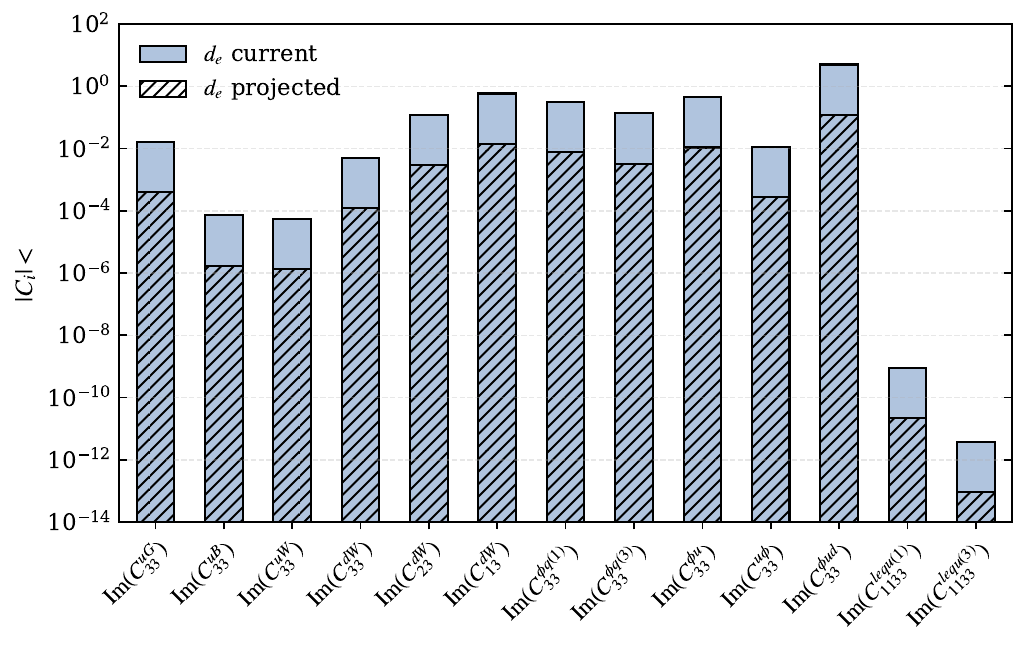}

    \vspace{0.5em}

    \includegraphics[width=1.0\linewidth]{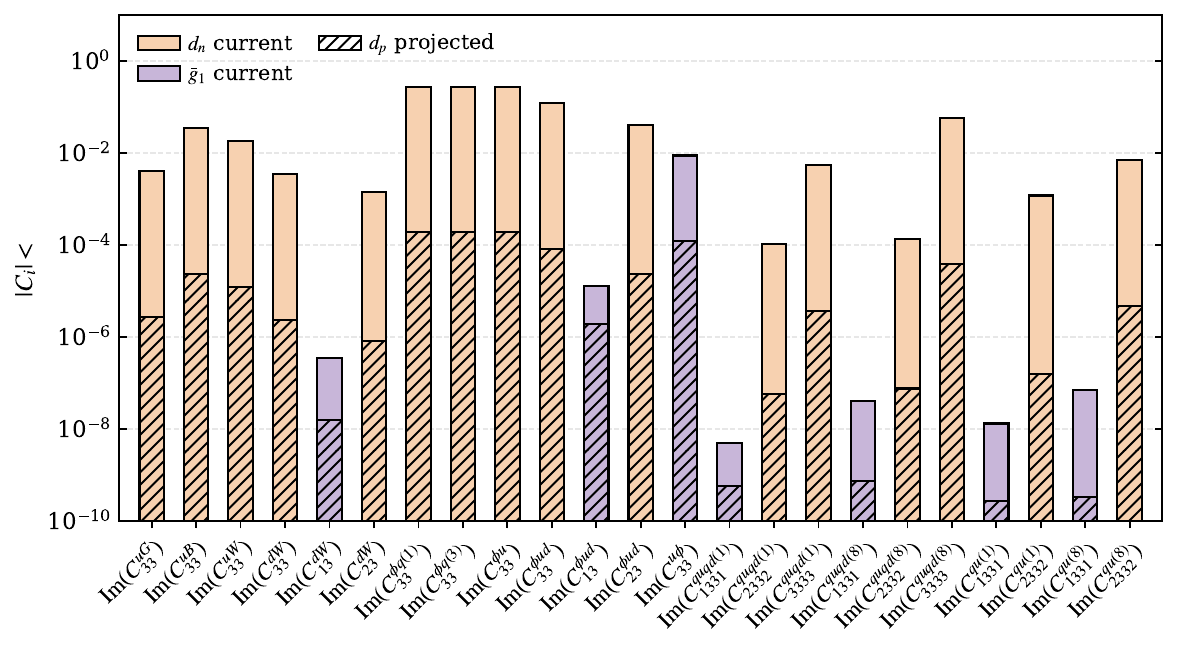}

    \caption{Current and projected upper bounds on $|\mathrm{Im}(C_i)|$ from
the electron EDM (upper panel) and hadronic EDMs (lower panel). The
couplings are defined at $\Lambda=1~\mathrm{TeV}$ and have mass
dimension $\mathrm{TeV}^{-2}$.}
    \label{fig:summary_EDM}
\end{figure*}
%---------------------------------
%\begin{figure*}[t!]
 %   \centering
  %  \includegraphics[width=1.0\linewidth]{Figure/dim6_NP_scale_bounds.pdf}
   % \caption{Lower bounds on the new-physics scale $\Lambda$ from the most
%stringent EDM constraint on each individual SMEFT WC, assuming
%$|C_i|\sim\mathcal{O}(1)$.}
 %   \label{fig:bound_Lambda}
%\end{figure*}
%---------------------------------
In this section, we summarise the most sensitive EDM observables for the different classes of SMEFT operators considered in this work and highlight the corresponding strongest constraints. The overall pattern of constraints obtained from leptonic and hadronic EDM measurements is illustrated in Fig.~\ref{fig:summary_EDM}. The hadronic EDM observables provide constraints on several of the same SMEFT operators that are also probed by the electron EDM. A clear complementarity emerges between the two classes of observables. Electron EDM measurements provide the strongest sensitivity to operators that directly generate leptonic dipoles through Barr-Zee diagrams or RGE-induced mixing, whereas hadronic EDM observables are particularly effective in probing operators that induce quark EDMs, CEDMs, and the Weinberg three-gluon operator. In the below items, we summarise the sensitivity of different operators to different EDM observables and the associated bounds which can be seen from the Fig.~\ref{fig:summary_EDM}. 

\begin{itemize} 
	
\item \textbf{\textit{Top-quark dipole couplings:}}
The electroweak dipole operators $\mathrm{Im}(C^{uB}_{33})$ and $\mathrm{Im}(C^{uW}_{33})$ are most strongly constrained by the electron EDM through double-logarithmically enhanced RGE effects, reaching sensitivities of $\mathcal{O}(10^{-4})\,\mathrm{TeV}^{-2}$. In contrast, the top chromodipole operator $\mathrm{Im}(C^{uG}_{33})$ is primarily constrained by hadronic EDM observables through the Weinberg operator, yielding bounds at the level of $\mathcal{O}(10^{-3})\,\mathrm{TeV}^{-2}$. Among the charged-current dipoles, $\mathrm{Im}(C^{dW}_{13})$ receives the strongest constraint through the pion--nucleon coupling $\bar g_1$, reaching $\mathcal{O}(10^{-7})\,\mathrm{TeV}^{-2}$, while $\mathrm{Im}(C^{dW}_{23})$ and $\mathrm{Im}(C^{dW}_{33})$ are constrained at the level of $\mathcal{O}(10^{-3})\,\mathrm{TeV}^{-2}$ through neutron EDM $d_n$.

\item \textbf{\textit{Top-current couplings:}} The neutral-current operators $\mathrm{Im}(C^{\phi q(1)}_{33})$ and $\mathrm{Im}(C^{\phi q(3)}_{33})$ are most strongly constrained by the electron EDM through Barr-Zee topologies, at the level of $\mathcal{O}(10^{-1})\,\mathrm{TeV}^{-2}$. The right-handed current operator $\mathrm{Im}(C^{\phi u}_{33})$ receives a comparable constraint from the pion--nucleon coupling $\bar g_1$, also at the level of $\mathcal{O}(10^{-1})\,\mathrm{TeV}^{-2}$. The charged-current operators $\mathrm{Im}(C^{\phi ud}_{13})$ and $\mathrm{Im}(C^{\phi ud}_{23})$ are dominantly constrained through hadronic observables, reaching sensitivities of $\mathcal{O}(10^{-5})\,\mathrm{TeV}^{-2}$ and $\mathcal{O}(10^{-2})\,\mathrm{TeV}^{-2}$, respectively.

\item \textbf{\textit{Top-Higgs Yukawa coupling:}} The coefficient $\mathrm{Im}(C^{u\phi}_{33})$ is one of the few flavor-conserving operators for which hadronic EDMs provide a slightly stronger constraint than the electron EDM. The dominant sensitivity originates from the pion--nucleon coupling $\bar g_1$, yielding a bound at the level of $\mathcal{O}(10^{-2})\,\mathrm{TeV}^{-2}$, while the electron EDM provides a comparable constraint through Barr-Zee topologies.

\item \textbf{\textit{Top four-fermion operators:}} The semileptonic operators $\mathrm{Im}(C^{lequ(3)}_{1133})$ and $\mathrm{Im}(C^{lequ(1)}_{1133})$ are overwhelmingly constrained by the electron EDM, with sensitivities reaching $\mathcal{O}(10^{-12})\,\mathrm{TeV}^{-2}$ and $\mathcal{O}(10^{-10})\,\mathrm{TeV}^{-2}$, respectively. In contrast, the four-quark operators $C^{quqd(1)}_{1331}$ and $C^{quqd(8)}_{1331}$ are most effectively probed by hadronic EDMs through the induced down-quark CEDM and the pion--nucleon coupling $\bar g_1$, yielding bounds at the level of $\mathcal{O}(10^{-9})\,\mathrm{TeV}^{-2}$ and $\mathcal{O}(10^{-8})\,\mathrm{TeV}^{-2}$. The corresponding strange-quark operators, $C^{quqd(1)}_{2332}$ and $C^{quqd(8)}_{2332}$, are constrained at the level of $\mathcal{O}(10^{-4})\,\mathrm{TeV}^{-2}$ through the neutron EDM.

\item \textbf{\textit{Top-FCNC couplings:}} EDM observables constrain products of FCNC couplings rather than individual couplings, since the relevant contributions arise through double insertions. The strongest constraints are obtained for products of up-sector FCNC couplings involving opposite chiralities, mainly through one-loop generation of light-quark EDMs and CEDMs, reaching sensitivities of $\mathcal{O}(10^{-8}-10^{-7})$. Additional Barr-Zee contributions involving one $tuH$ coupling and one $tuY$ coupling ($Y=\gamma,g,Z$) are constrained at the level of $\mathcal{O}(10^{-5}-10^{-4})$. In contrast, products involving charm-sector FCNC couplings are more weakly constrained because they enter predominantly through the charm-CEDM threshold correction and subsequent RGE evolution.

\end{itemize}

Overall, the comparison reveals a clear pattern. Electron EDM measurements provide the strongest constraints on operators that generate leptonic dipoles through Barr-Zee diagrams or RGE-enhanced electroweak dipole mixing, notably $C^{uB}_{33}$, $C^{uW}_{33}$, $C_{\phi q}^{(1,3)}$, $C^{\phi u}$, and the semileptonic operators $C_{lequ}^{(1,3)}$. In contrast, hadronic EDM observables dominate the sensitivity to operators that generate quark EDMs, chromoelectric dipole moments, and the Weinberg three-gluon operator, particularly $C^{uG}_{33}$, $C^{dW}_{13}$, $C^{dW}_{23}$, and several four-quark interactions. The combination of leptonic and hadronic EDM measurements therefore provides a highly complementary probe of the $CP$-violating top-quark sector, with different observables probing distinct operator structures and matching mechanisms.

%Overall, electron EDM measurements provide the leading sensitivity to electroweak dipole, neutral-current, and semileptonic operators, with the strongest bounds reaching $\mathcal{O}(10^{-12})\,\mathrm{TeV}^{-2}$. Hadronic EDM observables, particularly the neutron EDM and the pion--nucleon coupling $\bar g_1$, dominate the sensitivity to chromoelectric, gluonic, charged-current, top-Higgs, and four-quark interactions, yielding constraints as strong as $\mathcal{O}(10^{-9})\,\mathrm{TeV}^{-2}$ in several cases. Future proton-EDM measurements are expected to substantially strengthen the sensitivity to operators contributing through the Weinberg three-gluon operator, potentially improving the reach on several top-quark interactions by up to three orders of magnitude. Taken together, the complementary information provided by leptonic and hadronic EDM observables enables a comprehensive exploration of the $CP$-violating top-quark sector within the SMEFT framework.

%----------------------------------
\section{Dimension-Eight SMEFT Matching}
\label{sec:dim8_matching}
%----------------------------------
% The results presented in Table~\ref{tab:bound_dim8} demonstrate the remarkable sensitivity of EDM observables to dimension-eight SMEFT interactions. Since the operators listed in Table~\ref{tab:dim8_dim6_matching} generate, after electroweak symmetry breaking, the same low-energy structures as the dimension-six operators considered throughout this work, the corresponding EDM constraints can be straightforwardly reinterpreted as bounds on the associated dimension-eight WCs. For the naming convention of the dimension-eight operators and their WCs, we adopt the Murphy basis introduced in Ref.~\cite{Murphy:2020rsh}. The resulting limits on the relevant dimension-eight coefficients are summarized in Table~\ref{tab:bound_dim8}.

In view of the remarkable precision of EDM measurements, it is also
important to assess the extent to which these constraints can probe
higher-dimensional interactions.
 In this section, we therefore identify
the subset of dimension-eight SMEFT operators that, after electroweak
symmetry breaking, generate at tree level the dimension-six structures
relevant to the top-quark interactions studied in this work. For the
naming of the dimension-eight WCs, we adopt the convention of the Murphy
basis~\cite{Murphy:2020rsh}. The corresponding tree-level matching
relations are summarized in Table~\ref{tab:dim8_dim6_matching}. In Table~\ref{tab:bound_dim8}, we collect the bounds on the relevant
dimension-eight WCs, including those in each sector that yield the
strongest constraints on the imaginary parts of the corresponding
couplings.
\begin{table}[t!]
    \centering
    \renewcommand{\arraystretch}{1.3}
    \setlength{\tabcolsep}{3pt}
    \begin{tabular}{c c}
        \toprule
       \textbf{ Dimension-$6$ } & \textbf{Dimension-$8$ } \\
        \textbf{SMEFT WCs } & \textbf{SMEFT WCs }\\
        \midrule

        $C^{uG}_{pr}$
        &
        $\displaystyle
        \frac{v^2}{2\Lambda^2}
        C_{\underset{pr}{qu}GH^3}$
        \\[2mm]

        $C^{uB}_{pr}$
        &
        $\displaystyle
        \frac{v^2}{2\Lambda^2}
        C_{\underset{pr}{qu}BH^3}$
        \\[2mm]

        $C^{uW}_{pr}$
        &
        $\displaystyle
        \frac{v^2}{2\Lambda^2}
        \left(
        C_{\underset{pr}{qu}WH^3}^{(1)}
        -
        C_{\underset{pr}{qu}WH^3}^{(2)}
        \right)
        =
        \frac{v^2}{2\Lambda^2}
        C_{\underset{pr}{qu}WH^3}^{(-)}$
        \\[2mm]

        $C^{dW}_{pr}$
        &
        $\displaystyle
        \frac{v^2}{2\Lambda^2}
        \left(
        C_{\underset{pr}{qd}WH^3}^{(1)}
        +
        C_{\underset{pr}{qd}WH^3}^{(2)}
        \right)
        =
        \frac{v^2}{2\Lambda^2}
        C_{\underset{pr}{qd}WH^3}^{(+)}$
        \\[2mm]

        $C^{\phi q(1)}_{pr}$
        &
        $\displaystyle
        \frac{v^2}{2\Lambda^2}
        C^{(1)}_{\underset{pr}{q^2}H^4D}$
        \\[2mm]

        $C^{\phi q(3)}_{pr}$
        &
        $\displaystyle
        \frac{v^2}{\Lambda^2}
        C^{(2)}_{\underset{pr}{q^2}H^4D}$
        \\[2mm]

        $C^{\phi u}_{pr}$
        &
        $\displaystyle
        \frac{v^2}{2\Lambda^2}
        C^{(1)}_{\underset{pr}{u^2}H^4D}$
        \\[2mm]

        $C^{\phi ud}_{pr}$
        &
        $\displaystyle
        \frac{v^2}{2\Lambda^2}
        C_{\underset{pr}{ud}H^4D}$
        \\[2mm]

        $C^{u\phi}_{pr}$
        &
        $\displaystyle
        \frac{5v^2}{6\Lambda^2}
        C_{\underset{pr}{qu}H^5}$
        \\[2mm]

        $C^{lequ(1)}_{prst}$
        &
        $\displaystyle
        \frac{v^2}{2\Lambda^2}
        \left(
        C^{(1)}_{\underset{prst}{lequ}H^2}
        +
        C^{(2)}_{\underset{prst}{lequ}H^2}
        \right)$
        \\[2mm]

        $C^{lequ(3)}_{prst}$
        &
        $\displaystyle
        \frac{v^2}{2\Lambda^2}
        \left(
        C^{(3)}_{\underset{prst}{lequ}H^2}
        +
        C^{(4)}_{\underset{prst}{lequ}H^2}
        \right)$
        \\[2mm]

        $C^{quqd(1)}_{prst}$
        &
        $\displaystyle
        \frac{v^2}{2\Lambda^2}
        \left(
        C^{(1)}_{\underset{prst}{q^2ud}H^2}
        +
        C^{(2)}_{\underset{prst}{q^2ud}H^2}
        \right)$
        \\[2mm]

        $C^{quqd(8)}_{prst}$
        &
        $\displaystyle
        \frac{v^2}{2\Lambda^2}
        \left(
        C^{(3)}_{\underset{prst}{q^2ud}H^2}
        +
        C^{(4)}_{\underset{prst}{q^2ud}H^2}
        \right)$
        \\[2mm]

        $C^{qu(1)}_{prst}$
        &
        $\displaystyle
        \frac{v^2}{2\Lambda^2}
        \left(
        C^{(1)}_{\underset{prst}{q^2u^2}H^2}
        +
        C^{(2)}_{\underset{prst}{q^2u^2}H^2}
        \right)$
        \\[2mm]

        $C^{qu(8)}_{prst}$
        &
        $\displaystyle
        \frac{v^2}{2\Lambda^2}
        \left(
        C^{(3)}_{\underset{prst}{q^2u^2}H^2}
        +
        C^{(4)}_{\underset{prst}{q^2u^2}H^2}
        \right)$
        \\

        \bottomrule
    \end{tabular}
    \caption{Tree-level matching relations between the relevant dimension-eight
SMEFT WCs and the dimension-six coefficients considered
in our analysis.}
    \label{tab:dim8_dim6_matching}
\end{table}
%----------------------------------

%-----------------------------------
\begin{table*}[t!]
    \centering
    \renewcommand{\arraystretch}{1.3}
    \setlength{\tabcolsep}{8pt}
    \begin{tabular}{@{} c c c c c c@{}}
        \toprule
        \textbf{WCs $[\mathrm{TeV}^{-4}]$}
        & \textbf{Values}
        & \textbf{Observable} 
        & \textbf{WCs $[\mathrm{TeV}^{-4}]$}
        & \textbf{Values}
        & \textbf{Observable} \\
        \midrule

        $\mathrm{Im}\!\left(C_{\underset{33}{qu}GH^3}\right)$
        & $1.34 \times 10^{-1}$ & $d_n$ &  $\mathrm{Im}\!\left(
        C^{(1)}_{\underset{1133}{lequ}H^2}
        + C^{(2)}_{\underset{1133}{lequ}H^2}
        \right)$ & $2.92 \times 10^{-8}$ & $d_e$  \\

        $\mathrm{Im}\!\left(C_{\underset{33}{qu}BH^3}\right)$
        & $2.36 \times 10^{-3}$ & $d_e$ &  $\mathrm{Im}\!\left(
        C^{(3)}_{\underset{1133}{lequ}H^2}
        + C^{(4)}_{\underset{1133}{lequ}H^2}
        \right)$ & $1.23 \times 10^{-10}$ & $d_e$\\

        $\mathrm{Im}\!\left(C^{(-)}_{\underset{33}{qu}WH^3}\right)$
        & $1.22 \times 10^{-3}$ & $d_e$ & $\mathrm{Im}\!\left(
        C^{(3)}_{\underset{2233}{lequ}H^2}
        + C^{(4)}_{\underset{2233}{lequ}H^2}
        \right)$ & $3.28$ & $d_{\mu} $\\

        $\mathrm{Im}\!\left(C^{(+)}_{\underset{33}{qd}WH^3}\right)$
        & $8.79 \times 10^{-2}$ & $d_e$ & $\mathrm{Im}\!\left(
        C^{(3)}_{\underset{3333}{lequ}H^2}
        + C^{(4)}_{\underset{3333}{lequ}H^2}
        \right)$ & $1.77 \times 10^{1}$ & $d_{\tau} $\\

         $\mathrm{Im}\!\left(C^{(+)}_{\underset{23}{qd}WH^3}\right)$
        & $4.79\times 10^{-2}$ & $d_n$ & $\mathrm{Im}\!\left(
        C^{(1)}_{\underset{1331}{q^2ud}H^2}
        + C^{(2)}_{\underset{1331}{q^2ud}H^2}
        \right)$ & $1.68 \times 10^{-7}$ & $\bar{g}_1$ \\

         $\mathrm{Im}\!\left(C^{(+)}_{\underset{13}{qd}WH^3}\right)$
        & $1.16 \times 10^{-5}$ & $\bar{g}_1$ & $\mathrm{Im}\!\left(
        C^{(1)}_{\underset{2332}{q^2ud}H^2}
        + C^{(2)}_{\underset{2332}{q^2ud}H^2}
        \right)$ & $3.40 \times 10^{-3}$ & $d_n$\\

        $\mathrm{Im}\!\left(C^{(1)}_{\underset{33}{q^2}H^4D}\right)$
        & $9.15$ & $ d_n $ & $\mathrm{Im}\!\left(
        C^{(1)}_{\underset{3333}{q^2ud}H^2}
        + C^{(2)}_{\underset{3333}{q^2ud}H^2}
        \right)$ & $1.77 \times 10^{-1}$ & $d_n$\\

        $\mathrm{Im}\!\left(C^{(2)}_{\underset{33}{q^2}H^4D}\right)$
        & $1.41$ & $ \bar{g}_1 $ & $\mathrm{Im}\!\left(
        C^{(3)}_{\underset{1331}{q^2ud}H^2}
        + C^{(4)}_{\underset{1331}{q^2ud}H^2}
        \right)$ & $1.35 \times 10^{-6}$ & $\bar{g}_1$ \\

        $\mathrm{Im}\!\left(C^{(1)}_{\underset{33}{u^2}H^4D}\right)$
        & $9.22$ & $ d_n $ & $\mathrm{Im}\!\left(
        C^{(3)}_{\underset{2332}{q^2ud}H^2}
        + C^{(4)}_{\underset{2332}{q^2ud}H^2}
        \right)$ & $4.53 \times 10^{-3}$ & $d_n$ \\

        $\mathrm{Im}\!\left(C_{\underset{33}{ud}H^4D}\right)$
        & $4.03$ & $ d_n $ &  $\mathrm{Im}\!\left(
        C^{(3)}_{\underset{3333}{q^2ud}H^2}
        + C^{(4)}_{\underset{3333}{q^2ud}H^2}
        \right)$ & $1.87 $ & $d_n$\\

         $\mathrm{Im}\!\left(C_{\underset{32}{ud}H^4D}\right)$
        & $1.37$ & $ d_n $ & $\mathrm{Im}\!\left(
        C^{(1)}_{\underset{1331}{q^2u^2}H^2}
        + C^{(2)}_{\underset{1331}{q^2u^2}H^2}
        \right)$ & $4.39 \times 10^{-7}$ & $\bar{g}_1$\\

         $\mathrm{Im}\!\left(C_{\underset{31}{ud}H^4D}\right)$
        & $4.26 \times 10^{-4} $ & $ \bar{g}_1 $ & $\mathrm{Im}\!\left(
        C^{(1)}_{\underset{2332}{q^2u^2}H^2}
        + C^{(2)}_{\underset{2332}{q^2u^2}H^2}
        \right)$ & $3.93 \times 10^{-2}$ & $d_n$\\

        $\mathrm{Im}\!\left(C_{\underset{33}{qu}H^5}\right)$
        & $1.75 \times 10^{-1}$ & $ \bar{g}_1 $ & $\mathrm{Im}\!\left(
        C^{(3)}_{\underset{1331}{q^2u^2}H^2}
        + C^{(4)}_{\underset{1331}{q^2u^2}H^2}
        \right)$ & $2.38 \times 10^{-6}$ & $\bar{g}_1$  \\

        &&&$\mathrm{Im}\!\left(
        C^{(3)}_{\underset{2332}{q^2u^2}H^2}
        + C^{(4)}_{\underset{2332}{q^2u^2}H^2}
        \right)$
        & $2.35 \times 10^{-1}$ & $d_n$\\

        \bottomrule
    \end{tabular}
    \caption{Bounds on the dimension-eight SMEFT WCs from different EDM
    inputs. The coefficients are evaluated at $\Lambda=1~\mathrm{TeV}$ and
    have mass dimension $\mathrm{TeV}^{-4}$.}
    \label{tab:bound_dim8}
\end{table*}

%-----------------------------------
A clear pattern emerges from Table~\ref{tab:bound_dim8}. The strongest constraints are obtained for the semileptonic operators generating the tensor interaction $C^{lequ(3)}$. In particular, $\mathrm{Im}\!\left(C^{(3)}_{\underset{1133}{lequ}H^2}+C^{(4)}_{\underset{1133}{lequ}H^2}\right)$ is constrained at the level of $1.23\times10^{-10}\,\mathrm{TeV}^{-4}$ by the electron EDM, representing the strongest bound in the entire dimension-eight parameter space considered here. This mirrors the corresponding dimension-six analysis, where the electron EDM provided exceptional sensitivity to $C^{lequ(3)}_{1133}$ through dipole mixing and top-mass-enhanced contributions. The analogous operators involving second- and third-generation leptons are constrained less stringently, at the levels of $3.28~\mathrm{TeV}^{-4}$ and $1.77\times10^{1}~\mathrm{TeV}^{-4}$, respectively, reflecting the current hierarchy in experimental sensitivities among leptonic EDM observables.

Among the dipole operators, the electroweak top-dipole interactions receive the strongest bounds. As shown in Table~\ref{tab:bound_dim8}, the coefficients $\mathrm{Im}(C_{\underset{33}{qu}BH^3})$ and $\mathrm{Im}(C^{(-)}_{\underset{33}{qu}WH^3})$ are constrained to $2.36\times10^{-3}\,\mathrm{TeV}^{-4}$ and $1.22\times10^{-3}\,\mathrm{TeV}^{-4}$, respectively, with the electron EDM providing the dominant sensitivity. These limits originate from the same Barr-Zee and RGE-enhanced mechanisms responsible for the strong constraints on the corresponding dimension-six electroweak dipole operators. In contrast, the chromodipole operator $\mathrm{Im}(C_{\underset{33}{qu}GH^3})$ is constrained at the level of $1.34\times10^{-1}\,\mathrm{TeV}^{-4}$, predominantly through the neutron EDM. This reflects the special role of hadronic observables in probing gluonic sources of $CP$ violation through the Weinberg operator.

The charged-current sector exhibits a qualitatively different behavior. The strongest sensitivities arise for the coefficients $\mathrm{Im}(C^{(+)}_{\underset{13}{qd}WH^3})$ and $\mathrm{Im}(C_{\underset{31}{ud}H^4D})$, which are constrained at $1.16\times10^{-5}\,\mathrm{TeV}^{-4}$ and $4.26\times10^{-4}\,\mathrm{TeV}^{-4}$, respectively. In both cases the dominant observable is the CP-odd pion--nucleon coupling $\bar g_1$. Similar to the dimension-six analysis, these strong bounds arise from the efficient generation of light-quark CEDMs, which subsequently induce sizeable contributions to hadronic $CP$-violating observables.

A particularly noteworthy feature of Table~\ref{tab:bound_dim8} is the exceptional sensitivity to four-fermion operators. The combinations $\mathrm{Im}(C^{(1)}_{\underset{1331}{q^2ud}H^2}+C^{(2)}_{\underset{1331}{q^2ud}H^2})$, $\mathrm{Im}(C^{(3)}_{\underset{1331}{q^2ud}H^2}+C^{(4)}_{\underset{1331}{q^2ud}H^2})$, $\mathrm{Im}(C^{(1)}_{\underset{1331}{q^2u^2}H^2}+C^{(2)}_{\underset{1331}{q^2u^2}H^2})$, and $\mathrm{Im}(C^{(3)}_{\underset{1331}{q^2u^2}H^2}+C^{(4)}_{\underset{1331}{q^2u^2}H^2})$ are constrained at the level of $10^{-7}$--$10^{-6}\,\mathrm{TeV}^{-4}$, with $\bar g_1$ providing the dominant sensitivity. These results closely parallel the dimension-six bounds on the operators $C^{(1,8)}_{quqd}$ and $C^{(1,8)}_{qu}$, for which hadronic EDM observables were found to provide the leading constraints. The corresponding second-generation operators are bounded less stringently, typically at the level of $10^{-3}$--$10^{-1}\,\mathrm{TeV}^{-4}$, primarily through the neutron EDM.

The current-type operators $C^{(1)}_{\underset{33}{q^2}H^4D}$, $C^{(2)}_{\underset{33}{q^2}H^4D}$, $C^{(1)}_{\underset{33}{u^2}H^4D}$, and $C_{\underset{33}{ud}H^4D}$ are constrained at the level of ${\cal O}(1)$--${\cal O}(10)\,\mathrm{TeV}^{-4}$, as summarized in Table~\ref{tab:bound_dim8}. Relative to the dipole and four-fermion sectors, these weaker bounds can be traced to the fact that the corresponding dimension-six interactions contribute predominantly through two-loop Barr-Zee topologies and RGE-induced effects, leading to a reduced sensitivity of the EDM observables.

Overall, the pattern of constraints displayed in Table~\ref{tab:bound_dim8} closely mirrors that observed for the dimension-six operators. Leptonic observables, particularly the electron EDM, provide the strongest sensitivity to electroweak dipole and semileptonic interactions, yielding bounds as strong as ${\cal O}(10^{-10})\,\mathrm{TeV}^{-4}$. In contrast, hadronic observables, especially the neutron EDM and the pion-nucleon coupling $\bar g_1$, dominate the sensitivity to charged-current, four-quark, top-Higgs, and gluonic interactions. These results demonstrate that present EDM measurements already provide highly nontrivial constraints on dimension-eight SMEFT interactions and highlight the exceptional reach of precision EDM experiments into the parameter space of physics beyond the SM.

%-----------------------------------
\section{Conclusions}\label{sec:conclusion}
%---------------------------------
In this work, we have performed a comprehensive and systematic study of $CP$-violating top-quark interactions using a broad set of leptonic and hadronic EDM observables, consistently incorporating all relevant contributions up to two-loop order. We derived the complete matching relations connecting the relevant SMEFT WCs to the low-energy $CP$-violating observables, including one- and two-loop matching, Barr-Zee contributions, Weinberg-operator effects, operator mixing, renormalization-group evolution, and heavy-quark threshold corrections. For flavor-changing top-quark interactions, where the Wilson-coefficient structure becomes increasingly involved, we also employed a dimensionless effective-coupling parametrization to present the resulting constraints in a compact and transparent manner. Using current and projected sensitivities from leptonic, nucleon, atomic, molecular, and light-nuclear EDM measurements, we obtained bounds on the relevant SMEFT WCs under the single-coefficient assumption at a reference scale of $\Lambda=1~\mathrm{TeV}$, thereby providing a unified assessment of the present and future EDM sensitivity to $CP$-violating top-quark interactions.

An important feature of our analysis is the complementarity between
leptonic and hadronic EDMs. While the electron EDM provides particularly
strong constraints on several flavor-conserving top interactions through
Barr-Zee diagrams and RGE-enhanced contributions, hadronic EDMs provide
complementary sensitivity through light-quark EDMs and CEDMs, the
Weinberg operator, and heavy-quark threshold corrections. For
top-FCNC interactions, EDM observables generally constrain products of
couplings rather than individual coefficients. In particular, products
of opposite-chirality $tuX$ couplings can be probed down to the
$\mathcal{O}(10^{-8}-10^{-7})$ level, while the corresponding
$tuH$--$tuY$ combinations are constrained at the
$\mathcal{O}(10^{-5}-10^{-4})$ level, where $Y=\ga,g,H$ gauge bosons.

Finally, we extend our analysis to the dimension-eight SMEFT and identify
the operators that generate the anomalous top-quark interactions considered
in this work after electroweak symmetry breaking. We provide the
corresponding tree-level matching relations and derive their sensitivity
to the EDM observables, demonstrating that the high precision of EDM
measurements can also probe higher-dimensional top-quark interactions.

The electron and neutron EDMs probe complementary aspects of CP-violating top-quark interactions. Owing to its exceptional experimental precision and sensitivity to Barr-Zee and RGE-enhanced dipole contributions, the electron EDM provides the strongest constraints on electroweak top-dipole, top-current, and semileptonic operators, reaching sensitivities as strong as ${\cal O}(10^{-5})$-${\cal O}(10^{-12})~{\rm TeV}^{-2}$. In contrast, the neutron EDM is particularly sensitive to operators that generate quark chromoelectric dipole moments and the Weinberg three-gluon operator, yielding the leading constraints on the top chromodipole interaction and several charged-current operators. The combination of leptonic and hadronic EDM observables therefore provides a highly complementary probe of the CP-violating top-quark sector, with different observables dominating different regions of the SMEFT parameter space.

\begin{acknowledgments}
SK would like to thank Mr Vaibhav Pandey for useful discussions.
\end{acknowledgments}
%==================================
%          APPENDICES
%==================================
\newpage
\appendix
%--------------------------------------
\section{Loop-level matching contributions to EDMs}
\label{Append:loop_matching}
%------------------------------------
In this appendix, we collect the explicit loop-level matching contributions
to the EDM observables discussed in Sec.~\ref{sec:results}. These
expressions provide the matching relations between the relevant SMEFT
WCs and the WET coefficients that contribute to the EDMs,
which are subsequently used to derive the bounds presented in the main
text. It is worth noting that in most  of the cases the loop contributions are written in terms of ratios of heavy masses, defined as $ z_{t/H} = \frac{m_t^2}{m_H^2} $, $ z_{t/Z} = \frac{m_t^2}{m_Z^2} $ , $ z_{t/W}= \frac{m_t^2}{m_W^2} $ and $ z_{d_i/W} =\frac{m_{d_i}^2}{m_W^2} $,where $ d_i $ represents down-type quarks of the three generation.$ N_c$ indicates closed fermion loop factor , $N_c =3 (N_c=1)$ for closed quark(lepton) loop.
%-------------------------------
\subsection{Leptonic EDMs}\label{Append:leptonic EDM}
%-------------------------------

%----------------------------
\subsubsection{\textbf{Flavor-conserving top-quark interactions:}}\label{Append:leptonic EDM_FC}

%----------------------------

The one loop matching relation for lepton EDM arises through Fig.~\ref{fig:lepton_edm_contact},
%----------------------------------
\begin{align}\label{eq:op_lequ3}
    L^{e\gamma}_{pp} &= \frac{N_c Q_t m_t e}{2 \pi^2} \mathrm{ln}\left( \frac{\mu^2}{m_t^2} \right) \mathrm{Im}\left[C_{pp33}^{lequ(3)}\right].
\end{align}
%--------------------------------

Contributions from finite two-loop Barr-Zee topologies shown in
Figs.~\ref{dl_FCa}--\ref{dl_FCc}.
%-------------------------------------
\begin{subequations}\label{eq:BZ_de_flv_con}
\begin{align}\label{eq:BarrZee_top_cons_ga}
    &L^{e\gamma}_{pp} \left[\mathcal{M}^{\hyperref[dl_FCa]{a}}_{\gamma}+\mathcal{M}^{\hyperref[dl_FCb]{b}}_{\gamma}+\mathcal{M}^{\hyperref[dl_FCc]{c}}_{\gamma}\right]=
    \frac{e^2 N_c Q_t y_l v}{2 (4 \pi)^4 m_t}\, \nn \\ 
    &
    \times \text{Im} \Bigg[3 e v  Q_t C^{u \phi}_{33} g(z_{t/H}) +  \sqrt{2} m_t y_t\bigg( c_W C^{u B}_{33}+ s_W C^{u W}_{33}\bigg)\, \nn \\
    &
    \times  \left(\frac{f(z_{t/H})}{2}  - 3 g(z_{t/H})\right) \Bigg]  
\end{align}

\begin{align}\label{eq:BarrZee_top_cons_Z}
&L^{e\gamma}_{pp}\left[\mathcal{M}^{\hyperref[dl_FCa]{a}}_{Z}+\mathcal{M}^{\hyperref[dl_FCb]{b}}_{Z}+\mathcal{M}^{\hyperref[dl_FCc]{c}}_{Z}\right]= \frac{N_c Q_t g_V^l g_z m_t v y_l y_t }{8 (4 \pi)^4 (m_H^2-m_Z^2)}  \nn \\ 
& \times \text{Im} \Bigg[\frac{e v g}{c_W} \mathcal{I}_1~\left( C^{\phi q(1)}_{33}-C^{\phi q(3)}_{33}+C^{\phi u}_{33} \right) \nonumber \\
&
- \frac{3eg_V^u g_z v}{y_t}\mathcal{I}_2~ C^{u \phi}_{33}   
-\sqrt{2} m_t \bigg\{ e Q_t \mathcal{I}_3~ \left( s_W C^{u B}_{33}- c_W C^{u W}_{33}\right) \bigg. \nn  \\ 
& 
\left.- \frac{g_V^u g_z}{Q_t} \mathcal{I}_4~ \left( c_W C^{u B}_{33}+ s_W C^{u W}_{33}\right) \right\} \Bigg]
\end{align}
\end{subequations}
%------------------------------------
The functions $g(z)$ and $f(z)$ are given by,
%--------------------------------------------
\begin{subequations}
\begin{align}
    g(z_{t/H})&= \frac{z_{t/H}}{2} \int_0^1 \frac{dx}{x(1-x)-z_{t/H}} \text{ln}\left[\frac{x(1-x)}{z_{t/H}}\right]  \\
    f(z_{t/H})&= \frac{z_{t/H}}{2} \int_0^1 dx \frac{1-2x(1-x)}{x(1-x)-z_{t/H}}\text{ln}\left[\frac{x(1-x)}{z_{t/H}}\right] 
\end{align}
\end{subequations}
%--------------------------------------------
% where
% \begin{align}
% \omega_{t/H}&=\frac{m_t^2}{m_H^2},
% &
% \omega_{t/Z}&=\frac{m_t^2}{m_Z^2}.
% \end{align}

The functions $\mathcal{I}_n$ are given by
%-----------------
\begin{align}
    \mathcal{I}_n =\int_0^1   \frac{\mathcal{N}_n(x) dx}{x(1-x)} \bigg[ &\frac{m_Z^2}{(m_Z^2-m_x^2)}\mathrm{ln}\left( \frac{m_Z^2}{m_x^2}\right) \nn\\
    &- \frac{m_H^2}{(m_H^2-m_x^2)}\mathrm{ln}\left( \frac{m_H^2}{m_x^2}\right)\bigg]
\end{align}
%----------------
where

\begin{align}
m_x^2 &= \frac{m_t^2}{x(1-x)},  \nn \\
\mathcal{N}_1(x)&=1+x-2x^2,
&
\mathcal{N}_2(x)&=1-x,
\nonumber\\
\mathcal{N}_3(x)&=x^2-4x+3,
&
\mathcal{N}_4(x)&=x^2-1.
\end{align}

%----------------------------
\vspace{0.5em}
\subsubsection{\textbf{Flavor-violating charged-current interactions:}}\label{Append:leptonic EDM_FCCC}
%----------------------------
Contributions from finite two-loop Barr--Zee topologies shown in
Figs.~\ref{dl_FCd} and \ref{fig:BZ_lep_EDM_FCCC}.
% %-----------------------------

%-----------------------------
\begin{align}\label{eq:BarrZee_top_vio_W}
& L^{e\gamma}_{pp}\left[\mathcal{M}_{W}\right]= \frac{\sqrt{2 }N_c e^3 m_l q_{u_i}}{3(4 \pi)^4 m_W s_W^2 } V_{td_i} \, \nn   \\  
&
\times \mathrm{Im} \bigg[ \frac{m_t}{2}(C^{dW}_{i3}) \mathcal{F}_1 (z)+ m_b (C^{uW}_{i3}) (\mathcal{F}_2 (z) + \mathcal{F}_3 (z))\bigg] 
\end{align}

The required loop functions are given by,
% %----------------------

%----------------------------------
\begin{subequations}
\begin{align}
    &\mathcal{F}_1(z) = \int_0^1 \frac{(x^3-2x^2-x-2)dx}{[x(1-x)-x z_{d_i/W}-(1-x)z_{t/W}]}\, \nn \\
    &
    \times \left[ 1+\frac{(x z_{d_i/W} + (1-x) z_{t/W})}{(x(1-x)-x z_{d_i/W}-(1-x)z_{t/W})} \right.   \, \nn  \\
    &\hspace{2 cm}
    \left. \times \mathrm{ln} \left( \frac{x z_{d_i/W}+ (1-x) z_{t/W}}{x(1-x)} \right) \right] \, \\
    &
    \mathcal{F}_2(z) = \int_0^1 \frac{(3 z_{t/W}(1-x)+(x^3+x^2-2x))}{[x(1-x)-x z_{d_i/W}-(1-x)z_{t/W}]} dx\, \\
    &
    \mathcal{F}_3(z) = \int_0^1 \frac{(z_{t/W}x(1-x)^3 + z_{d_i/W}x^2(x^2+x-2))}{[x(1-x)-x z_{d_i/W}-(1-x)z_{t/W}]^2} dx\,  \nn \\
    & \hspace{2 cm}
    \times \mathrm{ln} \left( \frac{x z_{d_i/W}+ (1-x) z_{t/W}}{x(1-x)} \right) 
\end{align}
\end{subequations}
%----------------------------------

%----------------------------
\vspace{0.5em}
\subsubsection{\textbf{Flavor-changing neutral-current interactions}:}\label{Append:leptonic EDM_FCNC}
%----------------------------
Contributions from finite two-loop Barr-Zee topologies shown in
Figs.~\ref{fig:EDM_flavor_vio_neutral}.
%---------------------------------
\begin{subequations}\label{eq:matching_lepEDM_topFCNC}
    \begin{align}
    &L^{e\gamma}_{pp}\left[\mathcal{M}^{\hyperref[dl_FCNCa]{a}}_{\gamma}+\mathcal{M}^{\hyperref[dl_FCNCb]{b}}_{\gamma}+\mathcal{M}^{\hyperref[dl_FCNCc]{c}}_{\gamma}\right]=
    - \frac{e^3 N_c q_{u_i} y_l}{8 (4 \pi)^4 m_H^2 }  \nn \\ 
& \times \mathrm{Im}\bigg[2 m_{u_i}\bigg(\lambda_L^{u_it} \eta_R^{tu_i*}- \lambda_R^{u_it}\eta_L^{tu_i*}\bigg) \mathcal{G}_1(z) \nn\\
&+ m_t \bigg(\eta_L^{u_i t} \lambda_L^{t u_i*}-\lambda_R^{u_i t} \eta_R^{tu_i*}\bigg)\mathcal{G}_2(z)\bigg]\, \\
&L^{e\gamma}_{pp}\left[\mathcal{M}^{\hyperref[dl_FCNCa]{a}}_{Z}+\mathcal{M}^{\hyperref[dl_FCNCb]{b}}_{Z}+\mathcal{M}^{\hyperref[dl_FCNCc]{c}}_{Z}\right]= \frac{e^3 m_t N_c q_{u_i} y_l}{32(4\pi)^4 s_W^2 c_W^2 m_H^2} \, \nn \\
& \times \mathrm{Im}\bigg[ \left\{ 2 g_V^l \mathcal{G}_3(z_{t/H}) + g_A^l \mathcal{G}_4(z_{t/H}) \right\}\, \nn \\
& \times \left(X_R^{u_it} \eta_R^{tu_i*}- X_L^{u_it}\eta_L^{tu_i*}\right) - (g_A^l+2 g_V^l)\mathcal{G}_2(z_{t/H})\, \nn \\
& \times \left(\kappa_L^{u_it} \eta_L^{tu_i*}- \kappa_R^{u_it}\eta_R^{tu_i*}\right)\bigg]
    \end{align}
\end{subequations}
where rescaled couplings are presented in the Eq.~\eqref{eq:top_effective_matching},
%---------------------------------
The function $\mathcal{G}_n(z)$ is given by,
\begin{align}
    \mathcal{G}_n(z) &= \int_0^1 \frac{\mathcal{N}_n(x)dx}{x[x(1-x) - x z_{u_i/H}-(1-x)z_{t/H}]} \nn \\
    &\times \ln \left[ \frac{x z_{u_i/H} +(1-x) z_{t/H}}{x(1-x) } \right]
     % G_2(z) &= \int_0^1  \frac{(1-x)dx}{x[x(1-x) - x z_{u_i/H}-(1-x)z_{t/H}]} \nn\\
     % & \times\text{ln} \left[ \frac{x z_{u_i/H} +(1-x) z_{t/H}}{x(1-x)} \right]
\end{align}
%----------------------------------
where, 
%-----------------------
\begin{align}
\mathcal{N}_1(x)&=1,
&
\mathcal{N}_2(x)&=1-x, \nn \\
\mathcal{N}_3(x)&=1+x,
&
\mathcal{N}_4(x)&=1+x^2
\end{align}
%------------------------

where $z_{u_i/H} = \frac{m_{u_i}^2}{m_H^2},$ with $i=1,2$ corresponding to $u$ and $c$, respectively.

%-------------------------------
\subsection{Hadronic EDMs}\label{Append:Hadronic EDM}
%-------------------------------
\subsubsection{\textbf{One loop contributions}:}\label{Append:Hadronic EDM_oneloop}
%----------------------------
Contributions from one loop topologies shown in
Figs.~\ref{fig:EDM_double_a}-\ref{fig:CEDM_three_gluon}.
%----------------------------------
\begin{align}
\begin{split}
 & L_{u \gamma}=
\frac{e q_{u_i}}{32\pi^2m_t}
\Bigg\{C_F g_s^2(\mu)
\operatorname{Im}\left(
\xi_L^{u_it}\xi_R^{tu_i*}
\right)\ln\left(\frac{\mu^2}{m_t^2}\right)
\\
& 
+e^2\operatorname{Im}\left(\lambda_L^{u_it}\lambda_R^{tu_i*}\right)
\ln\left(\frac{\mu^2}{m_t^2}\right)
\Bigg\} +\frac{e q_{u_i}}{64\pi^2}
\frac{m_t}{M_H^2} \\
&
\times \Bigg\{
\frac{z_{t/H}-3}
     {2(z_{t/H}-1)^2}
+
\frac{\ln(z_{t/H})}
     {(z_{t/H}-1)^3}
\Bigg\}
\operatorname{Im}\left(
\eta_L^{u_it}\eta_R^{tu_i*}
\right)
\\
&+
\frac{g^2}{32\pi^2c_W^2}
\frac{e q_{u_i}}{m_t}
\Bigg[
\operatorname{Im}\left(
\kappa_L^{u_it}X_L^{tu_i*}
+X_R^{u_it}\kappa_R^{tu_i*}
+\kappa_L^{u_it}\kappa_R^{tu_i*}
\right)\\
& \times \ln\left(\frac{\mu^2}{m_t^2}\right) +\frac{\ln z_t}
{4(z_{t/Z}-1)^2}
\Bigg\{z_{t/Z}^2\operatorname{Im}\left(2X_L^{tu_i*}X_R^{u_it}\right.
\\
&
\left. +X_L^{u_it}\kappa_L^{tu_i*}
+X_R^{tu_i*}\kappa_R^{u_it}
\right)+2\operatorname{Im}\left(
X_L^{u_it}\kappa_L^{tu_i*}
+X_R^{tu_i*}\kappa_R^{u_it} \right. \\
&
\left. +\kappa_L^{u_it}\kappa_R^{tu_i*}
\right) +z_{t/Z}\operatorname{Im}\left(
2X_L^{tu_i*}X_R^{u_it}
+9X_L^{tu_i*}\kappa_L^{u_it}\right.
\\
&
\left. +9X_R^{u_it}\kappa_R^{tu_i*}
+10\kappa_L^{u_it}\kappa_R^{tu_i*}
\right)
\Bigg\}\\
&
+\frac{\ln z_{t/Z}}
       {2(z_{t/Z}-1)^3}
\Bigg\{
2z_{t/Z}^2
\operatorname{Im}\left(
X_L^{tu_i*}X_R^{u_it}
\right)
\\
&
+z_{t/Z}
\operatorname{Im}\left(
5X_L^{tu_i*}\kappa_L^{tu_i*}
+5X_R^{tu_i*}\kappa_R^{u_it}
+8\kappa_L^{u_it}\kappa_R^{tu_i*}
\right)
\\
&
+2\operatorname{Im}\left(
X_L^{u_it}\kappa_L^{tu_i*}
+X_R^{tu_i*}\kappa_R^{u_it}
+\kappa_L^{tu_i*}\kappa_R^{u_it}
\right)\Bigg\}\Bigg]
\end{split}
\label{Loop:up_EDM}
\end{align}
%-------------------------------------------
The expression for the LEFT operator $ L^{uG} $, from which we get CEDM, $ \tilde{d}_{u_i} $, of up and charm-type quarks is structurally analogous to that of the expression of $ L^{u\gamma} $, apart from a few modifications in coupling constant. Specifically for topologies shown in Fig.~\ref{fig:EDM_double_a}-\ref{fig:CEDM_three_gluon}, corresponding to two insertions of NP couplings, the factor $ e q_{u_i} $ does not contribute. In the case of Fig.~\ref{fig:EDM_double_a}, the color factor $ C_F=4/3  $ in the expression of $ L_{u\gamma} $ is replaced by $ - 1/(2N) $ for $ L^{uG} $. \\

For the single insertion diagrams depicted in Fig.~\ref{fig:qEDM_FCCC_a}-\ref{fig:Weinberg_EDM_FC1} the loop contributions will come in the following way,
%--------------------------------------
\begin{subequations}\label{FCCC_1loop_EDM_CEDM}
\begin{align}
    &L^{d\gamma} = \frac{e^2 g^2 v^2 }{64 \pi^2} \sum_{i \in \{ 1,2,3 \}}V_{td_i}~ \mathrm{Im }\Bigg[ \frac{1}{\sqrt{2}m_W}\left\{ \frac{8}{3} \mathrm{ln} \left( \frac{\mu^2}{m_t^2} \right) \right. \, \nn \\
    & \left. - \frac{9 z_{t/W}(z_{t/W}+1)}{(z_{t/W}-1)^2} + \frac{2 (3 z_{t/W}^2 -10 z_{t/W}-4)}{3 (z_{t/W}-1)^3}  \right\}   \left( C^{dW}_{i3}  \right)  \, \nn \\
    &+ \frac{z_{t/W}}{4 m_t} \left( C^{\phi ud}_{3i} \right)\left\{ \frac{2}{(z_{t/W}-1)^2}\right. \, \nn \\
    & 
    \left.+ \frac{z_{t/W}(3 z_{t/W}-4)}{3 (z_{t/W}-1)^3} \mathrm{ln}(z_{t/W}) \right\}\Bigg] \, \nn \\
    & 
    + \frac{e m_t q_t}{32 \pi^2} \mathrm{ln}\left( \frac{\mu^2}{m_t^2} \right)\Bigg[\left(C_{i33i}^{quqd(1)}\right)+C_F \mathrm{Im}\left(C_{i33i}^{quqd(8)}\right)\Bigg] \\
    \quad
    & L^{dG} = \frac{e^2 v^2 g_s(\mu)}{64 \pi^2 s_W^2}\sum_{i \in \{ 1,2,3 \}}V_{td_i}~ \mathrm{Im}\Bigg[ \frac{1}{\sqrt{2}m_W}\left( C^{dW}_{i3}  \right)\, \nn \\
    &
   \times \left\{ 4 ~\mathrm{ln}\left( \frac{\mu^2}{m_t^2} \right) +\frac{z_{t/W}^3 -10z_{t/W}^2 +11 z_{t/W} -2 }{(z_{t/W}-1)^3}  \right. \, \nn \\
   & 
   \left. + \frac{2 (5 z_{t/W}-2 )}{(z_{t/W}-1)^3} \mathrm{ln}(z_{t/W})\right\} +  \frac{1}{m_t} \left\{ \frac{z_{t/W}(z_{t/W}+1)}{(z_{t/W}-1)^2}\right. \, \nn \\
   &
   \left. -~ \frac{2 z_{t/W}}{(z_{t/W}-1)^3} ~\ln (z_{t/W}) \right\} \left( C^{\phi ud}_{3i} \right)\Bigg] \, \nn \\
   & - \frac{m_t g_s(\mu)}{32 \pi^2} \mathrm{ln}\left( \frac{\mu^2}{m_t^2} \right)\mathrm{Im} \left[\left(C_{i33i}^{quqd(1)}\right) - \frac{1}{2N} \left(C_{i33i}^{quqd(8)}\right)\right].
\end{align}
\end{subequations}
%----------------------------------

%---------------------------------------------
\begin{subequations}
\begin{align}
   &L^{u\gamma} = - \frac{e m_t q_t}{16 \pi^2}  \sum_{i \in \{ 1,2 \}}\mathrm{Im}\left[\left(C_{i33i}^{qu(1)}\right) + C_F \left(C_{i33i}^{qu(8)}\right)\right]\,, \\
   &L^{uG} = - \frac{m_t g_s(\mu)}{16 \pi^2}  \sum_{i \in \{ 1,2 \}} \mathrm{Im}\left[\left(C_{i33i}^{qu(1)}\right) - \frac{1}{2N} \left(C_{i33i}^{qu(8)}\right)\right].   
\end{align}
\end{subequations}
%------------------------------------------------
\begin{align}
     &L^{\tilde{G}} = \frac{\sqrt{2}g_s^2(\mu) v}{48 \pi^2 m _t} \mathrm{Im}\left[C_{33}^{uG}\right].\label{Append-eq:Weinberg_oneloop}
\end{align}
%---------------------------------
\vspace{0.5em}
\subsubsection{\textbf{Barr-Zee contributions}} \label{Append: Had_EDM_BZ}
%----------------------------------
Here we present the complete set of two-loop matching relations arises through all possible Bar-zee topologies shown in Fig.~\ref{fig:dq_FCa}-\ref{fig:dq_fcnc_c}. The single insertion diagrams Fig.~\ref{fig:dq_FCa}-\ref{fig:dq_FCCCa_BZ} gives contribution both in EDM and CEDM of light quarks. The matching relations for this case is structurally analogous to that of lepton EDM, apart from few key modifications. Firstly, for $ V_i = \gamma,Z  $ in Fig.~\ref{fig:dq_FCa}-\ref{fig:dq_FCc} the expressions of $ L^{q\gamma} $,($ q=u,d,s$) are analogous to eq.~\ref{eq:BZ_de_flv_con}. In this case the lepton yukawa term is replaced by the factor $ q_q y_q/2 $. The expressions of $ L^{q \gamma} $ corresponding to diagrams ~\ref{fig:dq_FCc}-\ref{fig:dq_FCCCa_BZ} are analogous to eq. \ref{eq:BarrZee_top_vio_W}, by replacing lepton mass with external quark mass.\\  
For the topologies of Fig.~\ref{fig:dq_fcnc_a}-\ref{fig:dq_fcnc_c}, corresponding to double insertion of NP couplings the matching relations are presented in terms of effective couplings. 
%---------------------------------
\begin{align}
    &L^{u\gamma}= \frac{N_c e }{8 (4 \pi)^4 (m_H^2-m_t^2) }\Bigg[ e^2 \bigg\{m_t q_{u_i} y_t \bigg(\frac{f(z_t)}{z_t} - \frac{f(z_{t/H})}{z_{t/H}}  \bigg)\, \nn \\
    & 
    +\frac{g^2 v}{4 \sqrt{2}} \left[7 \left(\frac{f(z_t)}{z_t} - \frac{f(z_{t/H})}{z_{t/H}}  \right)\right. \, \nn \\
    &
    \left. - 10\left(\frac{g(z_t)}{z_t} - \frac{g(z_{t/H})}{z_{t/H}}  \right)\right] \bigg\}\mathrm{Im} \left[\left(\eta_R^{ut}\lambda_L^{tu\,*}\right) -\left(\lambda_R^{tu\,*}\eta_L^{ut}\right) \right]  \, \nn \\
    &
    + \frac{g^2}{2} \bigg\{ \frac{g_V^u m_t y_t q_{u_i}}{s_W c_W^2 } \left\{  \frac{m_t^2}{(m_t^2-m_z^2)}\left(\frac{f(z_t)}{z_t} \right)  \right. \, \nn \\
    & 
    \left. -\frac{m_W^2}{m_H^2-m_Z^2} \left(\frac{f(z_{t/H})}{z_{t/H}} \right)\right. \, \nn \\
    &
    +\left. \frac{m_Z^2}{(m_H^2-m_Z^2)(m_t^2-m_Z^2)}\left(\frac{f(z_{t/Z})}{z_{t/Z}} \right)\right\}\, \nn \\
    & 
    +\frac{g^2 v}{2\sqrt{2}} \left[  \frac{m_t^2}{(m_t^2-m_Z^2)}\left(4 \frac{f(z_t)}{z_t} - 21 \frac{g(z_t)}{z_t}\right) \right.\, \nn \\
    &\left. - \frac{m_H^2}{(m_H^2-m_Z^2)}\left(4 \frac{f(z_{t/H})}{z_{t/H}} - 21 \frac{g(z_{t/H})}{z_{t/H}}\right) \right.\, \nn \\
    &
    \left. +\frac{m_z^2}{(m_H^2-m_Z^2)}\left(4 \frac{f(z_{t/Z})}{z_{t/Z}} - 21 \frac{g(z_{t/Z})}{z_{t/Z}}\right) \right] \ \bigg\}, \nn \\
    &
    \mathrm{Im}\left[ \eta_R^{ut}\kappa_L^{tu\,*}- \eta_L^{ut}\kappa_R^{tu\,*} + X_L^{ut} \eta_L^{tu\,*}-X_R^{ut} \eta_R^{tu\, *} \right] \Bigg]
\end{align}
%---------------------------------
%-------------------------------
\begin{align}
    &L^{uG}= \frac{N_c g_s^3 m_t T_F }{8 (4 \pi)^4 (m_H^2-m_t^2) }\bigg[   \bigg(\frac{f(z_t)}{z_t} - \frac{f(z_{t/H})}{z_{t/H}}  \bigg)\, \nn \\
    & \hspace{0.8 cm }
    \times \mathrm{Im} \left[\left(\eta_R^{ut}\xi_L^{tu\,*}\right) -\left(\xi_R^{tu\,*}\eta_L^{ut}\right) \right]\bigg] 
\end{align}

\subsection{Logarithmic Enhancement Contribution:}\label{Append:Logarithmic Enhancement Contribution}
%---------------------------------------
Matching relations for diagrams shown in Fig.~\ref{fig:RGE_induced}
%--------------------------------
\begin{align} \label{Append_eq:lepEDM_log_enhance}
    &L^{e\gamma} = \frac{m_l e}{16 \pi^2} \mathrm{Im}\Bigg[  \left\{ \frac{3}{2} + \mathrm{ln}\frac{\mu^2}{m_H^2}- \left(2 s_W^2-\frac{1}{2}\right) \right.\, \nn \\
    & \left. \times \frac{2 m_Z^2}{m_Z^2-m_H^2} ~\mathrm{ln}\frac{m_H^2}{m_Z^2}\right\} C^{\phi \tilde{W}}+\left\{ \frac{9}{2} + 3 ~\mathrm{ln}\frac{\mu^2}{m_H^2} \right.\, \nn \\
    & \left. +\left(2 s_W^2-\frac{1}{2}\right) \frac{2 m_Z^2}{m_Z^2-m_H^2}~ \mathrm{ln}\frac{m_H^2}{m_Z^2}\right\} C^{\phi \tilde{B}} \, \nn \\
    &+ \left\{ \left(2 s_W^2-\frac{1}{2}\right)\frac{2 c_W^2-1}{c_W s_W} .\frac{m_Z^2}{m_Z^2-m_H^2}~\mathrm{ln}\frac{m_H^2}{m_Z^2} \right.\, \nn \\
    & \left. -\frac{4 s_W^2+11}{4 c_W s_W} -\frac{c_W}{s_W} \mathrm{ln} \frac{\mu^2}{m_W^2} - \frac{1+2s_W^2}{2 c_W s_W} ~\mathrm{ln}\frac{\mu^2}{m_H^2} \right\} C^{\phi \tilde{W}B} \Bigg]
\end{align}
%------------------------------------
\begin{subequations}
    \begin{align}
        &L^{u\gamma} = -\frac{m_{u_i}e}{32\pi^2}\mathrm{Im}\Bigg[\left\{ \frac{5}{2} + \frac{5}{3}~ \mathrm{ln}\frac{\mu^2}{m_H^2}+ \left( \frac{4 s_W^2}{3} -\frac{1}{2} \right) \right.\, \nn \\
        &
        \left. \times \frac{2 m_Z^2}{m_Z^2-m_H^2} ~\mathrm{ln}\frac{m_H^2}{m_Z^2}\right\} C^{\phi \tilde{B}}+\left\{ \frac{3}{2} + \mathrm{ln}\frac{\mu^2}{m_H^2} \right.\, \nn \\
        & 
        \left. -\left( \frac{4 s_W^2}{3} -\frac{1}{2} \right) \frac{2 m_Z^2}{m_Z^2-m_H^2} ~\mathrm{ln}\frac{m_H^2}{m_Z^2}\right\}C^{\phi \tilde{W}}\, \nn \\
        & 
        +\left\{ \frac{2 c_{2W}+3}{4 c_Ws_W} +\frac{2 s_W^2+3}{6s_W c_W} ~ \mathrm{ln}\frac{\mu^2}{m_H^2} -\frac{c_W}{s_W} \mathrm{ln}\frac{\mu^2}{m_W^2} \right. \, \nn \\
        & 
        \left. + \left(\frac{4 s_W^2}{3}-\frac{1}{2}\right)\frac{2 c_W^2-1}{c_W s_W} \frac{m_Z^2}{m_Z^2-m_H^2}~\mathrm{ln}\frac{m_H^2}{m_Z^2} \right\}C^{\phi \tilde{W}B} \Bigg] \, \\
        &L^{d\gamma} = \frac{m_{d_i}e}{32\pi^2}\mathrm{Im}\Bigg[\left\{ \frac{1}{2} + \frac{1}{3}~ \mathrm{ln}\frac{\mu^2}{m_H^2}-\left( \frac{1}{2}-\frac{2 s_W^2}{3}\right) \right.\, \nn \\
        &
        \left. \times \frac{2 m_Z^2}{m_Z^2-m_H^2} ~\mathrm{ln}\frac{m_H^2}{m_Z^2}\right\} C^{\phi \tilde{B}}+\left\{ \frac{3}{2} + \mathrm{ln}\frac{\mu^2}{m_H^2} \right.\, \nn \\
        & 
        \left. -\left( \frac{1}{2}-\frac{2 s_W^2}{3}  \right) \frac{2 m_Z^2}{m_Z^2-m_H^2} ~\mathrm{ln}\frac{m_H^2}{m_Z^2}\right\}C^{\phi \tilde{W}}\, \nn \\
        & 
        -\left\{ \frac{2 c_{2W}+3}{4 c_Ws_W} +\frac{3-2 s_W^2}{6s_W c_W} ~ \mathrm{ln}\frac{\mu^2}{m_H^2} +\frac{c_W}{s_W} \mathrm{ln}\frac{\mu^2}{m_W^2} \right. \, \nn \\
        & 
        \left. + \left(\frac{1}{2} -\frac{4 s_W^2}{3}\right)\frac{2 c_W^2-1}{c_W s_W} \frac{m_Z^2}{m_Z^2-m_H^2}~\mathrm{ln}\frac{m_H^2}{m_Z^2} \right\}C^{\phi \tilde{W}B} \Bigg]
    \end{align}
\end{subequations}
%-----------------------------------
\begin{align}\label{Append_eq:qEDM_log_enhance}
    & L^{q G}= \frac{m_qg_s^2(\mu)}{16 \pi^2}\bigg[ 3+2 ~\mathrm{ln}\frac{\mu^2}{m_H^2} \bigg] \times \mathrm{Im}\left( C^{\phi \tilde{G}} \right)
\end{align}
%-----------------------------------
Here, $q$ can be any light quark.

%------------------------------------
\section{Methodology for Loop-Level Matching}
\label{appendix:loop_methodology}
%------------------------------------
In this appendix, we describe the methodology adopted in our analysis to
evaluate the relevant loop-level matching contributions. We encounter
three types of loop topologies: one-loop contributions, two-loop
Barr-Zee diagrams, and two-loop diagrams generating the Weinberg
operator. The one-loop contributions involve relatively straightforward
loop integrals, which we evaluate using our in-house code based on the
Passarino--Veltman reduction formalism. We have cross-checked these
results against the publicly available package \texttt{Package-X}~\cite{Patel:2016fam}. The
two-loop topologies are more involved and require a dedicated treatment;
we therefore provide the detailed calculation and methodology for these
contributions in the following subsections.
%-----------------------------------
\subsection{Barr-Zee Corrections}
\label{appendix:barr_zee}
%-----------------------------------

\begin{figure}[h!]
    \centering
    \begin{tikzpicture}
        \begin{feynman}
            % Main loop radius
            \def\r{1.0 cm} 
            \coordinate (O) at (0,0);
            
            % Outer circle (without the messy overlapping arrow)
            \draw[very thick] (O) circle (\r);
            
            % --- NEW: Inner clockwise arrow with 't' label ---
            % Draws an arc from 150 degrees to 30 degrees at radius 0.75cm
            \draw[->, thick] (150:0.75cm) arc (150:30:0.75cm) node[midway, below] {\(k\)};
            
            \vertex (A) at (90:\r);  
            \vertex [violet, square dot](B) at (210:\r) {};   
            \vertex [](C) at (330:\r);   
            
            % Extended leg distances proportionally
            \vertex[above=1.3 of A](F){\(\gamma\)};
            \vertex[below left=1.5cm and 0.8cm of B](G);
            \vertex[below right=1.5cm and 0.8cm of C](H);
            \vertex[left=0.8 cm of G](b){\(f\)};
            \vertex[right=0.8 cm of H](d){\(f\)};
            
            \diagram*{
                (A) --[very thick, boson, momentum=\(q_1\)](F),
                (G) --[violet, very thick, scalar, momentum=\(q_1+q_2\),edge label'=\(H\)](B),
                (C) --[very thick, boson, edge label'=\(\ga\), momentum=\(q_2\)](H),
                (b) --[very thick, fermion](G) --[very thick, fermion, edge label=\(f\)](H) --[very thick, fermion](d),
            };
        \end{feynman}
    \end{tikzpicture}
    \caption{Two-loop Barr-Zee topology contributing to the fermionic EDM.}
    \label{fig:Barr_Zee_ttH}
\end{figure}
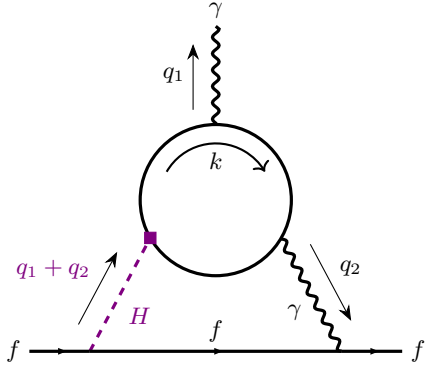
%---------------------------------
Although the Barr-Zee contribution is a genuine two-loop topology, its
calculation can be conveniently organized as two sequential one-loop
steps. We first evaluate the one-loop subdiagram associated with the
inner loop and express its result as an effective form factor. This
effective subdiagram is then inserted into the remaining one-loop
diagram, from which the complete two-loop contribution is obtained. 

For example, consider the subdiagram with an $H\gamma\gamma$
structure. After evaluating the amplitude of the inner subdiagram, its
Lorentz structure can be decomposed as
\begin{align}
    \Gamma^{\mu\nu}
    =
    A_{21}\left(q_2^\mu q_1^\nu-q_1\!\cdot\!q_2\,g^{\mu\nu}\right)
    +A_3\,i\epsilon^{\mu\nu\alpha\beta}
    q_{1\alpha}q_{2\beta},
\end{align}
where $q_1$ and $q_2$ denote the momenta of onshell and offshell photon
% of the two internal gauge bosons, $H$ and $\gamma$, 
respectively, as shown in
Fig.~\ref{fig:Barr_Zee_ttH}. The quantities $A_{21}$ and $A_3$ are the
corresponding form factors. After Feynman parametrization, followed by
Wick rotation and loop integration, the resulting expression takes the
form,
%--------------------------
\begin{align}\label{eq:BZ_inner_loop}
\mathcal{I}_{\rm inner}
=
\int_0^1 dx\,
\frac{
\begin{aligned}
&A_{21}(x)\left(q_2^\mu q_1^\nu
-q_1\!\cdot\!q_2\,g^{\mu\nu}\right) \\
&\quad
+A_3(x)\,i\epsilon^{\mu\nu\alpha\beta}
q_{1\alpha}q_{2\beta}
\end{aligned}
}{
q_2^2-M(x)^2
}.
\end{align}
%--------------------------
where $M(x)$ is defined in terms of the mass of the inner-loop particle,
$M(x)^2=m_t^2/[x(1-x)]$. It is clear from
Eq.~\eqref{eq:BZ_inner_loop} that the integral is free of the loop
momentum $k$, and its denominator takes a propagator-like form with a
modified mass $M(x)$.

The result obtained for the inner subdiagram can now be embedded into
the full Barr-Zee topology, after which the remaining loop integration
can be performed following the same procedure. Combining the propagator
denominators using Feynman parametrization, the three propagators of the
outer loop are combined with the effective propagator-like structure
arising from the inner subdiagram. The resulting parameter integrals
then take the standard form of the Barr-Zee loop functions, which can be
expressed in terms of the functions $f(z)$ and $g(z)$. It is worth noting that this standard
parametric form is obtained only after including the corresponding
topology with the fermion current flowing in the opposite (anticlockwise)
direction in the inner subloop.
  
The loop functions $f(z)$ and $g(z)$ are defined via the parametric integrals:
\begin{subequations}\label{eq:loop_functions_integrals}
\begin{align}
    f(z) &= \frac{z}{2}\int_0^1 dx\, \frac{1-2x(1-x)}{x(1-x)-z}\ln\left(\frac{x(1-x)}{z}\right), \\
    g(z) &= \frac{z}{2}\int_0^1 dx\, \frac{1}{x(1-x)-z}\ln\left(\frac{x(1-x)}{z}\right).
\end{align}
\end{subequations}

Here, the function $g(z)$ is not globally analytic, possessing a branch point at $z=1/4$. Above this threshold ($z > 1/4$), $g(z)$ can be expressed in closed analytic form in terms of the Clausen function ($\mathrm{Cl}_2(\theta)$)~\cite{clausen1832}. The explicit evaluation yields:
\begin{subequations}\label{eq:g_analytic_and_asymptotic}
\begin{align}
    g(z) &= \frac{2 z}{\sqrt{4 z-1}}\,\mathrm{Cl}_2\left(2 \sin^{-1}\left(\frac{1}{2\sqrt{z}}\right)\right), \quad \text{for } z > \frac{1}{4}\,,
\end{align}
whereas its limiting behavior in the asymptotic regimes is given by:
\begin{align}
    g(z) &\simeq 
    \begin{cases} 
        \frac{z}{2} \ln^2(z) & \text{for } z \ll 1 \,, \\[1.5ex]
        1 + \frac{1}{2} \ln(z) & \text{for } z \gg 1 \,.
    \end{cases}
\end{align}
\end{subequations}
%---------------------------

%-----------------------------------
\subsection{Weinberg-Operator Corrections}
\label{appendix:weinberg_corrections}
%-------------------------------------
\begin{figure}[h!]
    \centering
\subfloat[]{
		\begin{tikzpicture}
			\begin{feynman}
				\def\r{1.5 cm}
				\coordinate (O) at (0,0);
				\draw[very thick] (O) circle (\r);
                \draw[->, thick] (150:1.2cm) arc (150:30:1.2cm) node[midway, below] {\(t\)};
                
				\draw[fermion, very thick] (360:\r) arc (360:180:\r);
				\vertex (A) at (90:\r)  ;   
				\vertex (B) at (210:\r) ;   
				\vertex (C) at (330:\r) ;   
				\vertex (D) at (0:\r)   {}; 
				\vertex (E) at (180:\r) {};
				\vertex[above=1.0 of A](F){\(g~~\mu_3\)};
				\vertex[below left=1.0cm of B](G){\(g~~\mu_1\)};
				\vertex[below right=1.0cm of C](H){\(g~~\mu_2\)};
				\vertex[below=1.5cm of O](a1) ;
				\vertex[above right=1.5 cm of O](a2) ;
				\vertex[square dot, violet, at=(D)] ;
				\vertex[square dot, violet, at=(E)] {};
				\diagram*{
					(A) --[very thick, gluon, momentum=\(p_3\)](F),
					(G) --[very thick, gluon, momentum=\(p_1\)](B),
					(H) --[very thick, gluon, momentum=\(p_2\)](C),
					(D) --[scalar, very thick, violet,  momentum=\(k_1+k_2\)](E),
				};
			\end{feynman}
		\end{tikzpicture}
	}
    \caption{Diagram contributing to the Weinberg operator.}
    \label{fig:Weinberg_top_Higgs}
\end{figure}
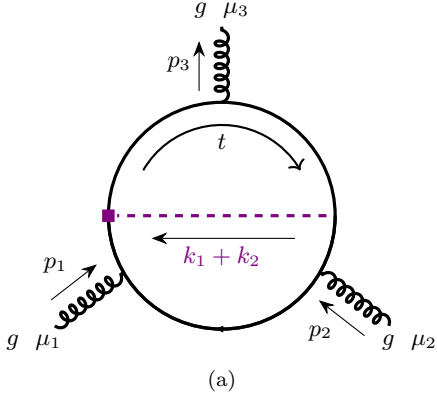
%-----------------------
Unlike the Barr-Zee contribution, the two-loop topology considered here cannot be factorized into two successive one-loop subdiagrams. The two loop momenta are correlated through the common propagator structure, as illustrated in Fig.~\ref{fig:Weinberg_top_Higgs}. We therefore treat this contribution as a genuine two-loop matching calculation. Expanding the full two-loop amplitude in the external gluon momenta, the matching coefficient can be extracted by isolating the appropriate CP-violating three-gluon Lorentz structure. The resulting effective amplitude can be written as
%-------------------------
\begin{align}
 &i\mathcal{M}_{\rm eff} =-\frac{2}{3} f^{abc} w ~\varepsilon^{\mu_1}(p_1) \varepsilon^{\mu_2}(p_2) \varepsilon^{\mu_3}(-p_1-p_2)\times \nn\\
 &\Big[(p_1-p_2)_{\mu_3}\varepsilon_{\mu_1\mu_2\sigma\rho}+2\left(p_{1\,\mu_2}\varepsilon_{\mu_1\mu_3\sigma\rho}+p_{2\,\mu_1}\varepsilon_{\mu_2\mu_3\sigma\rho}\right)\Big]p_1^{\sigma}p_2^{\rho}\,.  
\end{align}
%-------------------------
To simplify the calculation, it is convenient to select a single linearly independent $\mathrm{CP}$-violating Levi-Civita structure. We choose the routing of the internal loop momenta such that only three propagators in the denominators depend on the external momenta $p_1$ and $p_2$. Since these momenta are small compared to the heavy mass scales in the loop, we neglect terms of higher order in $p_i^2/M^2$, with $M\in\{m_t,M_H\}$.
%-------------------------
\begin{align}
    \frac{1}{(k_i+p)^2-M^2}=\frac{1}{k_i^2-M^2} &\Bigg[1-\frac{p^2+2(p.k_i)}{k_i^2-M^2}+\frac{4(p.k_i)^2}{(k_i^2-M^2)}\Bigg]\nn\\
    +\mathcal{O}\left(\frac{p^4}{M^4}\right)\,,
\end{align}
%-------------------------
Once the denominators are free of the external momenta $p_i$, odd-rank vacuum tensor integrals vanish by Lorentz symmetry. We isolate the desired $\mathrm{CP}$-violating structure by contracting the two loop amplitude $\mathcal{A}_{\mu_1\mu_2\mu_3}$ with
\begin{equation}
T^{\mu_1\mu_2\mu_3}
=
\epsilon^{\mu_1\mu_2\rho\sigma}
p_{1\rho}p_{2\sigma}(p_1-p_2)^{\mu_3}.
\end{equation}
The corresponding WC is then
\begin{equation}
w=\frac{T^{\mu_1\mu_2\mu_3}\mathcal{A}_{\mu_1\mu_2\mu_3}}{\mathcal{N}},
\quad
\mathcal{N}
=
T^{\mu_1\mu_2\mu_3}W_{\mu_1\mu_2\mu_3}
=
-4(p_1\!\cdot p_2)^3.
\end{equation}
The projected numerator is subsequently reduced to scalar polynomials using standard vacuum tensor reduction, separating the loop- and external-momentum scalar products, $(k_i\!\cdot k_j)$ and $(p_i\!\cdot p_j)$, respectively.

%----------------------------
\vspace{0.5em}
\textbullet~\textit{\textbf{Tensor Reductions}:}
\vspace{0.5em}
%----------------------------
The tensor reduction required in our calculation involves vacuum tensor
integrals up to rank six. We denote by $T_{mn}^{\mu_1\cdots\mu_{m+n}}$
a tensor containing $m$ powers of the loop momentum $k$ and $n$ powers
of the second loop momentum $q$, with all Lorentz indices left open.
The required tensor structures are
%-------------------------
\begin{subequations}
\label{eq:vacuum_tensor_reduction}
\begin{align}
T_{60}^{\mu\nu\alpha\beta\gamma\delta}
&=
\frac{(k^2)^3}{d(d+2)(d+4)}
\,\mathcal{S}^{(6)\,\mu\nu\alpha\beta\gamma\delta},
\\[1mm]
T_{51}^{\mu\nu\alpha\beta\gamma\delta}
&=
\frac{(k^2)^2(k\!\cdot q)}
{d(d+2)(d+4)}
\,\mathcal{S}^{(6)\,\mu\nu\alpha\beta\gamma\delta},
\\[1mm]
T_{42}^{\mu\nu\alpha\beta\gamma\delta}
&=
\frac{k^2\!\left[(d+3)k^2q^2-4(k\!\cdot q)^2\right]}
{d(d-1)(d+2)(d+4)}
\,\mathcal{S}_{42}^{(1)\,\mu\nu\alpha\beta\gamma\delta}
\nonumber\\
&\quad+
\frac{k^2\!\left[d(k\!\cdot q)^2-k^2q^2\right]}
{d(d-1)(d+2)(d+4)}
\,\mathcal{S}_{42}^{(2)\,\mu\nu\alpha\beta\gamma\delta},
\\[1mm]
T_{33}^{\mu\nu\alpha\beta\gamma\delta}
&=
\frac{(k\!\cdot q)
\left[(d+1)k^2q^2-2(k\!\cdot q)^2\right]}
{d(d-1)(d+2)(d+4)}
\,\mathcal{S}_{33}^{(1)\,\mu\nu\alpha\beta\gamma\delta}
\nonumber\\
&\quad+
\frac{(k\!\cdot q)
\left[(d+2)(k\!\cdot q)^2-3k^2q^2\right]}
{d(d-1)(d+2)(d+4)}
\,\mathcal{S}_{33}^{(2)\,\mu\nu\alpha\beta\gamma\delta},
\end{align}

\begin{align}
T_{40}^{\mu\nu\alpha\beta}
&=
\frac{(k^2)^2}{d(d+2)}
\,\mathcal{S}^{(4)\,\mu\nu\alpha\beta},
\\[1mm]
T_{31}^{\mu\nu\alpha\beta}
&=
\frac{k^2(k\!\cdot q)}{d(d+2)}
\,\mathcal{S}^{(4)\,\mu\nu\alpha\beta},
\\[1mm]
T_{22}^{\mu\nu\alpha\beta}
&=
\frac{(d+1)k^2q^2-2(k\!\cdot q)^2}
{d(d-1)(d+2)}
\,g^{\mu\nu}g^{\alpha\beta}
\nonumber\\
&\quad+
\frac{d(k\!\cdot q)^2-k^2q^2}
{d(d-1)(d+2)}
\left(
g^{\mu\alpha}g^{\nu\beta}
+
g^{\mu\beta}g^{\nu\alpha}
\right),
\end{align}

\begin{align}
T_{30}^{\mu\nu\alpha}&=0,
\qquad
T_{21}^{\mu\nu\alpha}=0,
\\[1mm]
T_{20}^{\mu\nu}
&=
\frac{k^2}{d}\,g^{\mu\nu},
\qquad
T_{11}^{\mu\nu}
=
\frac{k\!\cdot q}{d}\,g^{\mu\nu}.
\end{align}
\end{subequations}
%-----------------------------
Here, $\mathcal{S}^{(4)}$ and $\mathcal{S}^{(6)}$ denote the completely
symmetrized sums over all inequivalent pairings of the corresponding
Lorentz indices,
%----------------------------
\begin{subequations}
\begin{align}
\mathcal{S}^{(4)\,\mu\nu\alpha\beta}
&=
g^{\mu\nu}g^{\alpha\beta}
+g^{\mu\alpha}g^{\nu\beta}
+g^{\mu\beta}g^{\nu\alpha},
\end{align}
%---------------------------
\begin{align}
\mathcal{S}^{(6)\,\mu\nu\alpha\beta\gamma\delta}
={}&
g^{\mu\nu}g^{\alpha\beta}g^{\gamma\delta}
+g^{\mu\nu}g^{\alpha\gamma}g^{\beta\delta}
+g^{\mu\nu}g^{\alpha\delta}g^{\beta\gamma}
\nonumber\\
&+g^{\mu\alpha}g^{\nu\beta}g^{\gamma\delta}
+g^{\mu\alpha}g^{\nu\gamma}g^{\beta\delta}
+g^{\mu\alpha}g^{\nu\delta}g^{\beta\gamma}
\nonumber\\
&+g^{\mu\beta}g^{\nu\alpha}g^{\gamma\delta}
+g^{\mu\beta}g^{\nu\gamma}g^{\alpha\delta}
+g^{\mu\beta}g^{\nu\delta}g^{\alpha\gamma}
\nonumber\\
&+g^{\mu\gamma}g^{\nu\alpha}g^{\beta\delta}
+g^{\mu\gamma}g^{\nu\beta}g^{\alpha\delta}
+g^{\mu\gamma}g^{\nu\delta}g^{\alpha\beta}
\nonumber\\
&+g^{\mu\delta}g^{\nu\alpha}g^{\beta\gamma}
+g^{\mu\delta}g^{\nu\beta}g^{\alpha\gamma}
+g^{\mu\delta}g^{\nu\gamma}g^{\alpha\beta}.
\end{align}
%---------------------------
\begin{align}
\mathcal{S}_{42}^{(1)\,\mu\nu\alpha\beta\gamma\delta}
&=
\left(
g^{\mu\nu}g^{\alpha\beta}
+g^{\mu\alpha}g^{\nu\beta}
+g^{\mu\beta}g^{\nu\alpha}
\right)g^{\gamma\delta},
\\[1mm]
\mathcal{S}_{42}^{(2)\,\mu\nu\alpha\beta\gamma\delta}
&=
g^{\gamma\mu}g^{\delta\nu}g^{\alpha\beta}
+g^{\gamma\mu}g^{\delta\alpha}g^{\nu\beta}
+g^{\gamma\mu}g^{\delta\beta}g^{\nu\alpha}
\nonumber\\
&\quad
+g^{\gamma\nu}g^{\delta\mu}g^{\alpha\beta}
+g^{\gamma\nu}g^{\delta\alpha}g^{\mu\beta}
+g^{\gamma\nu}g^{\delta\beta}g^{\mu\alpha}
\nonumber\\
&\quad
+g^{\gamma\alpha}g^{\delta\mu}g^{\nu\beta}
+g^{\gamma\alpha}g^{\delta\nu}g^{\mu\beta}
+g^{\gamma\alpha}g^{\delta\beta}g^{\mu\nu}
\nonumber\\
&\quad
+g^{\gamma\beta}g^{\delta\mu}g^{\nu\alpha}
+g^{\gamma\beta}g^{\delta\nu}g^{\mu\alpha}
+g^{\gamma\beta}g^{\delta\alpha}g^{\mu\nu},
\\[2mm]
\mathcal{S}_{33}^{(1)\,\mu\nu\alpha\beta\gamma\delta}
&=
g^{\mu\nu}g^{\beta\gamma}g^{\alpha\delta}
+g^{\mu\nu}g^{\beta\delta}g^{\alpha\gamma}
+g^{\mu\nu}g^{\gamma\delta}g^{\alpha\beta}
\nonumber\\
&\quad
+g^{\mu\alpha}g^{\beta\gamma}g^{\nu\delta}
+g^{\mu\alpha}g^{\beta\delta}g^{\nu\gamma}
+g^{\mu\alpha}g^{\gamma\delta}g^{\nu\beta}
\nonumber\\
&\quad
+g^{\nu\alpha}g^{\beta\gamma}g^{\mu\delta}
+g^{\nu\alpha}g^{\beta\delta}g^{\mu\gamma}
+g^{\nu\alpha}g^{\gamma\delta}g^{\mu\beta},
\\[2mm]
\mathcal{S}_{33}^{(2)\,\mu\nu\alpha\beta\gamma\delta}
&=
g^{\mu\beta}g^{\nu\gamma}g^{\alpha\delta}
+g^{\mu\beta}g^{\nu\delta}g^{\alpha\gamma}
+g^{\mu\gamma}g^{\nu\beta}g^{\alpha\delta}
\nonumber\\
&\quad
+g^{\mu\gamma}g^{\nu\delta}g^{\alpha\beta}
+g^{\mu\delta}g^{\nu\beta}g^{\alpha\gamma}
+g^{\mu\delta}g^{\nu\gamma}g^{\alpha\beta}.
\end{align}
\end{subequations}
%---------------------------------------
\vspace{0.5em}
\textbullet~\textit{\textbf{Feynman Parametrization}:}
\vspace{0.5em}
%----------------------------
The resulting tensor-reduced integrals can be written in terms of the
following scalar vacuum integrals:
%----------------------------
\begin{align}
 &\mathcal{W}_{\{00;10;01;11\}}(\alpha,\beta,\gamma;a,b,c)\equiv  \int \frac{d^D k_1}{(2\pi)^D}\int \frac{d^D k_2}{(2\pi)^D}  \,,\nn\\
 & \times \frac{\{1;~k_1\cdot k_1;~k_2\cdot k_2;~ k_1 \cdot k_2\}}{\left(k_1^2-a^2\right)^{\alpha} \left(k_2^2-b^2\right)^{\beta} \left((k_1+k_2)^2-c^2\right)^{\gamma}}
\end{align}
We further eliminate the scalar products of the loop momenta in favour
of the propagator denominators,
%----------------------------
\begin{align}
    k_1\cdot k_1&=\mathcal{P}_1+a^2\,, \,\, k_2\cdot k_2=\mathcal{P}_2+b^2\,,\nn\\
    k_1\cdot k_2&=\frac{1}{2}\left(\mathcal{P}_3-\mathcal{P}_1-\mathcal{P}_2+c^2-a^2-b^2\right)\,.
\end{align}
%----------------------------
Finally, all contributions are reduced to the standard scalar
two-loop vacuum integral
%---------------------------
\begin{align}
    \mathcal{W}(\alpha,\beta,\ga)=\int \frac{d^D k_1}{(2\pi)^D}\int \frac{d^D k_2}{(2\pi)^D} \frac{1}{\mathcal{P}_1^{\alpha}\mathcal{P}_2^{\beta}\mathcal{P}_3^{\ga}}
\end{align}
%--------------------------
We further re-express the $\mathcal{W}$ functions using the standard
Feynman parametrisation, combining two denominators at a time. Solving
the resulting master integrals yields the desired analytic form of the
Weinberg loop function $h(r)$~\cite{PhysRevLett.63.2333}.
%-------------------------
\begin{widetext}
\begin{align}
    \mathcal{W}(\al,\be,\ga,a,b,c) &=\frac{(-1)^{D/2-\al-\ga}}{(4\pi)^D} \frac{\Gamma (\al+\be+\ga-D)}{\Gamma(\al) \Gamma(\be)\Gamma(\ga)}
    \times \int_0^1 dx\int_0^1 dy~ x^{D/2-\ga-1}(1-x)^{\be+1-D/2} y^{\al-1}\,,\nn\\
    &\times (1-y)^{\be+\ga-D/2-1}\left(a^2 y+(1-y)\frac{b^2 x+(1-x)c^2}{x(1-x)}\right)^{D-\al-\be-\ga}\,.
\end{align}
\end{widetext}
%-------------------------
We adopt the master integrals from Ref.~\cite{Adams:2015gva, Davydychev:1992mt}, in terms of which all
the resulting integrals can be reduced to four independent master
integrals, $\mathcal{W}(1,1,1)$, $\mathcal{W}(1,0,1)$,
$\mathcal{W}(1,1,0)$, and $\mathcal{W}(0,1,1)$. It is important to note
that each diagram admits three equivalent permutations, obtained by
rotating the internal scalar propagator by $120^\circ$. Including these
permutations is crucial for obtaining a finite result, as it removes the
divergences arising in the individual parametric integrals.

%--------------------------
\newpage
\bibliography{biblio}% Produces the bibliography via BibTeX.

\end{document}